\documentclass[aps,preprint,tightenlines,superscriptaddress,byrevtex,nofootinbib]{revtex4}
\usepackage{ifthen} 
\usepackage{etoolbox} 
\usepackage{forarray} 

\newboolean{pdflatex}
\setboolean{pdflatex}{true} 

\newboolean{articletitles}
\setboolean{articletitles}{true} 

\newboolean{uprightparticles}
\setboolean{uprightparticles}{false} 

\newboolean{inbibliography}
\setboolean{inbibliography}{false} 

\usepackage{tikz}
\usetikzlibrary{arrows}
\usetikzlibrary{calc}
\usetikzlibrary{chains}
\usetikzlibrary{circuits.logic.IEC,circuits.ee.IEC}
\usetikzlibrary{decorations.pathmorphing}   
\usetikzlibrary{decorations.markings}
\usetikzlibrary{fit}
\usetikzlibrary{positioning}
\usetikzlibrary{shadows}
\usetikzlibrary{shapes}
\usetikzlibrary{shapes.geometric}
\usetikzlibrary{through}
\usetikzlibrary{trees}
\usetikzlibrary{mindmap}
\usetikzlibrary{patterns,fadings}
\usetikzlibrary{intersections}
\usetikzlibrary{plotmarks}

\usepackage{pgfplots}
\pgfplotsset{compat=1.13}

\pgfdeclarelayer{background}
\pgfdeclarelayer{foreground}
\pgfsetlayers{background,main,foreground}

\usepackage{adjustbox}
\usepackage{graphicx} 
\usepackage[abs]{overpic}

\usepackage{microtype}
\usepackage{placeins} 
\usepackage{lineno}  
\usepackage{csquotes} 
\usepackage{xspace} 
\usepackage{scalefnt}

\usepackage{xcolor}
\definecolor{nice_green}{rgb}{0.08,0.72,0.08}
\definecolor{nice_yellow}{rgb}{1,0.56,0}
\definecolor{nice_red}{rgb}{0.72,0.08,0.08}
\definecolor{nice_blue}{rgb}{0.08,0.08,0.72}
\definecolor{nice_purple}{rgb}{0.72,0.08,0.72}
\definecolor{nice_turquoise}{rgb}{0.08,0.72,0.72}
\usepackage{color}
\graphicspath{{./figs/}} 

\usepackage{siunitx}
\usepackage{amsmath} 
\usepackage{amssymb}
\usepackage{amsfonts}
\usepackage{amsbsy}
\usepackage{upgreek} 
\usepackage{fancyvrb} 
\usepackage{xfrac}

\usepackage{booktabs}
\usepackage{longtable}
\usepackage{multirow}
\usepackage{lscape}

\usepackage{hyperref}    
\usepackage[all]{hypcap} 
\usepackage{orcidlink}

\usepackage{cleveref}
\crefname{chapter}{Ch.\@}{Chs.\@}
\crefname{section}{Sec.\@}{Secs.\@}
\crefname{subsection}{Sec.\@}{Secs.\@}
\crefname{appendix}{Appendix\@}{Appendices\@}
\crefname{figure}{Fig.\@}{Figs.\@}
\crefname{table}{Table\@}{Tables\@}
\crefname{equation}{Eq.\@}{Eqs.\@}

\def\babar  {\mbox{BABAR}\xspace}
\def\belle  {\mbox{Belle}\xspace}

\ifthenelse{\boolean{uprightparticles}}%
{

 \def\Pmu         {\ensuremath{\upmu}\xspace}                 
 \def\Pnu         {\ensuremath{\upnu}\xspace}                 
                  
 \def\Ppi         {\ensuremath{\uppi}\xspace}

 \def\Ptau        {\ensuremath{\uptau}\xspace}

 \def\Ppsi        {\ensuremath{\uppsi}\xspace}

 \def\PDelta      {\ensuremath{\Delta}\xspace}                 
 \def\PXi      {\ensuremath{\Xi}\xspace}                 
 \def\PLambda      {\ensuremath{\Lambda}\xspace}                 
 \def\PSigma      {\ensuremath{\Sigma}\xspace}                 
 \def\POmega      {\ensuremath{\Omega}\xspace}                 
 \def\PUpsilon      {\ensuremath{\Upsilon}\xspace}

 \def\PB      {\ensuremath{\mathrm{B}}\xspace}                 
                  
 \def\PD      {\ensuremath{\mathrm{D}}\xspace}

 \def\PJ      {\ensuremath{\mathrm{J}}\xspace}                 
 \def\PK      {\ensuremath{\mathrm{K}}\xspace}

 \def\Pb      {\ensuremath{\mathrm{b}}\xspace}                 
 \def\Pc      {\ensuremath{\mathrm{c}}\xspace}                 
                  
 \def\Pe      {\ensuremath{\mathrm{e}}\xspace}

 \def\Pq      {\ensuremath{\mathrm{q}}\xspace}                 
                  
 \def\Ps      {\ensuremath{\mathrm{s}}\xspace}                 
                  
 \def\Pu      {\ensuremath{\mathrm{u}}\xspace}

}
{

 \def\Pmu         {\ensuremath{\mu}\xspace}                 
 \def\Pnu         {\ensuremath{\nu}\xspace}                 
                  
 \def\Ppi         {\ensuremath{\pi}\xspace}

 \def\Ptau        {\ensuremath{\tau}\xspace}

 \def\Ppsi        {\ensuremath{\psi}\xspace}                 
                  
 \mathchardef\PDelta="7101
 \mathchardef\PXi="7104
 \mathchardef\PLambda="7103
 \mathchardef\PSigma="7106
 \mathchardef\POmega="710A
 \mathchardef\PUpsilon="7107
                  
 \def\PB      {\ensuremath{B}\xspace}                 
                  
 \def\PD      {\ensuremath{D}\xspace}

 \def\PJ      {\ensuremath{J}\xspace}                 
 \def\PK      {\ensuremath{K}\xspace}

 \def\Pb      {\ensuremath{b}\xspace}                 
 \def\Pc      {\ensuremath{c}\xspace}                 
                  
 \def\Pe      {\ensuremath{e}\xspace}

 \def\Pq      {\ensuremath{q}\xspace}                 
                  
 \def\Ps      {\ensuremath{s}\xspace}                 
                  
 \def\Pu      {\ensuremath{u}\xspace}

}

\def\electron   {{\ensuremath{\Pe}}\xspace}
\def\ep         {{\ensuremath{\Pe^+}}\xspace}
 
\def\epem       {{\ensuremath{\Pe^+\Pe^-}}\xspace}

\def\muon       {{\ensuremath{\Pmu}}\xspace}
\def\mup        {{\ensuremath{\Pmu^+}}\xspace}
\def\mun        {{\ensuremath{\Pmu^-}}\xspace} 

\def\taup       {{\ensuremath{\Ptau^+}}\xspace}
\def\taum       {{\ensuremath{\Ptau^-}}\xspace}

\def\lepton     {{\ensuremath{\ell}}\xspace}
\def\ellm       {{\ensuremath{\ell^-}}\xspace}
\def\ellp       {{\ensuremath{\ell^+}}\xspace}
\def\ellell     {\ensuremath{\ell^+ \ell^-}\xspace}

\def\neu        {{\ensuremath{\Pnu}}\xspace}
\def\neub       {{\ensuremath{\overline{\Pnu}}}\xspace}
\def\neue       {{\ensuremath{\neu_e}}\xspace}

\def\neum       {{\ensuremath{\neu_\mu}}\xspace}
\def\neumb      {{\ensuremath{\neub_\mu}}\xspace}
\def\neut       {{\ensuremath{\neu_\tau}}\xspace}
\def\neutb      {{\ensuremath{\neub_\tau}}\xspace}
\def\neul       {{\ensuremath{\neu_\ell}}\xspace}
\def\neulb      {{\ensuremath{\neub_\ell}}\xspace}

\def\quark     {{\ensuremath{\Pq}}\xspace}
\def\quarkbar  {{\ensuremath{\overline \quark}}\xspace}
\def\qqbar     {{\ensuremath{\quark\quarkbar}}\xspace}
\def\uquark    {{\ensuremath{\Pu}}\xspace}

\def\squark    {{\ensuremath{\Ps}}\xspace}

\def\cquark    {{\ensuremath{\Pc}}\xspace}

\def\bquark    {{\ensuremath{\Pb}}\xspace}

\def\pion   {{\ensuremath{\Ppi}}\xspace}
\def\piz    {{\ensuremath{\pion^0}}\xspace}

\def\pip    {{\ensuremath{\pion^+}}\xspace}
\def\pim    {{\ensuremath{\pion^-}}\xspace}
\def\pipi  {\ensuremath{\pion^+\pion^-}\xspace}

\def\kaon    {{\ensuremath{\PK}}\xspace}
\def\Kbar    {{\kern 0.2em\overline{\kern -0.2em \PK}{}}\xspace}

\def\Kp      {{\ensuremath{\kaon^+}}\xspace}
\def\Km      {{\ensuremath{\kaon^-}}\xspace}

\def\KpKm    {{\ensuremath{\Kp \kern -0.16em \Km}}\xspace}
\def\KS      {{\ensuremath{\kaon^0_{\mathrm{ \scriptscriptstyle S}}}}\xspace}
\def\KL      {{\ensuremath{\kaon^0_{\mathrm{ \scriptscriptstyle L}}}}\xspace}

\def\Dbar    {{\kern 0.2em\overline{\kern -0.2em \PD}{}}\xspace}
\def\D       {{\ensuremath{\PD}}\xspace}

\def\Dz      {{\ensuremath{\D^0}}\xspace}
\def\Dzb     {{\ensuremath{\Dbar{}^0}}\xspace}
\def\Dp      {{\ensuremath{\D^+}}\xspace}
\def\Dm      {{\ensuremath{\D^-}}\xspace}

\def\Dstar   {{\ensuremath{\D^*}}\xspace}
\def\Dstarb  {{\ensuremath{\Dbar{}^*}}\xspace}
\def\Dstarz  {{\ensuremath{\D^{*0}}}\xspace}
\def\Dstarzb {{\ensuremath{\Dbar{}^{*0}}}\xspace}
\def\Dstarp  {{\ensuremath{\D^{*+}}}\xspace}
\def\Dstarm  {{\ensuremath{\D^{*-}}}\xspace}

\def\Dstst   {{\ensuremath{\D^{**}}}\xspace}
\def\Dststb  {{\ensuremath{\Dbar{}^{**}}}\xspace}
\def\Dststzb {{\ensuremath{\Dbar{}^{**0}}}\xspace}
\def\Dststm  {{\ensuremath{\D^{**-}}}\xspace}

\def\B       {{\ensuremath{\PB}}\xspace}
\def\Bbar    {{\ensuremath{\kern 0.18em\overline{\kern -0.18em \PB}{}}}\xspace}

\def\BBbar   {{\ensuremath{\PB\Bbar}}\xspace}

\def\Bzb     {{\ensuremath{\Bbar{}^0}}\xspace}
\def\Bu      {{\ensuremath{\B^+}}\xspace}

\def\Bd      {{\ensuremath{\B^0}}\xspace}

\def\jpsi     {{\ensuremath{{\PJ\mskip -3mu/\mskip -2mu\Ppsi\mskip 2mu}}}\xspace}

  \def\Y#1S{\ensuremath{\PUpsilon{(#1S)}}\xspace}

\def\FourS {{\Y4S}}

\def\Lbar        {{\ensuremath{\kern 0.1em\overline{\kern -0.1em\PLambda}}}\xspace}

\def\BF         {{\ensuremath{\mathcal{B}}}\xspace}

\def\BR         {\BF}
\newcommand{\decay}[2]{\ensuremath{#1\!\to #2}\xspace}         

\def\to                 {\ensuremath{\rightarrow}\xspace}

\def\qsq       {{\ensuremath{q^2}}\xspace}

\def\Vub   {{\ensuremath{V_{\uquark\bquark}^{}}}\xspace}
\def\Vcb   {{\ensuremath{V_{\cquark\bquark}^{}}}\xspace}

\def\AT#1     {\ensuremath{A_{\mathrm{T}}^{#1}}\xspace}           

\def\C#1      {\ensuremath{\mathcal{C}_{#1}}\xspace}                       
\def\Cp#1     {\ensuremath{\mathcal{C}_{#1}^{'}}\xspace}                    
\def\Ceff#1   {\ensuremath{\mathcal{C}_{#1}^{\mathrm{(eff)}}}\xspace}        
\def\Cpeff#1  {\ensuremath{\mathcal{C}_{#1}^{'\mathrm{(eff)}}}\xspace}       
\def\Ope#1    {\ensuremath{\mathcal{O}_{#1}}\xspace}                       
\def\Opep#1   {\ensuremath{\mathcal{O}_{#1}^{'}}\xspace}                    

\newcommand{\tev}{\ensuremath{\mathrm{\,Te\kern -0.1em V}}\xspace}
\newcommand{\gev}{\ensuremath{\mathrm{\,Ge\kern -0.1em V}}\xspace}
\newcommand{\mev}{\ensuremath{\mathrm{\,Me\kern -0.1em V}}\xspace}
\newcommand{\kev}{\ensuremath{\mathrm{\,ke\kern -0.1em V}}\xspace}
\newcommand{\ev}{\ensuremath{\mathrm{\,e\kern -0.1em V}}\xspace}
\newcommand{\gevc}{\ensuremath{{\mathrm{\,Ge\kern -0.1em V\!/}c}}\xspace}
\newcommand{\mevc}{\ensuremath{{\mathrm{\,Me\kern -0.1em V\!/}c}}\xspace}
\newcommand{\gevcc}{\ensuremath{{\mathrm{\,Ge\kern -0.1em V\!/}c^2}}\xspace}
\newcommand{\gevgevcccc}{\ensuremath{{\mathrm{\,Ge\kern -0.1em V^2\!/}c^4}}\xspace}
\newcommand{\mevcc}{\ensuremath{{\mathrm{\,Me\kern -0.1em V\!/}c^2}}\xspace}
\newcommand{\evcc}{\ensuremath{{\mathrm{\,e\kern -0.1em V\!/}c^2}}\xspace}

\def\invfb   {\ensuremath{\mbox{\,fb}^{-1}}\xspace}

\newcommand{\stat}{\ensuremath{\mathrm{\,(stat)}}\xspace}
\newcommand{\syst}{\ensuremath{\mathrm{\,(syst)}}\xspace}

\def\gsim{{~\raise.15em\hbox{$>$}\kern-.85em
          \lower.35em\hbox{$\sim$}~}\xspace}
\def\lsim{{~\raise.15em\hbox{$<$}\kern-.85em
          \lower.35em\hbox{$\sim$}~}\xspace}

\def\pt         {\mbox{$p_{\mathrm{ T}}$}\xspace}

\def\degrees{\ensuremath{^{\circ}}\xspace}

\def\JPsi         {\ensuremath{\jpsi}\xspace}

\def\BdToDstpi          {\ensuremath{\decay{\Bd}{\Dstarm\pip}}\xspace}

\def\BdToDstmunu        {\ensuremath{\decay{\Bd}{\Dstarm\mup\neum}}\xspace}

\def\DstpToDpizero      {\ensuremath{\decay{\Dstarp}{\Dp\piz}}\xspace}

\def\DstmToDzeropi      {\ensuremath{\decay{\Dstarm}{\Dzb\pim}}\xspace}

\def\DstzToDzpi         {\ensuremath{\decay{\Dstarz}{\Dz\piz}}\xspace}

\def\DzToKSpi     {\ensuremath{\decay{\Dz}{\KS\piz}}\xspace}

\def\DststToDorDstpi  {\ensuremath{\decay{\Dstst}{\D^{(*)}\pi}}\xspace}

\def\DststToDstpi   {\ensuremath{\decay{\Dstst}{\Dstar\pi}}\xspace}
\def\DonezToDpipi   {\ensuremath{\decay{\Donez}{\Dz\pipi}}\xspace}
\def\DonezbToDpipi    {\ensuremath{\decay{\Donezb}{\Dzb\pipi}}\xspace}
\def\DonemToDpipi   {\ensuremath{\decay{\Donem}{\Dm\pipi}}\xspace}
\def\DoneToDpipi    {\ensuremath{\decay{\Done}{\D\pipi}}\xspace}

\def\BToDststlnu    {\ensuremath{\decay{\PB}{\Dststb\ellp\neul}}\xspace}
\def\BToDorDstnpilnu    {\ensuremath{\decay{\PB}{\Dbar{}^{(*)}\text{n}\pion\ellp\neul}}\xspace}
\def\BToDlnu        {\ensuremath{\decay{\PB}{\Dbar\ellp\neul}}\xspace}
\def\BToDstlnu        {\ensuremath{\decay{\PB}{\Dstarb\ellp\neul}}\xspace}
\def\BToDorDstlnu     {\ensuremath{\decay{\PB}{\Dbar{}^{(*)}\ellp\neul}}\xspace}

\def\BToDpilnu        {\ensuremath{\decay{\PB}{\Dbar\pi\ellp\neul}}\xspace}
\def\BToDorDstpilnu     {\ensuremath{\decay{\PB}{\Dbar{}^{(*)}\pi\ellp\neul}}\xspace}
\def\BToDstpilnu      {\ensuremath{\decay{\PB}{\Dstarb\pi\ellp\neul}}\xspace}
\def\BToDpipilnu      {\ensuremath{\decay{\PB}{\Dbar\pipi\ellp\neul}}\xspace}
\def\BToDorDstpipilnu {\ensuremath{\decay{\PB}{\Dbar{}^{(*)}\pipi\ellp\neul}}\xspace}

\def\BToDstpipilnu      {\ensuremath{\decay{\PB}{\Dstarb\pipi\ellp\neul}}\xspace}
\def\BdToDststlnu     {\ensuremath{\decay{\Bd}{\Dststm\ellp\neul}}\xspace}

\def\BuToDststlnu     {\ensuremath{\decay{\Bu}{\Dststzb\ellp\neul}}\xspace}

\def\BdToDzpilnu      {\ensuremath{\decay{\Bd}{\Dzb\pim\ellp\neul}}\xspace}
\def\BdToDzpienu      {\ensuremath{\decay{\Bd}{\Dzb\pim\ep\neue}}\xspace}
\def\BdToDzpimunu     {\ensuremath{\decay{\Bd}{\Dzb\pim\mup\neum}}\xspace}
\def\BuToDmpilnu    {\ensuremath{\decay{\Bu}{\Dm\pip\ellp\neul}}\xspace}
\def\BuToDmpienu    {\ensuremath{\decay{\Bu}{\Dm\pip\ep\neue}}\xspace}
\def\BuToDmpimunu   {\ensuremath{\decay{\Bu}{\Dm\pip\mup\neum}}\xspace}
\def\BdToDstpilnu     {\ensuremath{\decay{\Bd}{\Dstarzb\pim\ellp\neul}}\xspace}
\def\BdToDstpienu     {\ensuremath{\decay{\Bd}{\Dstarzb\pim\ep\neue}}\xspace}
\def\BdToDstpimunu      {\ensuremath{\decay{\Bd}{\Dstarzb\pim\mup\neum}}\xspace}
\def\BuToDstpilnu   {\ensuremath{\decay{\Bu}{\Dstarm\pip\ellp\neul}}\xspace}
\def\BuToDstpienu   {\ensuremath{\decay{\Bu}{\Dstarm\pip\ep\neue}}\xspace}
\def\BuToDstpimunu    {\ensuremath{\decay{\Bu}{\Dstarm\pip\mup\neum}}\xspace}
\def\BdToDmlnu        {\ensuremath{\decay{\Bd}{\Dm\ellp\neul}}\xspace}
\def\BdToDmmunu       {\ensuremath{\decay{\Bd}{\Dm\mup\neum}}\xspace}
\def\BdToDmenu        {\ensuremath{\decay{\Bd}{\Dm\ep\neue}}\xspace}
\def\BdToDstlnu       {\ensuremath{\decay{\Bd}{\Dstarm\ellp\neul}}\xspace}
\def\BdToDstmunu      {\ensuremath{\decay{\Bd}{\Dstarm\mup\neum}}\xspace}
\def\BdToDstenu       {\ensuremath{\decay{\Bd}{\Dstarm\ep\neue}}\xspace}
\def\BuToDzlnu        {\ensuremath{\decay{\Bu}{\Dzb\ellp\neul}}\xspace}
\def\BuToDzmunu       {\ensuremath{\decay{\Bu}{\Dzb\mup\neum}}\xspace}
\def\BuToDzenu        {\ensuremath{\decay{\Bu}{\Dzb\ep\neue}}\xspace}
\def\BuToDstlnu       {\ensuremath{\decay{\Bu}{\Dstarzb\ellp\neul}}\xspace}
\def\BuToDstmunu      {\ensuremath{\decay{\Bu}{\Dstarzb\mup\neum}}\xspace}
\def\BuToDstenu       {\ensuremath{\decay{\Bu}{\Dstarzb\ep\neue}}\xspace}

\def\BdToDzstarlnu    {\ensuremath{\decay{\Bd}{\Dzstarm\ellp\neul}}\xspace}
\def\BuToDzstarlnu    {\ensuremath{\decay{\Bu}{\Dzstarzb\ellp\neul}}\xspace}
\def\BToDonelnu     {\ensuremath{\decay{\PB}{\Doneb\ellp\neul}}\xspace}
\def\BdToDonelnu    {\ensuremath{\decay{\Bd}{\Donem\ellp\neul}}\xspace}
\def\BuToDonelnu    {\ensuremath{\decay{\Bu}{\Donezb\ellp\neul}}\xspace}
\def\BToDoneprimelnu  {\ensuremath{\decay{\PB}{\Dprimeoneb\ellp\neul}}\xspace}
\def\BdToDoneprimelnu {\ensuremath{\decay{\Bd}{\Dprimeonem\ellp\neul}}\xspace}
\def\BuToDoneprimelnu {\ensuremath{\decay{\Bu}{\Dprimeonezb\ellp\neul}}\xspace}

\def\BdToDtwostarlnu  {\ensuremath{\decay{\Bd}{\Dtwostarm\ellp\neul}}\xspace}
\def\BuToDtwostarlnu  {\ensuremath{\decay{\Bu}{\Dtwostarzb\ellp\neul}}\xspace}
\def\BdToDorDstpipilnu  {\ensuremath{\decay{\Bd}{\D^{(*)-}\pipi\ellp\neul}}\xspace}
\def\BdToDmpipilnu      {\ensuremath{\decay{\Bd}{\Dm\pipi\ellp\neul}}\xspace}
\def\BdToDmpipimunu     {\ensuremath{\decay{\Bd}{\Dm\pipi\mup\neum}}\xspace}
\def\BdToDmpipienu      {\ensuremath{\decay{\Bd}{\Dm\pipi\ep\neue}}\xspace}
\def\BdToDstpipilnu     {\ensuremath{\decay{\Bd}{\Dstarm\pipi\ellp\neul}}\xspace}
\def\BdToDstpipimunu  {\ensuremath{\decay{\Bd}{\Dstarm\pipi\mup\neum}}\xspace}
\def\BdToDstpipienu   {\ensuremath{\decay{\Bd}{\Dstarm\pipi\ep\neue}}\xspace}
\def\BuToDorDstpipilnu  {\ensuremath{\decay{\Bu}{\D^{(*)}\pipi\ellp\neul}}\xspace}
\def\BuToDzpipilnu    {\ensuremath{\decay{\Bu}{\Dzb\pipi\ellp\neul}}\xspace}
\def\BuToDzpipimunu   {\ensuremath{\decay{\Bu}{\Dzb\pipi\mup\neum}}\xspace}
\def\BuToDzpipienu    {\ensuremath{\decay{\Bu}{\Dzb\pipi\ep\neue}}\xspace}
\def\BuToDstpipilnu   {\ensuremath{\decay{\Bu}{\Dstarzb\pipi\ellp\neul}}\xspace}
\def\BuToDstpipimunu  {\ensuremath{\decay{\Bu}{\Dstarzb\pipi\mup\neum}}\xspace}
\def\BuToDstpipienu   {\ensuremath{\decay{\Bu}{\Dstarzb\pipi\ep\neue}}\xspace}
\def\BToDsttaunu    {\ensuremath{\decay{\PB}{\Dstarb\taup\neutb}}\xspace}
\def\BToDorDsttaunu   {\ensuremath{\decay{\PB}{\Dbar{}^{(*)}\taup\neutb}}\xspace}

\def\bToc               {\ensuremath{\decay{\bquark}{\cquark}}\xspace}

\def\Bsig         {\ensuremath{\PB_\text{sig}}\xspace}
\def\Btag         {\ensuremath{\PB_\text{tag}}\xspace}

\def\Dpi                {\ensuremath{\PD\pi}\xspace}
\def\Dstpi              {\ensuremath{\Dstar\pi}\xspace}
\def\Dpipi              {\ensuremath{\PD\pi\pi}\xspace}

\def\Dzstarz      {\ensuremath{\D_0^{*0}}\xspace}
\def\Dzstarzb     {\ensuremath{\Dbar{}_0^{*0}}\xspace}

\def\Dzstarm      {\ensuremath{\D_0^{*-}}\xspace}
\def\Dzstar       {\ensuremath{\D_0^{*}}\xspace}

\def\Donez        {\ensuremath{\D_1^0}\xspace}
\def\Donezb       {\ensuremath{\Dbar{}_1^0}\xspace}

\def\Donem        {\ensuremath{\D_1^-}\xspace}
\def\Done       {\ensuremath{\D_1}\xspace}
\def\Doneb        {\ensuremath{\Dbar_1}\xspace}
\def\Dprimeonez     {\ensuremath{\D^{\prime\,0}_1}\xspace}
\def\Dprimeonezb    {\ensuremath{\Dbar{}^{\prime\,0}_1}\xspace}

\def\Dprimeonem     {\ensuremath{\D^{\prime\,-}_1}\xspace}
\def\Dprimeone      {\ensuremath{\D^\prime_1}\xspace}
\def\Dprimeoneb     {\ensuremath{\Dbar{}^\prime_1}\xspace}
\def\Dtwostarz      {\ensuremath{\D_2^{*0}}\xspace}
\def\Dtwostarzb     {\ensuremath{\Dbar{}_2^{*0}}\xspace}

\def\Dtwostarm      {\ensuremath{\D_2^{*-}}\xspace}
\def\Dtwostar     {\ensuremath{\D_2^{*}}\xspace}

\DeclareMathAlphabet      {\mathbfit}{OML}{cmm}{b}{it}
\newcommand{\vect}[1]{\ensuremath{\vec{#1}}\xspace}

\newcommand{\Emiss}{\ensuremath{E_\text{miss}}\xspace}
\newcommand{\pmiss}{\ensuremath{p_\text{miss}}\xspace}

\newcommand{\asymbol}[3]{
\ensuremath{#1
^{\ForEach{;}{
\ifboolexpr{test {\ifnumcomp{\the\thislevelcount}{=}{1}}}
  {\text{\thislevelitem}}
  {, \text{\thislevelitem}}
}{#2}}
_{\ForEach{;}{
\ifboolexpr{test {\ifnumcomp{\the\thislevelcount}{=}{1}}}
  {\text{\thislevelitem}}
  {, \text{\thislevelitem}}
}{#3}}
\xspace}
}

\newcommand{\pdf}[2]{\asymbol{\mathcal{P}}{#1}{#2}}

\newcommand{\effc}[2]{\asymbol{\epsilon}{#1}{#2}}
\newcommand{\yield}[2]{\asymbol{N}{#1}{#2}}

\DeclareSIUnit\clight{\text{\ensuremath{c}}}

\DeclareSIUnit[per-mode=symbol]\eVc{\eV\per\clight}
\DeclareSIUnit[per-mode=symbol]\keVc{\kilo\eV\per\clight}
\DeclareSIUnit[per-mode=symbol]\MeVc{\mega\eV\per\clight}
\DeclareSIUnit[per-mode=symbol]\GeVc{\giga\eV\per\clight}
\DeclareSIUnit[per-mode=symbol]\TeVc{\tera\eV\per\clight}

\DeclareSIUnit[per-mode=symbol]\eVcc{\eV\per\square\clight}
\DeclareSIUnit[per-mode=symbol]\keVcc{\kilo\eV\per\square\clight}
\DeclareSIUnit[per-mode=symbol]\MeVcc{\mega\eV\per\square\clight}
\DeclareSIUnit[per-mode=symbol]\GeVcc{\giga\eV\per\square\clight}
\DeclareSIUnit[per-mode=symbol]\TeVcc{\tera\eV\per\square\clight}

\def\BRBdToDzpil {0.360}
\def\BRBdToDzpilstat {0.018}
\def\BRBdToDzpilsyst {0.011}

\def\BRBuToDmpil {0.378}
\def\BRBuToDmpilstat {0.013}
\def\BRBuToDmpilsyst {0.017}

\def\BRBdToDstpil {0.551}
\def\BRBdToDstpilstat {0.024}
\def\BRBdToDstpilsyst {0.017}

\def\BRBuToDstpil {0.530}
\def\BRBuToDstpilstat {0.019}
\def\BRBuToDstpilsyst {0.025}

\def\BRBdToDmpipil {0.145}
\def\BRBdToDmpipilstat {0.018}
\def\BRBdToDmpipilsyst {0.013}

\def\BRBuToDzpipil {0.173}
\def\BRBuToDzpipilstat {0.014}
\def\BRBuToDzpipilsyst {0.013}

\def\BRBdToDstpipil {0.051}
\def\BRBdToDstpipilstat {0.021}
\def\BRBdToDstpipilsyst {0.009}

\def\BRBuToDstpipil {0.070}
\def\BRBuToDstpipilstat {0.015}
\def\BRBuToDstpipilsyst {0.008}

\def\BRBdToDzpilRatio {7.24}
\def\BRBdToDzpilstatRatio {0.36}
\def\BRBdToDzpilsystRatio {0.12}

\def\BRBuToDmpilRatio {6.78}
\def\BRBuToDmpilstatRatio {0.24}
\def\BRBuToDmpilsystRatio {0.15}

\def\BRBdToDstpilRatio {11.10}
\def\BRBdToDstpilstatRatio {0.48}
\def\BRBdToDstpilsystRatio {0.20}

\def\BRBuToDstpilRatio {9.50}
\def\BRBuToDstpilstatRatio {0.33}
\def\BRBuToDstpilsystRatio {0.27}

\def\BRBdToDmpipilRatio {2.91}
\def\BRBdToDmpipilstatRatio {0.37}
\def\BRBdToDmpipilsystRatio {0.25}

\def\BRBuToDzpipilRatio {3.10}
\def\BRBuToDzpipilstatRatio {0.26}
\def\BRBuToDzpipilsystRatio {0.21}

\def\BRBdToDstpipilRatio {1.03}
\def\BRBdToDstpipilstatRatio {0.43}
\def\BRBdToDstpipilsystRatio {0.18}

\def\BRBuToDstpipilRatio {1.25}
\def\BRBuToDstpipilstatRatio {0.27}
\def\BRBuToDstpipilsystRatio {0.15}
\usepackage{dcolumn}  

\begin{document}

\title{First observation of $\decay{\PB}{\Doneb(\to\Dbar\pipi)\ellp\neul}$ and measurement of the $\BToDorDstpilnu$ and $\BToDorDstpipilnu$ branching fractions with hadronic tagging at Belle}

\noaffiliation
  \author{F.~Meier\,\orcidlink{0000-0002-6088-0412}} 
  \author{A.~Vossen\,\orcidlink{0000-0003-0983-4936}} 
  \author{I.~Adachi\,\orcidlink{0000-0003-2287-0173}} 
  \author{K.~Adamczyk\,\orcidlink{0000-0001-6208-0876}} 
  \author{H.~Aihara\,\orcidlink{0000-0002-1907-5964}} 
  \author{S.~Al~Said\,\orcidlink{0000-0002-4895-3869}} 
  \author{D.~M.~Asner\,\orcidlink{0000-0002-1586-5790}} 
  \author{H.~Atmacan\,\orcidlink{0000-0003-2435-501X}} 
  \author{T.~Aushev\,\orcidlink{0000-0002-6347-7055}} 
  \author{R.~Ayad\,\orcidlink{0000-0003-3466-9290}} 
% ...
\collaboration{The Belle Collaboration}

\begin{abstract}

We report measurements of the ratios of branching fractions for $\BToDorDstpilnu$ and
$\BToDorDstpipilnu$ relative to $\BToDstlnu$ decays with $\lepton = e, \muon$. These results are obtained from a data sample that contains $772 \times 10^6~B\bar{B}$ pairs
collected near the $\FourS$ resonance with the Belle detector at the
KEKB asymmetric energy $\epem$ collider.
Fully reconstructing both \PB mesons in the event, we obtain
\begin{align*}
\frac{\mathcal{B}(\BdToDzpilnu)}{\mathcal{B}(\BdToDstlnu)} &= \num[parse-numbers=false]{(\BRBdToDzpilRatio\pm\BRBdToDzpilstatRatio\pm\BRBdToDzpilsystRatio)}\%\ ,\\
\frac{\mathcal{B}(\BuToDmpilnu)}{\mathcal{B}(\BuToDstlnu)} &= \num[parse-numbers=false]{(\BRBuToDmpilRatio\pm\BRBuToDmpilstatRatio\pm\BRBuToDmpilsystRatio)}\%\ ,\\
\frac{\mathcal{B}(\BdToDstpilnu)}{\mathcal{B}(\BdToDstlnu)} &= \num[parse-numbers=false]{(\BRBdToDstpilRatio\pm\BRBdToDstpilstatRatio\pm\BRBdToDstpilsystRatio)}\%\ ,\\
\frac{\mathcal{B}(\BuToDstpilnu)}{\mathcal{B}(\BuToDstlnu)} &= \num[parse-numbers=false]{(\BRBuToDstpilRatio\pm\BRBuToDstpilstatRatio\pm\BRBuToDstpilsystRatio)}\%\ ,\\
\frac{\mathcal{B}(\BdToDmpipilnu)}{\mathcal{B}(\BdToDstlnu)} &= \num[parse-numbers=false]{(\BRBdToDmpipilRatio\pm\BRBdToDmpipilstatRatio\pm\BRBdToDmpipilsystRatio)}\%\ ,\\
\frac{\mathcal{B}(\BuToDzpipilnu)}{\mathcal{B}(\BuToDstlnu)} &= \num[parse-numbers=false]{(\BRBuToDzpipilRatio\pm\BRBuToDzpipilstatRatio\pm\BRBuToDzpipilsystRatio)}\%\ ,\\
\frac{\mathcal{B}(\BdToDstpipilnu)}{\mathcal{B}(\BdToDstlnu)} &= \num[parse-numbers=false]{(\BRBdToDstpipilRatio\pm\BRBdToDstpipilstatRatio\pm\BRBdToDstpipilsystRatio)}\%\ ,\\
\frac{\mathcal{B}(\BuToDstpipilnu)}{\mathcal{B}(\BuToDstlnu)} &= \num[parse-numbers=false]{(\BRBuToDstpipilRatio\pm\BRBuToDstpipilstatRatio\pm\BRBuToDstpipilsystRatio)}\%\ ,
\end{align*}
where the uncertainties are statistical and systematic, respectively.
These are the most precise measurements of these branching fraction ratios to date.
The invariant mass spectra of the \Dpi, \Dstpi, and \Dpipi systems are
studied, and the branching fraction products
\begin{align*}
\BR(\BdToDtwostarlnu) \times \BR(\decay{\Dtwostarm}{\Dzb\pim}) &= \num[parse-numbers=false]{(0.157\pm0.015\pm0.005)}\%\ , \\
\BR(\BuToDzstarlnu) \times \BR(\decay{\Dzstarzb}{\Dm\pip})     &= \num[parse-numbers=false]{(0.054\pm0.022\pm0.005)}\%\ , \\
\BR(\BuToDtwostarlnu) \times \BR(\decay{\Dtwostarzb}{\Dm\pip}) &= \num[parse-numbers=false]{(0.163\pm0.011\pm0.007)}\%\ , \\
\BR(\BdToDonelnu) \times \BR(\decay{\Donem}{\Dstarzb\pim}) &= \num[parse-numbers=false]{(0.306\pm0.050\pm0.028)}\%\ , \\
\BR(\BdToDoneprimelnu) \times \BR(\decay{\Dprimeonem}{\Dstarzb\pim}) &= \num[parse-numbers=false]{(0.206\pm0.068\pm0.025)}\%\ , \\
\BR(\BdToDtwostarlnu) \times \BR(\decay{\Dtwostarm}{\Dstarzb\pim}) &= \num[parse-numbers=false]{(0.051\pm0.040\pm0.010)}\%\ , \\
\BR(\BuToDonelnu) \times \BR(\decay{\Donezb}{\Dstarm\pip}) &= \num[parse-numbers=false]{(0.249\pm0.023\pm0.014)}\%\ , \\
\BR(\BuToDoneprimelnu) \times \BR(\decay{\Dprimeonezb}{\Dstarm\pip}) &= \num[parse-numbers=false]{(0.138\pm0.036\pm0.008)}\%\ , \\
\BR(\BuToDtwostarlnu) \times \BR(\decay{\Dtwostarzb}{\Dstarm\pip}) &= \num[parse-numbers=false]{(0.137\pm0.026\pm0.009)}\%\ , \\
\BR(\BdToDonelnu) \times \BR(\DonemToDpipi) &= \num[parse-numbers=false]{(0.102\pm0.013\pm0.009)}\%\ , \\
\BR(\BuToDonelnu) \times \BR(\DonezbToDpipi) &= \num[parse-numbers=false]{(0.105\pm0.011\pm0.008)}\%\ ,
\end{align*}
are extracted. This is the first observation of the decays \BToDonelnu with \DoneToDpipi.
\end{abstract}

\maketitle

{\renewcommand{\thefootnote}{\fnsymbol{footnote}}}
\setcounter{footnote}{0}

\section{Introduction}
\label{sec:introduction}

Semileptonic decays of \PB mesons are an important tool for precision
measurements of the Cabibbo-Kobayashi-Maskawa matrix elements \Vcb
and \Vub~\cite{Cabibbo:1963yz,CKM}. The latest determinations of $|\Vcb|$ from inclusive
semileptonic \decay{\B}{X_c\ellp\neul} decays, with $X_c$ being a charmed hadronic
state that is not explicitly reconstructed, differ from those using the exclusive
semileptonic decays \BToDlnu and \BToDstlnu by about \SI{2.4}{\sigma}~\cite
{PDG2022}. The measured sum of the exclusive \BToDorDstlnu, \BToDorDstpilnu,
and \decay{\Bu}{\D^{(*)-}_\squark\Kp\ellp\neul} rates accounts for only \num
{85\pm2}\%~\cite{PDG2022} of the inclusive rate for
semileptonic \PB decays to charm final states.

Semileptonic decays of \PB
mesons can also be used for other precision tests of the electroweak sector
of the standard model, such as lepton flavor universality. An example is the
ratio $R(D^{(\ast)})$ of the branching fractions $\BR($\BToDorDsttaunu$)$ and
$\BR(\BToDorDstlnu)$ ($\lepton = e, \muon$), for which a persistent \SI{3}
{\sigma} deviation between the standard model expectation~\cite
{Fajfer:2012vx} and the combined experimental results~\cite
{HFLAV} from \babar~\cite{Lees:2013uzd,Lees:2012xj}, \belle~\cite
{Huschle:2015rga,Hirose:2016wfn,Belle:2019rba}, and LHCb~\cite
{Aaij:2015yra,Aaij:2017deq} has been observed.
Important backgrounds in these processes are the decays \BToDorDstpipilnu
and \BToDorDstpilnu. The former accounts for part of the missing exclusive
rate described above. The latter proceeds predominantly
via \BToDststlnu, \DststToDorDstpi, where the \Dstst is an orbitally excited
(L = 1) charmed meson. The \Dstst mass-spectrum contains two doublets of
states that have light-quark total angular momenta of $j_q = \frac 12$ and $j_q
= \frac 32$~\cite{Isgur:1991wq}. The spin-0 state \Dzstar can only decay
to \Dpi and the spin-1 states \Done and \Dprimeone only via \DststToDstpi.
The spin-2 state \Dtwostar can decay both into \Dpi and \Dstpi. The \Dstst
masses are not far from threshold. Since the $j_q = \frac 32$ states
(\Done and \Dtwostar) have a significant D-wave component, these states are
narrow and were observed with a typical width of about \SI{20}{\MeVcc}~\cite
{Liventsev:2007rb,Aubert:2008zc,Aubert:2008ea}. On the other hand, the states
with $j_q = \frac 12$ decay mainly via S-wave and are therefore expected to
be broad resonances with a width of several hundred \si{\MeVcc}~\cite
{Isgur:1991wq,Leibovich:1997em}. The decay rate of semileptonic \PB decays to
the $j_q = \frac 12$ states is observed to be similar to the rate to the $j_q
= \frac 32$ doublet, while model calculations predict a substantially smaller
rate to the $j_q = \frac 12$ doublet~\cite{LeYaouanc:2021ggx}.

The decay modes with one charged pion in the final state have been measured
by \babar~\cite{Aubert:2008ea} and in a previous \belle analysis~\cite
{Vossen:2018zeg}. For the \BToDorDstpipilnu channel so far only a \babar
result~\cite{Lees:2015eya} with limited statistical precision is available.
The results of these three measurements are listed in \cref
{tab:oldB2DststlnuResults}.
\begin{table}[ht]
\caption{Previous results of \BToDorDstpilnu and \BToDorDstpipilnu branching fraction measurements
 by \babar~\cite{Aubert:2008ea,Lees:2015eya} and \belle~\cite{Vossen:2018zeg}. The first uncertainty is statistical, the second systematic, and the third comes from the branching fraction of the normalization mode.}
\centering
\resizebox{\textwidth}{!}{%
\setlength{\tabcolsep}{20pt}
\begin{tabular}{lcc}
\toprule
Decay mode        & \babar  & \belle \\
\midrule
\BdToDzpilnu      & \num[parse-numbers=false]{(0.43 \pm 0.08 \pm 0.03)}\% & \num[parse-numbers=false]{(0.405 \pm 0.036 \pm 0.041)}\%  \\
\BuToDmpilnu      & \num[parse-numbers=false]{(0.42 \pm 0.06 \pm 0.03)}\% & \num[parse-numbers=false]{(0.455 \pm 0.027 \pm 0.039)}\%  \\
\BdToDstpilnu     & \num[parse-numbers=false]{(0.48 \pm 0.08 \pm 0.04)}\% & \num[parse-numbers=false]{(0.646 \pm 0.053 \pm 0.052)}\%  \\
\BuToDstpilnu     & \num[parse-numbers=false]{(0.59 \pm 0.05 \pm 0.04)}\% & \num[parse-numbers=false]{(0.603 \pm 0.043 \pm 0.038)}\%  \\
\BdToDmpipilnu          &   \num[parse-numbers=false]{(0.127 \pm 0.039 \pm 0.026 \pm 0.007)}\%    &   - \\
\BuToDzpipilnu          &   \num[parse-numbers=false]{(0.161 \pm 0.030 \pm 0.018 \pm 0.008)}\%    &   - \\
\BdToDstpipilnu         &   \num[parse-numbers=false]{(0.138 \pm 0.039 \pm 0.030 \pm 0.003)}\%    &   - \\
\BuToDstpipilnu         &   \num[parse-numbers=false]{(0.080 \pm 0.040 \pm 0.023 \pm 0.003)}\%    &   - \\
\bottomrule
\end{tabular}%
}%
\label{tab:oldB2DststlnuResults}
\end{table}
The current measurement improves upon the
aforementioned \belle result by using a new method to reconstruct (\enquote
{tag}) the other \PB in the event, known as the Full Event Interpretation
(FEI) algorithm~\cite{FEI}, and by providing a result for the
\BToDorDstpipilnu channel as well.

\section{Experimental Apparatus and data}

The Belle detector is a large-solid-angle magnetic spectrometer. Its innermost
component is a silicon vertex detector (SVD). A 50-layer central drift
chamber (CDC) provides tracking and charged particle identification (PID) information using specific ionization
measurements. An array of
aerogel threshold Cherenkov counters (ACC), a
barrel-like arrangement of time-of-flight scintillation counters (TOF) in the
central part, and an electromagnetic calorimeter (ECL) comprised of CsI
(Tl) crystals provide further PID information. These detector components are
located inside a super-conducting solenoid coil that provides a \SI{1.5}
{T} magnetic field. The iron return yoke located outside of the coil is
instrumented to detect \KL mesons and to identify muons (KLM). The detector's
$z$-axis is defined to be anti-parallel to the \ep beam. More details about the
detector can be found in Ref.~\cite{Belle}.

Electron candidates are identified using the ratio between the energy deposited in the ECL
and their track momentum, the ECL shower shape, the matching between
the track and the ECL cluster, the energy loss in the CDC, and the number of
photoelectrons in the ACC~\cite{ElectronPID}. Muons are identified based on their penetration
range and transverse scattering in the KLM~\cite{MuonPID}. Charged kaons and pions
are identified by a combination of the energy loss in the CDC, the Cherenkov
light in the ACC, and the time of flight in the TOF.

A data sample corresponding to an integrated luminosity of $L_{\rm on} = \SI{711}{\invfb}$, collected with the
Belle detector at the KEKB asymmetric-energy $\epem$ collider~\cite
{KEKB} operating at the $\FourS$ resonance at $\sqrt{s} = \SI{10.58}
{\gev}$, is used for the measurement. The sample contains \num{772e6} \BBbar
pairs. A further data sample corresponding to an integrated luminosity of $L_{\rm off} = \SI{89}{\invfb}$ taken
slightly below the resonance, at $\sqrt{s} = \SI{10.52}{\gev}$, is used for background
templates. These two samples are referred to as the on-resonance and off-resonance samples, respectively.

We use a sample of simulated \BBbar background Monte Carlo (MC) events generated
with EvtGen~\cite{Lange:2001uf}. This sample has six times more events than the \belle
collision data. The full detector simulation is based on GEANT3~\cite
{Geant3}. Final-state radiation is simulated with the PHOTOS package~\cite
{PHOTOS}. The $\BToDstlnu$ decays are simulated using the HQET2 model~\cite{HQET} of
EvtGen. For the $\BToDorDstpilnu$ decay modes, dedicated MC samples
of \num{73e6} events for each of five transitions (via \Dzstar and \Dtwostar
for \BToDpilnu, and via \Done, \Dprimeone, and \Dtwostar for \BToDstpilnu) are
generated with the ISGW2 model~\cite
{Scora:1995ty}. A signal MC sample of \num{50e6} events of \BToDpipilnu is
used for simulating the \BToDonelnu decay modes. An MC sample of \num{25e6} events
of \BToDstpipilnu is used for simulating the \BToDoneprimelnu decay modes. Both are simulated
with the ISGW2 model.

Data-MC efficiency differences due to a variety of sources are corrected. A
more detailed description can be found in \cref{sec:systematics}.

\section{Measurement overview}

The $\BdToDzpilnu$ branching ratio relative to \BdToDstlnu is measured,
\begin{equation}
\frac{\BR(\BdToDzpilnu)}{\BR(\BdToDstlnu)} = \frac{\yield{}{sig}(\BdToDzpilnu) / \effc{}{sig}(\BdToDzpilnu)}{\yield{}{sig}(\BdToDstlnu) / \effc{}{sig}(\BdToDstlnu)} \ ,
\label{eq:overview:BRratio}
\end{equation}
which reduces systematic uncertainties due to data-MC differences and
external branching fraction values of the charm modes as they largely cancel
in the ratio. Similar expressions are used for the other measured decays.
Here $\yield{}{sig}$ is the number of signal candidates and $\effc{}{sig}$ is
the corresponding signal efficiency. Branching fractions of \BToDorDstnpilnu (n = \num{1}, \num{2}) are
also reported, after multiplying by the $\BR(\BuToDstlnu) = \num{5.58\pm0.22}
\%$ and $\BR(\BdToDstlnu) = \num{4.97\pm0.12}\%$ averages from the Particle Data Group (PDG)~\cite{PDG2022}.

\section{Event selection}

The Belle data are converted into the Belle~II format~\cite{B2BII}, and the
particle and event reconstruction is performed within the \texttt
{basf2} framework~\cite{basf2,basf2-zenodo} of the Belle~II experiment. 

\subsection{Common selection requirements}

\begin{figure}[hbt]
\hspace*{\fill}
\begin{overpic}
[width=0.32\textwidth]{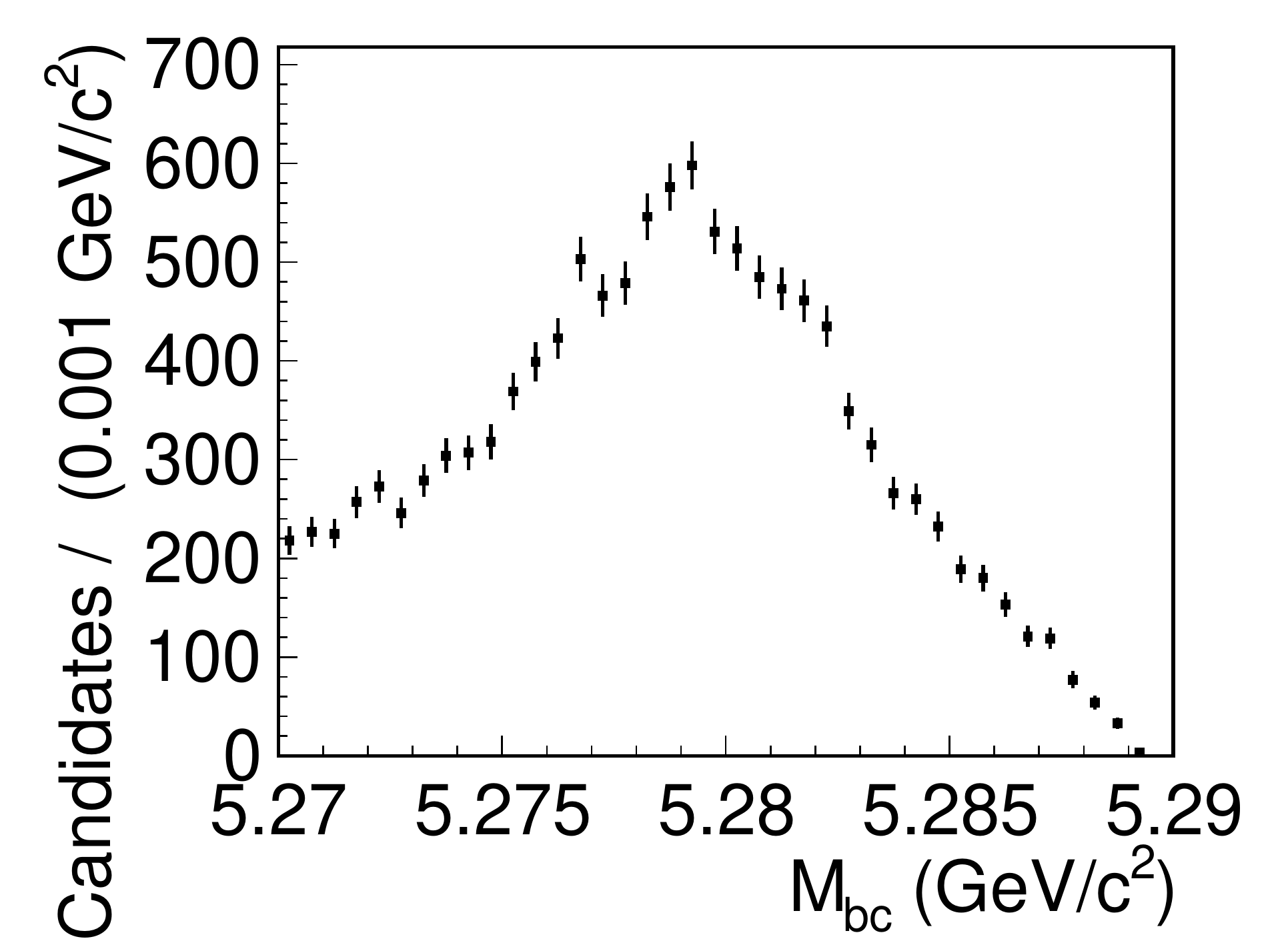}
\put(35,90){Belle}
\end{overpic}
\hfill
\begin{overpic}
[width=0.32\textwidth]{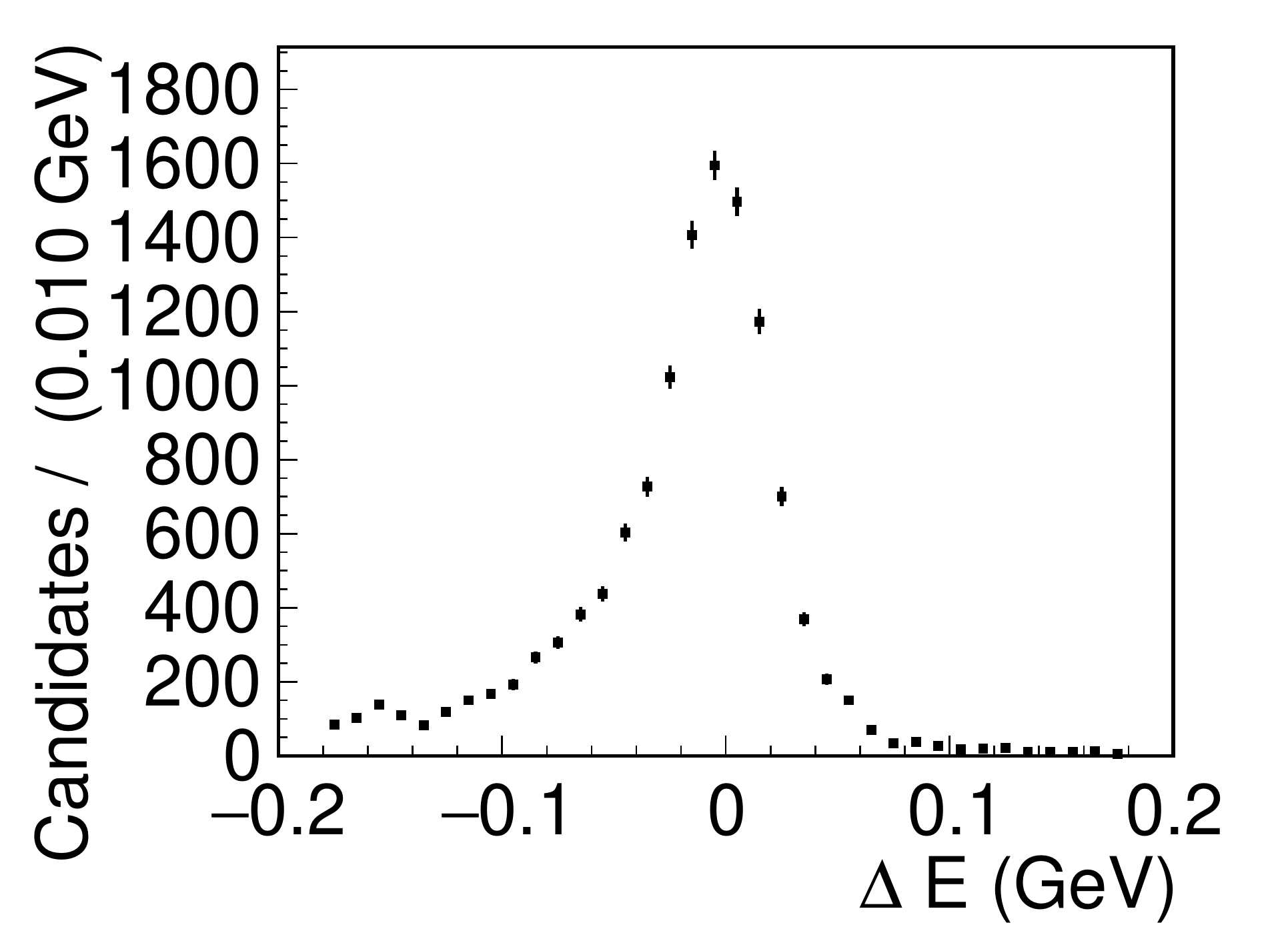}
\put(35,90){Belle}
\end{overpic}
\hfill
\begin{overpic}
[width=0.32\textwidth]{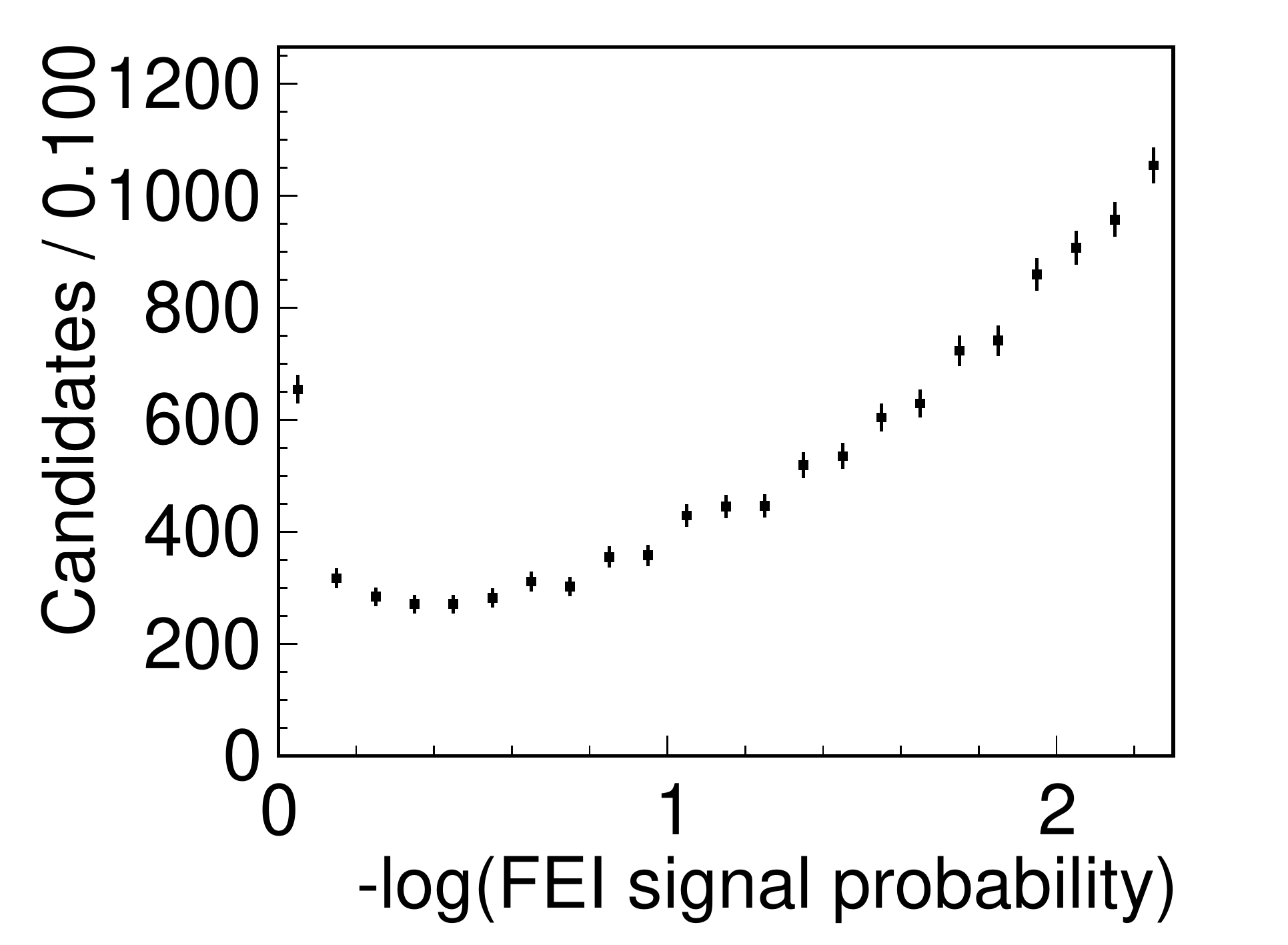}
\put(35,90){Belle}
\end{overpic}
\hspace*{\fill}
\caption{Distribution of beam-constrained mass (left), energy difference
 (middle), and FEI signal probability (right) of charged \Btag meson
 candidates for the data after the full selection. The error bars show the
 statistical uncertainties.}
\label{fig:selection:Mbc_sigProb}
\end{figure}

In each event two \PB meson candidates are reconstructed. One of the \PB meson
candidates (\Btag) is reconstructed with FEI. The FEI algorithm follows a
hierarchical approach. Final-state particle candidates are combined to
intermediate particles until the final \Btag candidates are formed. More
than \num{100} explicit decay channels, leading to $\mathcal{O}(\num
{10000})$ distinct decay chains are reconstructed. For each final-state
particle and for each decay channel of an intermediate particle, a
multivariate classifier is trained which estimates the probability that each
decay chain correctly describes the true process. In this analysis only
hadronically reconstructed decay chains are considered. The \Btag meson
candidates are required to have a beam-constrained mass $M_{\rm bc} = \sqrt
{\left(E_{\rm c.m.} / c^2\right)^2 - \big(\vect{P}_{\Btag} / c\big)^2} > \SI
{5.27}{\GeVcc}$, and an energy difference $\Delta E = E_{\Btag} - E_
{\rm c.m.}$ within \SI{\pm180}{\mev}. Here $E_{\rm c.m.}$ is half of the
center-of-mass (c.m.) energy of the beams, and $\vect{P}_{\Btag}$ and $E_
{\Btag}$ are the momentum and energy of the \Btag meson in the c.m. frame,
respectively. The FEI signal probability of \Btag candidates is required to be
greater than \num{0.5}\%. Distributions of $M_{\rm bc}$, $\Delta E$,
and the signal probability are shown in \cref{fig:selection:Mbc_sigProb}. We
take into account that the composition of decay modes reconstructed by
FEI differs between data and MC. The ratio between the relative abundance in
each decay mode is used to correct this effect.

The other \PB meson candidate (\Bsig) is reconstructed in the decays of
interest. The first selection step of the \Bsig reconstruction is the
requirement of at least one electron or muon candidate in the event. For both
lepton types, the lepton is required to have a minimum momentum of $p > \SI
{300}{\MeVc}$. The lepton's point of closest approach to the KEKB interaction
point (IP) is required to be within $|dz| < \SI{2}{cm}$ of the IP along the
detector axis and within $dr < \SI{0.5}{cm}$ in the transverse plane.

The polar angle of muon candidate tracks is required to be within the range $\SI{45}
{\degrees} < \theta_\mu < \SI{145}{\degrees}$ to ensure that the tracks enter the
KLM. Electron tracks need to be within the CDC acceptance $\SI{17}
{\degrees} < \theta_\electron < \SI{150}{\degrees}$. This implies
that the track is within the ECL acceptance.

The likelihood ratio $\mathcal{R}_\mu = \mathcal{L}_\mu / (\mathcal
{L}_\mu + \mathcal{L}_{\rm hadron})$, where $\mathcal{L}_\mu$ and $\mathcal
{L}_{\rm hadron}$ are the likelihoods for muons and charged hadrons, is required to be
greater than \num{0.9} for muon candidates. This selection has an average efficiency
of \num{89}\% with a pion misidentification rate of \num{1.4}\%
for muons with momenta between \num{1} and \SI{3}{\GeVc}~\cite{MuonPID}. For electron
candidates the likelihood ratio $\mathcal{R}_\electron$ is required to be greater
than \num{0.8}. This requirement has an average efficiency of \num{92}\% at a pion
misidentification rate of \num{0.25}\% for electrons with momenta between \num{1} and \SI{3}{\GeVc}~\cite{ElectronPID}.

The four-momentum of the closest photon that is within a \SI{5}{\degrees} cone
around an electron's momentum direction is added to that of the electron candidate
to correct for bremsstrahlung. The photon's energy is required to be greater
than \SIlist{50;75;100}{\mev} for the barrel ($\SI{32.2}{\degree} < \theta_\gamma < \SI{128.7}{\degree}$), forward ($\SI{12.4}{\degree} < \theta_\gamma < \SI{31.4}{\degree}$) and backward end cap ($\SI{130.7}{\degree} < \theta_\gamma < \SI{155.1}{\degree}$)
region of the ECL, respectively.

Kaons and pions are identified using the ratio $\mathcal{R}_{\kaon/\pion} = \mathcal
{L}_\kaon / (\mathcal{L}_\kaon + \mathcal{L}_\pion)$ between the combined
ACC, TOF, and CDC likelihood for a kaon and the sum of the kaon and pion
likelihoods~\cite{BellePID}. Kaons (pions) are required to have $\mathcal{R}_
{\kaon/\pion} > \num{0.6}$ ($\mathcal{R}_
{\kaon/\pion} < \num{0.4}$), which has an average efficiency of \num{92}\% (\num{93.5}\%).
Kaon and pion candidate tracks must satisfy $dr < \SI{2}{cm}$ and
$|dz| < \SI{5}{cm}$.

Neutral kaon candidates are reconstructed from \pipi pairs. The invariant mass of \KS
candidates is required to be in the range \SIrange[]{482}{514}{\MeVcc}, which is
about \SI{4}{\sigma} around the nominal mass, where \si{\sigma} corresponds to the mass resolution.
For low- ($p < \SI{0.5}{\GeVc}$), medium-
($0.5 \leq p \leq \SI{1.5}{\GeVc}$), and high-momentum ($p > \SI{1.5}
{\GeVc}$) \KS candidates, we require that the pion daughters have $dr > \SI{0.05}{cm}$, $\SI{0.03}{cm}$, and
$\SI{0.02}{cm}$, respectively. The angle in the transverse plane between the
vector from the interaction point to the \KS vertex and the \KS flight
direction is required to be less than $\SI{0.3}{rad}$, $\SI{0.1}{rad}$, and
$\SI{0.03}{rad}$ for low-, medium-, and high-momentum candidates, respectively;
the separation distance along the beam axis of the two pion trajectories at
their point of closest approach is required to be below $\SI{0.8}{cm}$, $\SI{1.8}{cm}$, and $\SI
{2.4}{cm}$, respectively. For medium- (high-) momentum \KS candidates, we
require the flight length in the transverse plane to be greater than $\SI
{0.08}{cm}$ ($\SI{0.22}{cm}$). Finally, a mass-constrained vertex fit of
the \KS candidate must converge.

Neutral pion candidates are reconstructed from pairs of photons, which must
satisfy the same region-dependent energy requirements as the photons
considered for the bremsstrahlung correction described above. The diphoton
invariant mass is required to be between \num{120} and \SI{150}{\MeVcc}, which
corresponds to about \SI{5}{\sigma} around the nominal mass. A
mass-constrained fit of the two photons is required to converge. Photons are
not allowed to be shared between \piz candidates. To eliminate duplicates,
all \piz candidates of an event are sorted according to the most energetic
daughter photon (and then, if needed, the second most energetic daughter).
Any \piz candidate that shares photons with one that appears earlier in this
list is removed.

Charged kaons, charged and neutral pions, and \KS mesons are combined to form neutral
and charged \PD meson candidates. A total of \num{10} hadronic \Dz modes with the final states $\Km\pip$, 
$\Km\pip\piz$, $\Km\pip\pip\pim$, $\KS\pipi$, $\Kp\Km$, $\KS\piz$, $\KS\pipi\piz$,
$\pipi$, $\Km\pipi\pip\piz$, and $\pipi\piz$, and \num{9} hadronic \Dp modes with the final states
$\KS\pip$, $\KS\pipi\pip$, $\Km\pip\pip$, $\Km\Kp\pip$, $\Km\pip\pip\piz$, $\KS\pip\piz$,
$\KS\Kp$, $\pip\piz$, and $\pipi\pip$ are considered.
For \PD final states with at least one \piz the \PD-candidate invariant mass
is required to be within \SI{\pm25}{\MeVcc} of the nominal value~\cite{PDG2022}, while the requirement for
all other modes is \SI{\pm15}{\MeVcc}, which corresponds to about \SI{3}{\sigma}.
A global decay chain fit~\cite{TreeFit} is performed for all \D modes except
for \DzToKSpi. In these fits, mass constraints are applied to the
\PD candidate as well as to \KS and \piz candidates. If the fit fails, the
candidate is discarded.

Neutral \Dz meson candidates are combined with \piz candidates to form \Dstarz
candidates. The mass difference between the \Dstarz and the \Dz candidates is restricted
to be between \num{138.9} and \SI{145.5}{\MeVcc}, and a global decay chain fit
with mass constraints on the \Dstarz, \Dz, \KS, and \piz must converge.
Similarly, \Dstarp meson candidates are formed from combinations of \Dp and
\piz as well as \Dz and \pip. The invariant mass of the \Dstarp candidates
is allowed to deviate from the nominal mass by no more than \SI{3}
{\MeVcc}. Again, a global decay chain fit is performed with mass constraints
on the \Dstarp, \Dz or \Dp, \KS, and \piz.

\subsection{Specific event selection of \texorpdfstring{\BToDorDstpilnu}{B -> D(*)pilnu} decays}

By combining one $D^{(*)}$ meson candidate, one lepton
candidate, and one charged pion candidate, \PB meson candidates are formed. The invariant mass $M(\Dpi)$ is
required to be below \SI{2.8}{\GeVcc}, as the potential \Dstst states are
expected to be at lower masses. We also require $M(\Dpi)$ to be above \SI
{2.05}{\GeVcc} to suppress \BToDstlnu contributions.

\subsection{Specific event selection of \texorpdfstring{\BToDorDstpipilnu}{B -> D(*)pipilnu} decays}

Further \PB meson candidates are formed from $D^{(*)}$ meson candidates, one lepton
candidate, and two oppositely-charged pion candidates. The PID requirement for the muons
is tightened to $\mathcal{R}_\mu > \num{0.97}$, which implicitly also removes
all muon candidates with momenta lower than \SI{500}{\MeVc}. To suppress the
background from hadronically decaying \PB meson events, the missing
momentum \pmiss of the event is required to be greater than \SI{200}
{\MeVc}. Here $\pmiss = |\vect{p}(\epem) - \vect{p}(\Btag) - \vect{p}(D^{
(*)}) - \vect{p}(\pi_1) - \vect{p}(\pi_2) - \vect{p}(\lepton)|$ is the
difference between the total momentum of the initial colliding beam particles
and the combined momentum of all visible particles measured in the center-of-mass frame.
Analogously, the missing energy \Emiss is defined as the energy difference
between the center-of-mass energy and the sum over the energies of the \Bsig
and \Btag candidates.

To suppress \Dstarm contributions to the final state in \BuToDzpipilnu, a veto is implemented: the combined
invariant mass of the neutral \PD meson and the pion with the opposite charge
to that of the \PB meson is required to be above \SI{2.05}{\GeVcc}. The contamination
from \BuToDstpilnu with \DstmToDzeropi is reduced by \num{50}\% with
this veto. However, the pions used in the reconstruction of the \Bsig meson
candidate can also arise from the decay of the \Btag meson. Therefore, a
second veto is implemented: the invariant mass of each \pip used in the \Btag
reconstruction combined with the signal \Dz is required to be greater
than \SI{2.05}{\GeVcc}.

\begin{figure}[hbt]
\hspace*{\fill}
\includegraphics[width=0.48\textwidth]{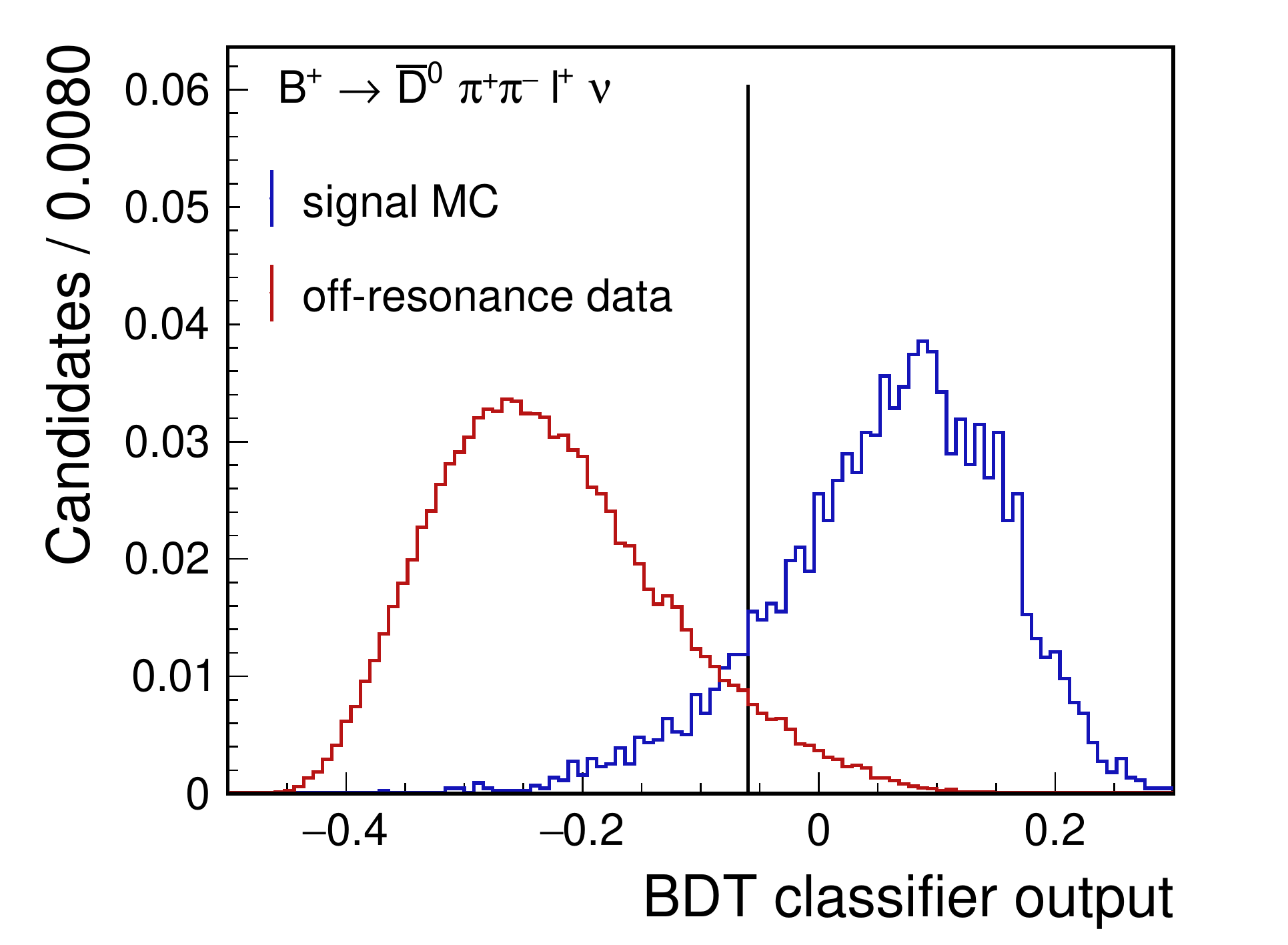}\hfill
\includegraphics[width=0.48\textwidth]{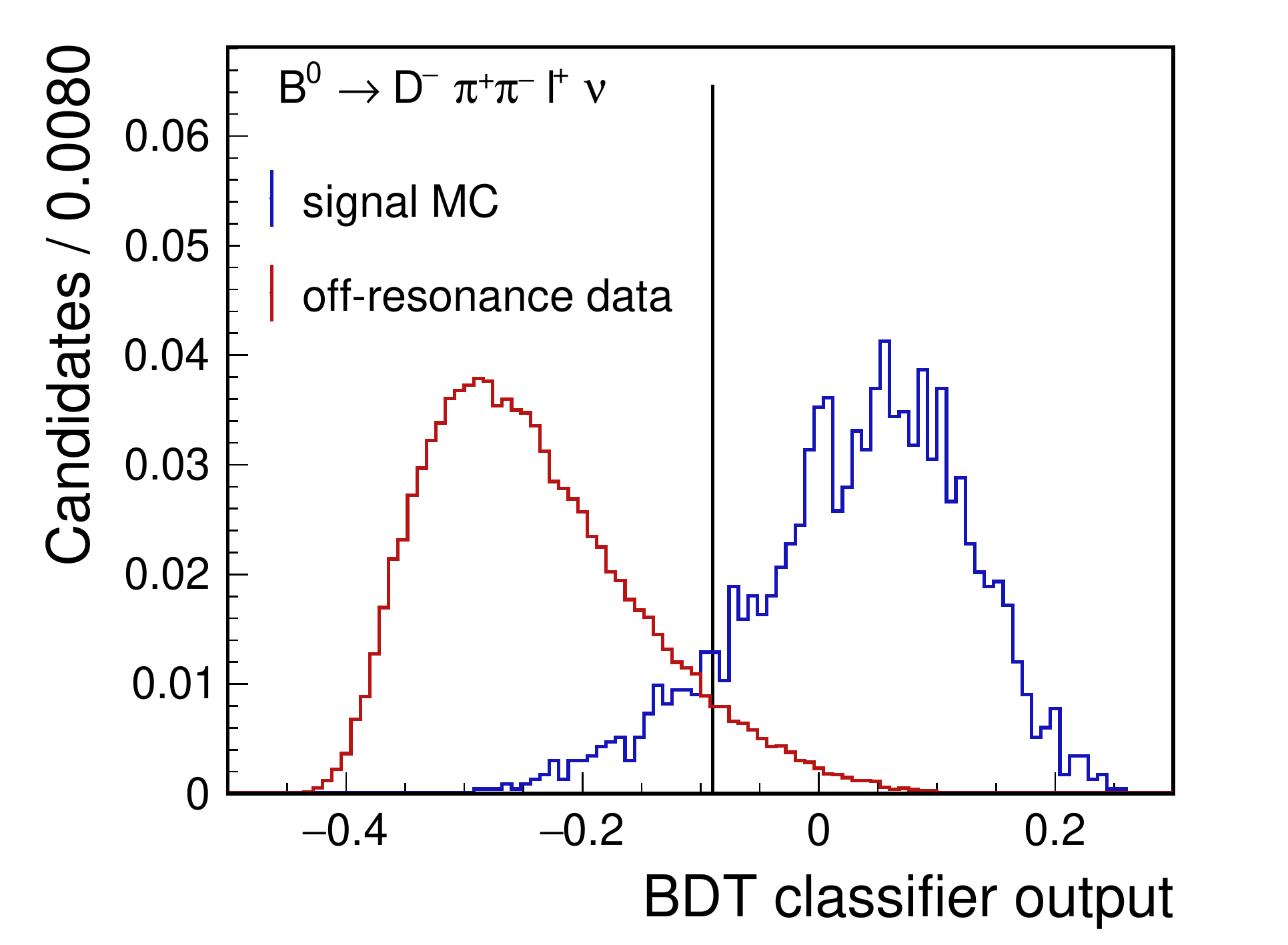}
\hspace*{\fill}
\caption{Distribution of BDT output classifier for \BuToDzpipilnu
 (left) and \BdToDmpipilnu (right). Candidates to the right of the vertical line are retained.}
\label{fig:selection:bdt}
\end{figure}

The \BToDpipilnu mode has much more background than the \BToDstlnu
and \BToDorDstpilnu modes. In order to increase the sensitivity of this
channel, a boosted decision tree (BDT)~\cite{BDT} is used to further reduce
the background. The following 25 input variables are used in the BDT:
$E_\text{extra}$, the unaccounted energy in the ECL; $R_1 - R_4$, the ratios of
the first, second, third, and fourth to the zeroth Fox-Wolfram moments~\cite
{Fox-Wolfram}; $H_0 - H_4$, the harmonic moments of zeroth to fourth order with
respect to the thrust axis (Chapter~9.3 of Ref.~\cite{BFactories}); $C_0 - C_8$, the momentum flow in
nine cones of \SI{10}{\degrees} around the thrust axis~\cite{CLEO}; the
sphericity and the aplanarity of the event(Chapter~9.3 of Ref.~\cite{BFactories}); the thrust
value of the event and the cosine of the polar angle of the thrust axis(Chapter~9.3 of Ref.~\cite{BFactories}); the number of tracks used in the \Btag reconstruction; the
number of neutral clusters used in the \Btag reconstruction. The BDT is
trained with signal MC simulations and off-resonance data, as most
of the remaining background originates from $\decay{\epem}{\qqbar}$ ($q$ = $u$, $d$, $s$, $c$) \enquote
{continuum} events. The signal MC is divided into $N$ subsamples each containing the number of
expected candidates in the full Belle dataset based on the
branching fraction results of the \babar measurement~\cite{Lees:2015eya}. For
each subsample an individual BDT is trained using the other $N-1$ subsamples
such that the size of the training sample is maximized while keeping it
independent from the sample that the BDT is applied to and therefore avoiding
bias. Separate BDTs are trained for the \Bu and \Bd modes. The distribution
of all BDT output classifiers combined is shown in \cref{fig:selection:bdt}.
The BDT output variable is required to be greater than \num
{-0.06} for \BuToDorDstpipilnu and greater than \num
{-0.09} for \BdToDorDstpipilnu, which maximizes the ratio between the signal
yield and its uncertainty from a fit to MC samples.

\subsection{\texorpdfstring{$\FourS$}{Y(4S)} selection}

A total of \num {12} \Bu modes and \num{12} \Bd modes are reconstructed.
Each \Bsig candidate and \Btag candidate are combined to form an $\FourS$
candidate. In the combinations the electric charge must be conserved but the flavor of two neutral \PB mesons is allowed to be the same. Candidates with tracks that are not assigned to the $\FourS$ candidate
are rejected. In events that contain more than one $\FourS$ candidate, a
single candidate is selected, as follows. Firstly, the \Btag candidate with
the highest FEI signal probability is selected. If multiple \Bsig candidates
remain, the \Dstar mode is preferred over the \PD mode, since otherwise an
additional \piz candidate would be left in the event. In some events,
candidates are reconstructed in both the one-pion and the two-pion modes.
As the distribution of $U = \Emiss - \pmiss c$ is used for signal extraction,
if $U$ is between \SIrange[range-phrase=\,and\,]{-0.1}{0.1}{\gev} for at
least one candidate in both decay modes, all candidates in the event are
rejected. If there are still multiple candidates, only the one with the
smallest difference between the $\D^{(*)}$ candidate mass and the nominal
mass is retained.

For each signal MC mode, the efficiency is taken to be the fraction of
correctly reconstructed candidates. A weighted average of the efficiencies
based on the relative abundances of the \Dstst state reported by the
PDG~\cite{PDG2022} is taken as the final efficiency value. The ratios between
the efficiencies of the signal and normalization modes
\begin{equation}
R(\Bu) = \frac{\epsilon(\BuToDststlnu)}{\epsilon(\BuToDstlnu)} \quad \text{and} \quad R(\Bd) = \frac{\epsilon(\BdToDststlnu)}{\epsilon(\BdToDstlnu)}
\end{equation}
are given in \cref{tab:selection:efficiencyratio}.

\begin{table}[ht]
\caption{Ratios between the selection efficiencies of the signal and normalization modes. The uncertainty is the MC sample statistical uncertainty.}
\centering
\sisetup{table-format = 1.4(1), separate-uncertainty}
\setlength{\tabcolsep}{20pt}
\begin{tabular}{lSS}
\toprule
            & {Electron mode} & {Muon mode}   \\
\midrule
\BdToDzpilnu      & 1.194\pm0.012 & 1.139\pm0.010 \\
\BuToDmpilnu      & 0.518\pm0.004 & 0.482\pm0.004 \\
\BdToDstpilnu     & 1.156\pm0.009 & 1.094\pm0.009 \\
\BuToDstpilnu     & 0.4040\pm0.0026 & 0.3824\pm0.0027 \\
\BdToDmpipilnu      & 0.450\pm0.015 & 0.389\pm0.014 \\
\BuToDzpipilnu      & 0.288\pm0.007 & 0.264\pm0.007 \\
\BdToDstpipilnu     & 0.286\pm0.016 & 0.270\pm0.017 \\
\BuToDstpipilnu     & 0.220\pm0.008 & 0.179\pm0.008 \\
\bottomrule
\end{tabular}
\label{tab:selection:efficiencyratio}
\end{table}

\section{Extraction of signal yields}

The number of signal candidates is determined with an unbinned extended maximum
likelihood fit of $U = \Emiss - \pmiss\,c$. The probability density
function (PDF) used to describe the $U$ distribution is constructed from
templates based on the MC.

\subsection{Fit of \texorpdfstring{\BToDorDstlnu}{B -> D(*) l nu} sample}
\label{sec:fitting:btodstlnu}

For the fit of the $\BToDorDstlnu$ sample the total PDF consists of four (three)
components
\begin{align}
\pdf{$\Dbar$\electron}{} &= \yield{$\Dbar$\electron}{sig} \pdf{$\Dbar$\electron}{sig} + \frac{\effc{$\Dbar$\electron}{MC}}{\effc{$\Dbar$\electron}{MC}+\effc{\Dstarb\electron}{MC}} \yield{\Dstarb\electron}{sig} \pdf{$\Dbar$\electron}{fd} + f^{\Dbar\electron}_{\rm bkg} \yield{$\Dbar$\electron}{bkg} \pdf{$\Dbar$\electron}{\BBbar} + \big(1 - f^{\Dbar\electron}_{\rm bkg}\big) \yield{$\Dbar$\electron}{bkg} \pdf{$\Dbar$\lepton}{cont},\\
\pdf{$\Dbar$\muon}{} &= \yield{$\Dbar$\muon}{sig} \pdf{$\Dbar$\muon}{sig} + \frac{\effc{$\Dbar$\muon}{MC}}{\effc{$\Dbar$\muon}{MC}+\effc{\Dstarb\muon}{MC}}\yield{\Dstarb\muon}{sig} \pdf{$\Dbar$\muon}{fd} + f^{\Dbar\muon}_{\rm bkg} \yield{$\Dbar$\muon}{bkg} \pdf{$\Dbar$\muon}{\BBbar} + \big(1 - f^{\Dbar\muon}_{\rm bkg}\big) \yield{$\Dbar$\muon}{bkg} \pdf{$\Dbar$\lepton}{cont}, \\
\pdf{\Dstarb\electron}{} &= \frac{\effc{\Dstarb\electron}{MC}}{\effc{$\Dbar$\electron}{MC}+\effc{\Dstarb\electron}{MC}} \yield{\Dstarb\electron}{sig} \pdf{\Dstarb\electron}{sig} + f^{\Dstarb\electron}_{\rm bkg} \yield{\Dstarb\electron}{bkg} \pdf{\Dstarb\electron}{\BBbar} + \big(1 - f^{\Dstarb\electron}_{\rm bkg}\big) \yield{\Dstarb\electron}{bkg} \pdf{\Dstarb\lepton}{cont}, \\
\pdf{\Dstarb\muon}{} &= \frac{\effc{\Dstarb\muon}{MC}}{\effc{$\Dbar$\muon}{MC} + \effc{\Dstarb\muon}{MC}} \yield{\Dstarb\muon}{sig} \pdf{\Dstarb\muon}{sig} + f^{\Dstarb\muon}_{\rm bkg} \yield{\Dstarb\muon}{bkg} \pdf{\Dstarb\muon}{\BBbar} + \big(1 - f^{\Dstarb\muon}_{\rm bkg}\big) \yield{\Dstarb\muon}{bkg} \pdf{\Dstarb\lepton}{cont},
\end{align}
where \pdf{$\Dbar{}^{(*)}\lepton$}{sig} is the signal PDF and \pdf{$\Dbar$\lepton}
{fd}, \pdf{$\Dbar{}^{(*)}\lepton$}{\BBbar}, and \pdf{$\Dbar{}^{(*)}\lepton$}{cont} are
the PDFs describing the feeddown, \BBbar background, and continuum
background, respectively. Feeddown describes a contribution from $\BToDstlnu$
that shows up in the $\BToDlnu$ modes if the neutral pion of a \DstzToDzpi or a \DstpToDpizero
decay is missed in the reconstruction. Due to the missing \piz it is shifted
to higher values in the $U$ distribution. Thus, this contribution can be
separated and used to improve the sensitivity of the branching fraction
measurement.

The fraction of the \BBbar component among the total background, $f_{\rm bkg}$,
is constrained to the values estimated in simulation. A simultaneous fit of
$\BToDlnu$ and $\BToDstlnu$ is performed, where the total $\BToDstlnu$
yield $\yield{\Dstarb\lepton}{sig}$ is determined as the sum of the signal
and feeddown components, which are related via their efficiencies $\effc{}
{MC}$. The templates used to construct the PDFs are created with \num
{125} bins between \num{-0.5} and \SI{2}{\gev}. Separate PDFs are used for the
electron and muon modes except for the continuum PDF, which is created from
the combined sample of the two modes as their distributions are statistically
compatible with each other.

The width of the signal peak in the $U$ distribution differs between data and
MC, even after all known corrections are applied. To compensate for this
effect, the signal PDFs are constructed by convolving the signal-MC templates
with a Gaussian whose mean and width are floating in the fit to data.
Independent widths are used for the electron and muon modes. The
fitted \BToDorDstlnu signal and background yields are listed in \cref
{tab:fitting:datayields_0pi} and the corresponding plots are shown
in \crefrange{fig:fitting:bdtodlnu}{fig:fitting:butodstlnu}. In \cref{appendix}
the fit results of the mean and width of the Gaussian are listed. 

\begin{figure}[ht]
\hspace*{\fill}
\begin{overpic}
[width=0.48\textwidth]{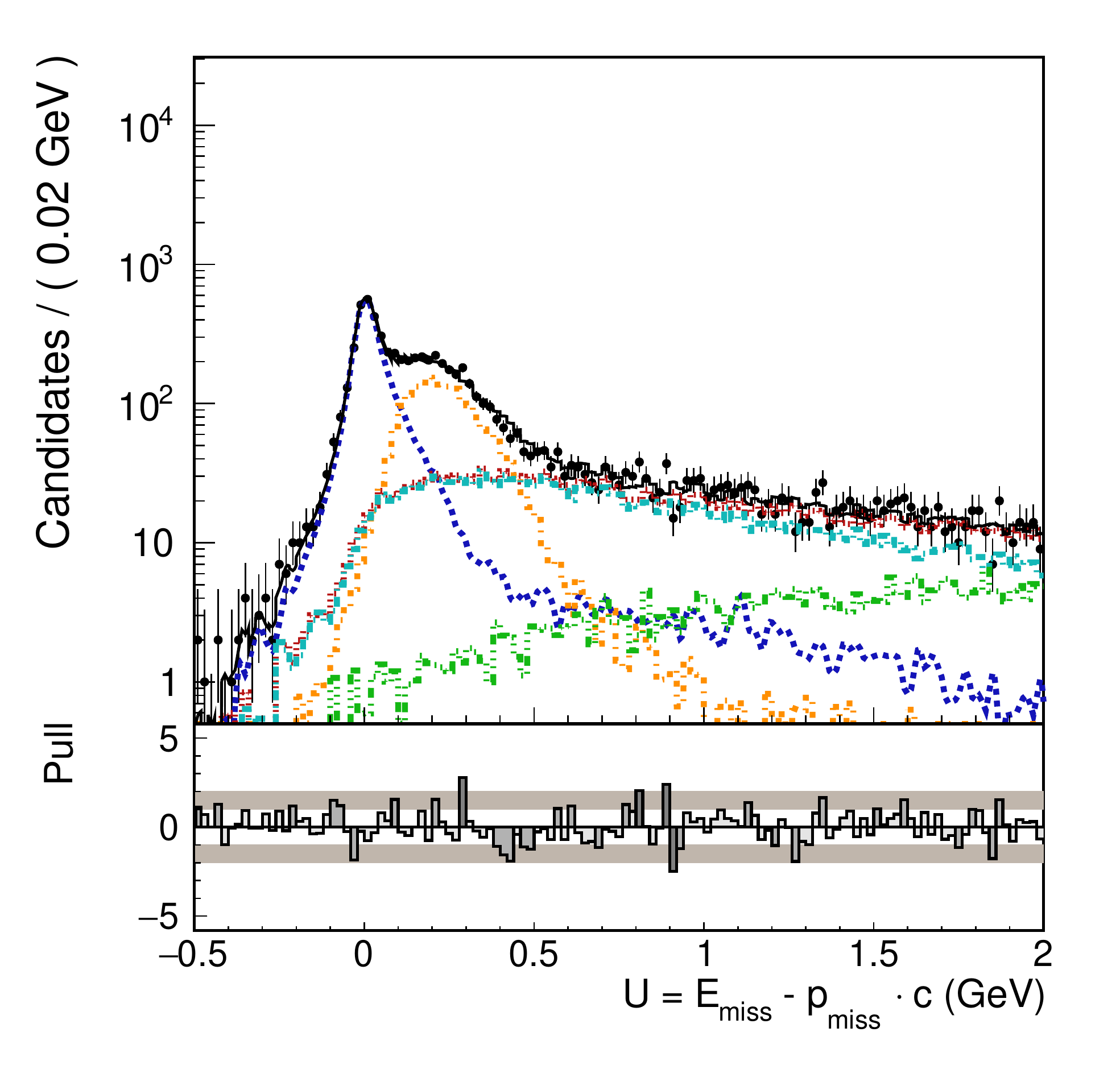}
\put(45,195){Belle}
\put(45,180){\BdToDmenu}
\put(135,190){\textcolor{black}{\rule{5mm}{2pt}}\,\,\scalefont{0.8}Total}
\put(135,178){\textcolor{nice_blue}{\rule{5mm}{2pt}}\,\,\scalefont{0.8}Signal}
\put(135,166){\textcolor{nice_yellow}{\rule{5mm}{2pt}}\,\,\scalefont{0.8}Feeddown}
\put(135,154){\textcolor{nice_red}{\rule{5mm}{2pt}}\,\,\scalefont{0.8}Background}
\put(145,144){\textcolor{nice_green}{\rule{5mm}{2pt}}\,\,\scalefont{0.8}Continuum}
\put(145,134){\textcolor{nice_turquoise}{\rule{5mm}{2pt}}\,\,\scalefont{0.8}\BBbar}
\end{overpic}
\hfill
\begin{overpic}
[width=0.48\textwidth]{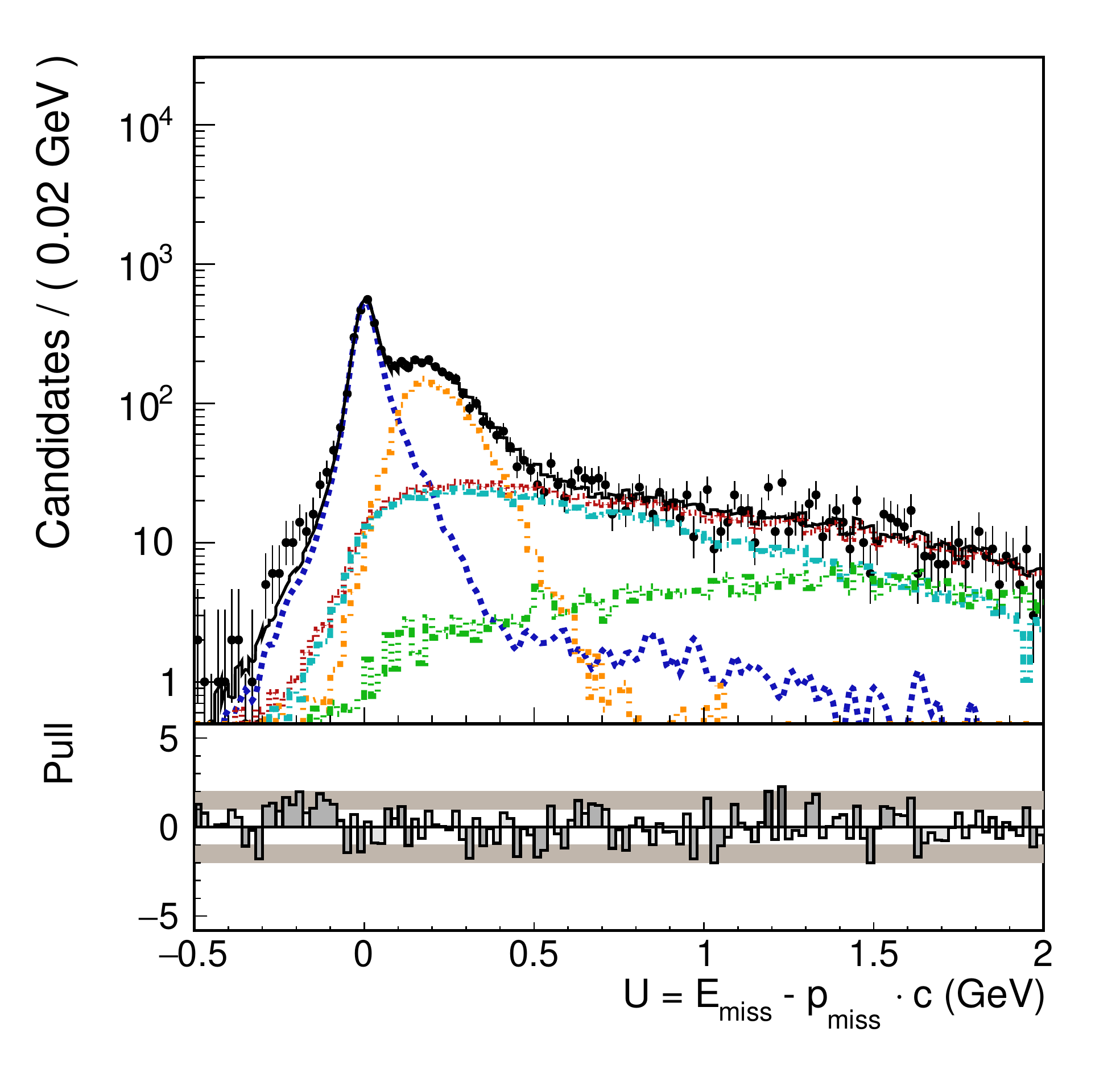}
\put(45,195){Belle}
\put(45,180){\BdToDmmunu}
\put(135,190){\textcolor{black}{\rule{5mm}{2pt}}\,\,\scalefont{0.8}Total}
\put(135,178){\textcolor{nice_blue}{\rule{5mm}{2pt}}\,\,\scalefont{0.8}Signal}
\put(135,166){\textcolor{nice_yellow}{\rule{5mm}{2pt}}\,\,\scalefont{0.8}Feeddown}
\put(135,154){\textcolor{nice_red}{\rule{5mm}{2pt}}\,\,\scalefont{0.8}Background}
\put(145,144){\textcolor{nice_green}{\rule{5mm}{2pt}}\,\,\scalefont{0.8}Continuum}
\put(145,134){\textcolor{nice_turquoise}{\rule{5mm}{2pt}}\,\,\scalefont{0.8}\BBbar}
\end{overpic}
\hspace*{\fill}
\caption{Distribution of $\Emiss - \pmiss\,c$ of $\BdToDmenu$ (left) and
 $\BdToDmmunu$ (right) for the data. The MC shapes, normalized according to
 the result of the fit, are also shown.}
\label{fig:fitting:bdtodlnu}
\end{figure}

\begin{figure}[ht]
\hspace*{\fill}
\begin{overpic}
[width=0.48\textwidth]{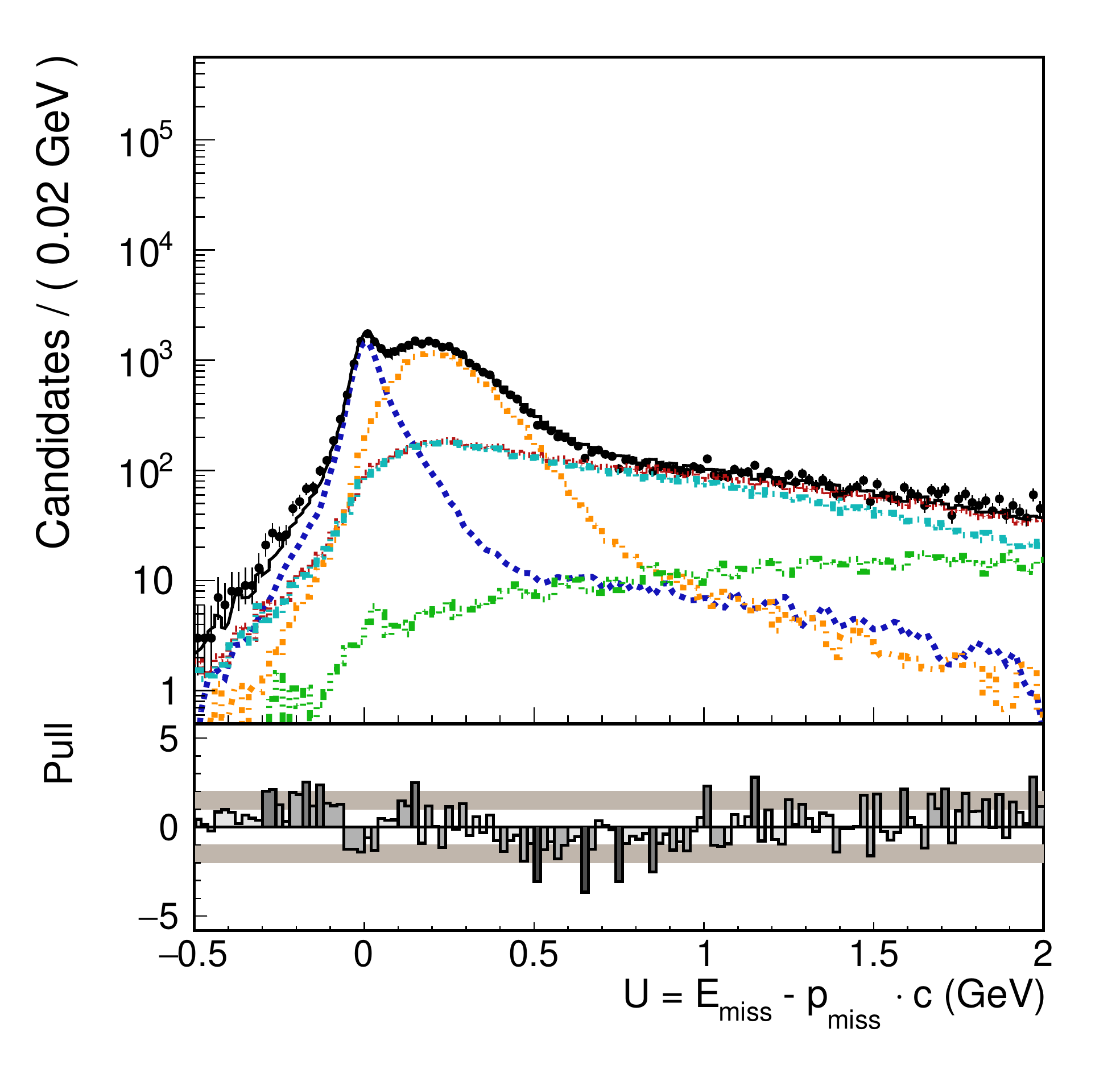}
\put(45,195){Belle}
\put(45,180){\BuToDzenu}
\put(135,190){\textcolor{black}{\rule{5mm}{2pt}}\,\,\scalefont{0.8}Total}
\put(135,178){\textcolor{nice_blue}{\rule{5mm}{2pt}}\,\,\scalefont{0.8}Signal}
\put(135,166){\textcolor{nice_yellow}{\rule{5mm}{2pt}}\,\,\scalefont{0.8}Feeddown}
\put(135,154){\textcolor{nice_red}{\rule{5mm}{2pt}}\,\,\scalefont{0.8}Background}
\put(145,144){\textcolor{nice_green}{\rule{5mm}{2pt}}\,\,\scalefont{0.8}Continuum}
\put(145,134){\textcolor{nice_turquoise}{\rule{5mm}{2pt}}\,\,\scalefont{0.8}\BBbar}
\end{overpic}
\hfill
\begin{overpic}
[width=0.48\textwidth]{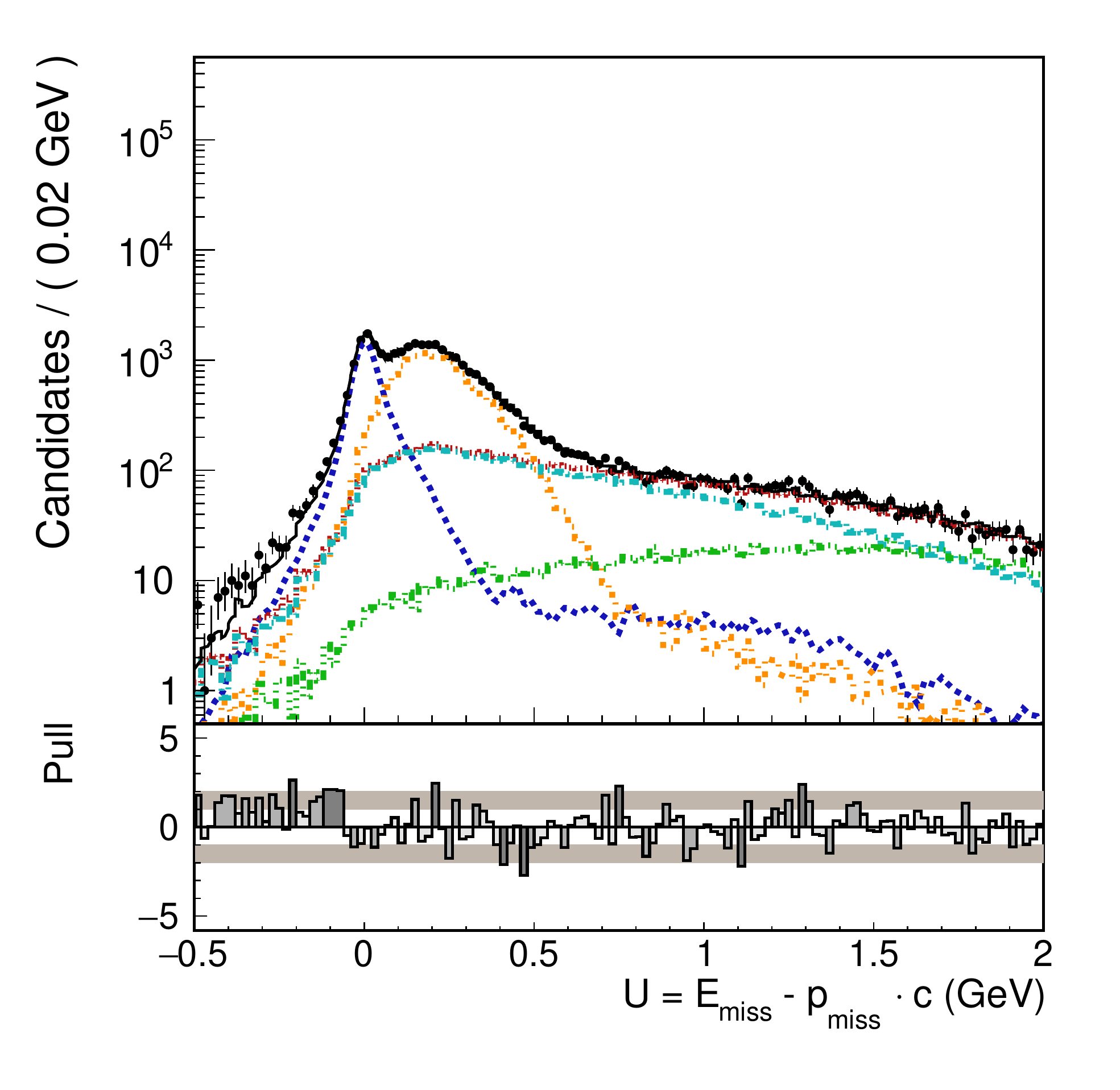}
\put(45,195){Belle}
\put(45,180){\BuToDzmunu}
\put(135,190){\textcolor{black}{\rule{5mm}{2pt}}\,\,\scalefont{0.8}Total}
\put(135,178){\textcolor{nice_blue}{\rule{5mm}{2pt}}\,\,\scalefont{0.8}Signal}
\put(135,166){\textcolor{nice_yellow}{\rule{5mm}{2pt}}\,\,\scalefont{0.8}Feeddown}
\put(135,154){\textcolor{nice_red}{\rule{5mm}{2pt}}\,\,\scalefont{0.8}Background}
\put(145,144){\textcolor{nice_green}{\rule{5mm}{2pt}}\,\,\scalefont{0.8}Continuum}
\put(145,134){\textcolor{nice_turquoise}{\rule{5mm}{2pt}}\,\,\scalefont{0.8}\BBbar}
\end{overpic}
\hspace*{\fill}
\caption{Distribution of $\Emiss - \pmiss\,c$ of $\BuToDzenu$ (left) and
 $\BuToDzmunu$ (right) for the data. The MC shapes, normalized according to
 the result of the fit, are also shown.}
\label{fig:fitting:butodlnu}
\end{figure}

\begin{figure}[ht]
\hspace*{\fill}
\begin{overpic}
[width=0.48\textwidth]{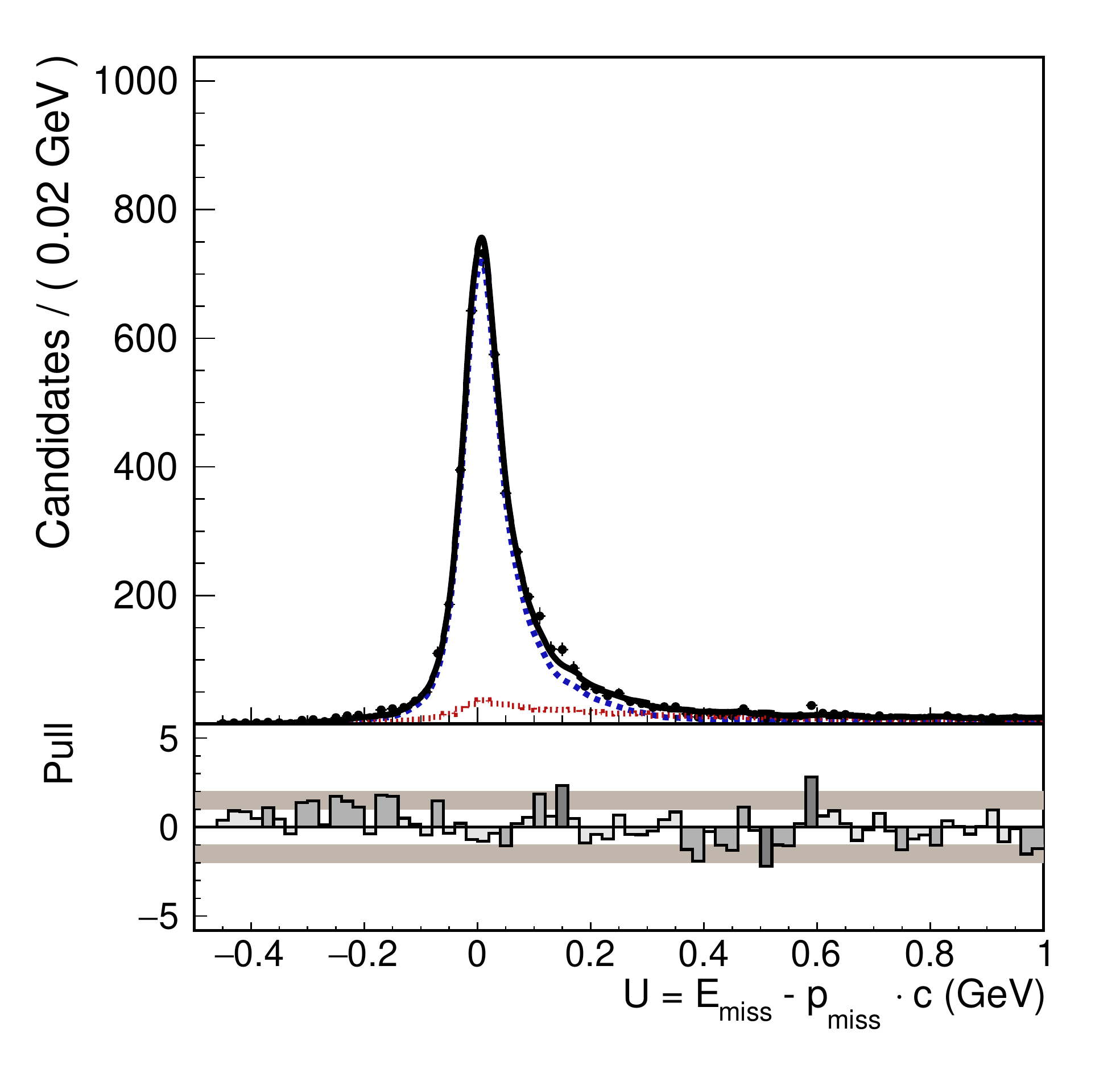}
\put(45,180){Belle}
\put(110,180){\BdToDstenu}
\put(120,165){\textcolor{black}{\rule{5mm}{2pt}}\,\,Total}
\put(120,150){\textcolor{nice_blue}{\rule{5mm}{2pt}}\,\,Signal}
\put(120,135){\textcolor{nice_red}{\rule{5mm}{2pt}}\,\,Background}
\end{overpic}
\hfill
\begin{overpic}
[width=0.48\textwidth]{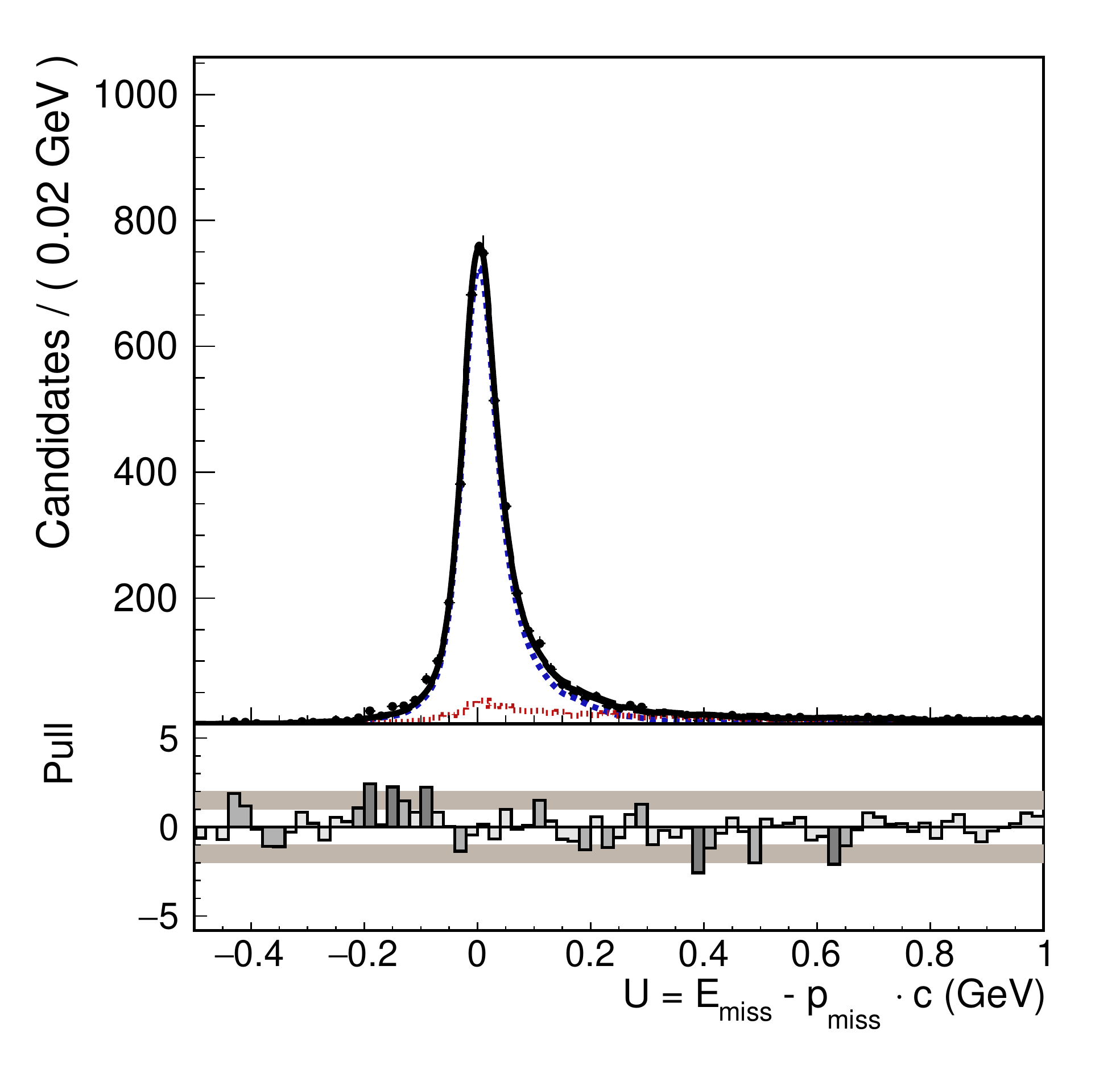}
\put(45,180){Belle}
\put(110,180){\BdToDstmunu}
\put(120,165){\textcolor{black}{\rule{5mm}{2pt}}\,\,Total}
\put(120,150){\textcolor{nice_blue}{\rule{5mm}{2pt}}\,\,Signal}
\put(120,135){\textcolor{nice_red}{\rule{5mm}{2pt}}\,\,Background}
\end{overpic}
\hspace*{\fill}
\caption{Distribution of $\Emiss - \pmiss\,c$ of $\BdToDstenu$ (left) and
 $\BdToDstmunu$ (right) for the data. The MC shapes, normalized according to
 the result of the fit, are also shown.}
\label{fig:fitting:bdtodstlnu}
\end{figure}

\begin{figure}[ht]
\hspace*{\fill}
\begin{overpic}
[width=0.48\textwidth]{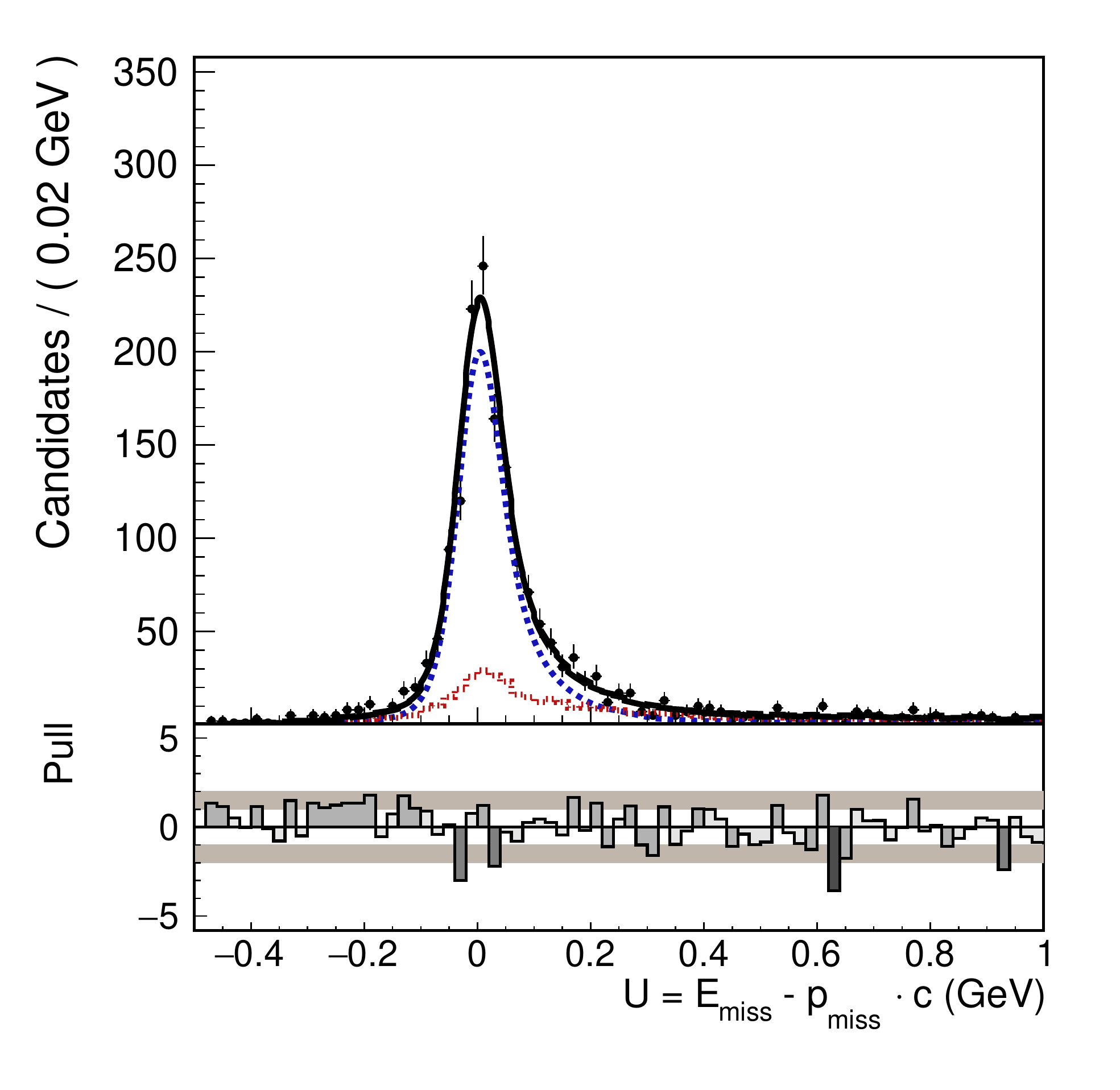}
\put(45,180){Belle}
\put(105,180){\BuToDstenu}
\put(120,165){\textcolor{black}{\rule{5mm}{2pt}}\,\,Total}
\put(120,150){\textcolor{nice_blue}{\rule{5mm}{2pt}}\,\,Signal}
\put(120,135){\textcolor{nice_red}{\rule{5mm}{2pt}}\,\,Background}
\end{overpic}
\hfill
\begin{overpic}
[width=0.48\textwidth]{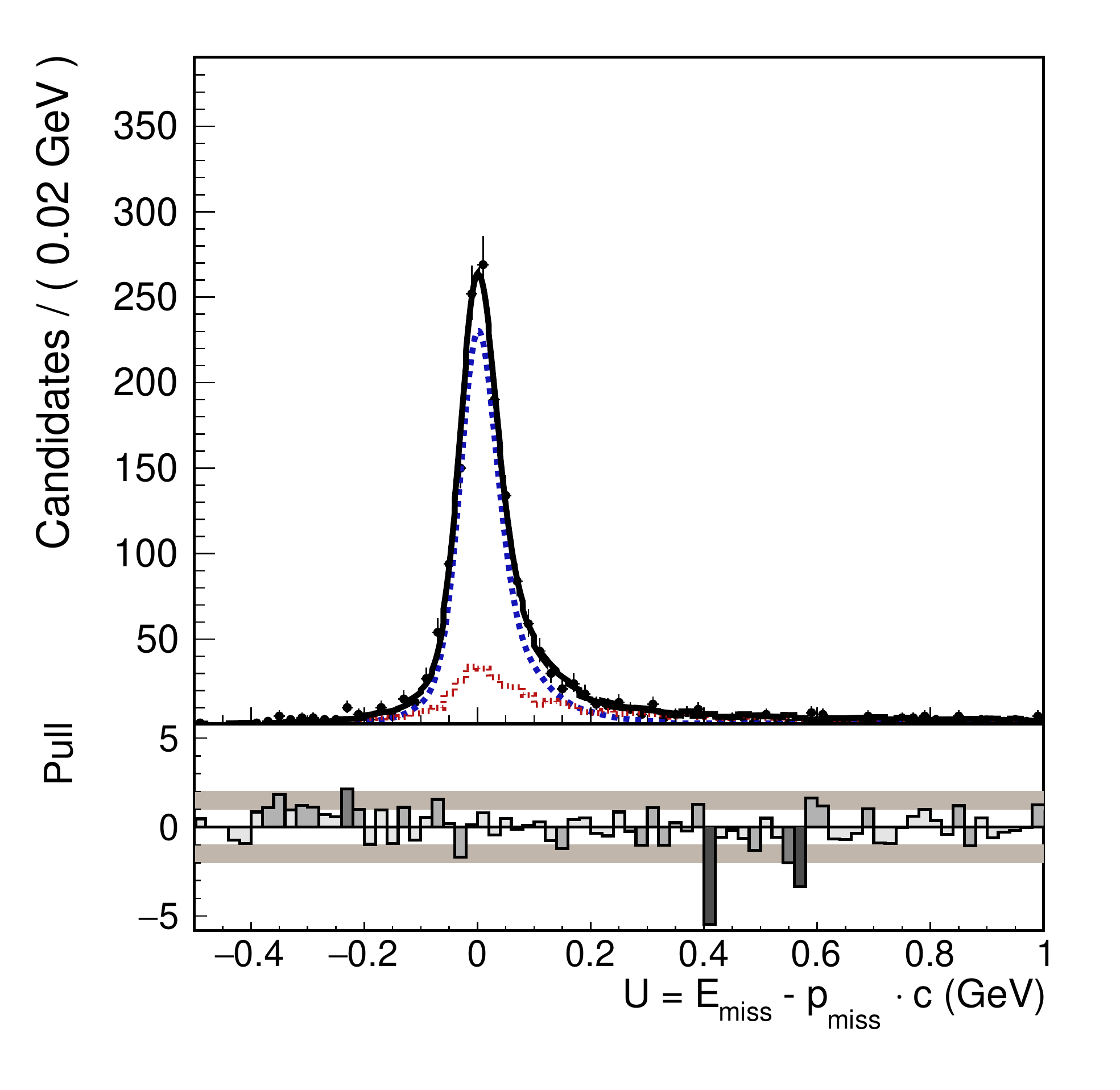}
\put(45,180){Belle}
\put(105,180){\BuToDstmunu}
\put(120,165){\textcolor{black}{\rule{5mm}{2pt}}\,\,Total}
\put(120,150){\textcolor{nice_blue}{\rule{5mm}{2pt}}\,\,Signal}
\put(120,135){\textcolor{nice_red}{\rule{5mm}{2pt}}\,\,Background}
\end{overpic}
\hspace*{\fill}
\caption{Distribution of $\Emiss - \pmiss\,c$ of $\BuToDstenu$ (left) and
 $\BuToDstmunu$ (right) for the data. The MC shapes, normalized according to
 the result of the fit, are also shown.}
\label{fig:fitting:butodstlnu}
\end{figure}

\begin{table}[ht]
\caption{Fitted signal and background yields of the normalization
 modes \BToDorDstlnu in the full \belle data sample. The quoted uncertainties
 are statistical only.}
\centering
\begin{tabular}{lS[table-format = 3.0(2)]S[table-format = 3.0(2)]S[table-format = 3.0(2)]S[table-format = 3.0(2)]}
\toprule
                        &   \multicolumn{2}{c}{Signal}          &   \multicolumn{2}{c}{Background}  \\
                        &   {electron mode} &   {muon mode}     &   {electron mode} &   {muon mode} \\
\midrule
\BdToDmlnu              &   {$3154\pm\phantom{1}67$} &   {$2723\pm\phantom{1}60$}       &   {$2097\pm\phantom{1}60$}       &   {$1696\pm\phantom{1}52$}   \\
\BuToDzlnu              &   8974\pm136      &   7752\pm124      &   9840\pm149      &   8548\pm133  \\
\BdToDstlnu             &   6271\pm102      &   {$5624\pm\phantom{1}91$}   &   {$\phantom{1}925\pm\phantom{1}54$}        &   {$\phantom{1}742\pm\phantom{1}44$}    \\
\BuToDstlnu             &   19940\pm200     &   18045\pm183     &   {$\phantom{1}523\pm\phantom{1}35$}        &   {$\phantom{1}508\pm\phantom{1}34$}    \\
\bottomrule
\end{tabular}
\label{tab:fitting:datayields_0pi}
\end{table}

\subsection{Fit of \texorpdfstring{\BToDorDstpilnu}{B -> D(*) pi l nu} and \texorpdfstring{\BToDorDstpipilnu}{B -> D(*) pi pi l nu} samples}
\label{sec:fitting:btodststlnu}

A simultaneous fit to the $U$ distribution of 16 categories splitting the full
sample according to the \B flavor mode (\Bd vs \Bu), the \D mode
(\Dz/\Dp vs \Dstarz/\Dstarp), the number of pion daughters (\Dpi vs \Dpipi),
and the lepton mode (\electron vs \muon) is performed. This allows several background sources to
be constrained directly from the data, as described
below. All templates are constructed with \SI{120}{bins} in the range \SIrange{-1}{2}
{\gev}.

The \BuToDmpilnu fit PDF consists of five components: signal, feeddown,
misreconstructed \BToDststlnu background, other \BBbar, and continuum:
\begin{equation}
\begin{split}
\pdf{\Dm\pip\ellp}{} = \yield{\Dm\pip\ellp}{sig} \pdf{\Dm\pip\ellp}{sig} + \frac{\effc{\Dm\pip\ellp}{MC}}{\effc{\Dm\pip\ellp}{MC}+\effc{\Dstarm\pip\ellp}{MC}} \yield{\Dstarm\pip\ellp}{sig} \pdf{\Dm\pip\ellp}{fd}\\
+ \sum_{\rm i = 1}^{14} \yield{}{\Dstst, i} \pdf{\Dm\pip\ellp}{\Dstst, i} + \yield{\Dm\pip\ellp}{\BBbar} \pdf{\Dm\pip\ellp}{\BBbar} + \frac{L_{\rm on}}{L_{\rm off}} \pdf{\Dm\pip\ellp}{off}\ .
\end{split}
\end{equation}
The signal template \pdf{\Dm\pip\ellp}{sig} is obtained from signal MC, in
which the \Dpi is produced in \Dzstarz decay \num{62}\% of the time, and in \Dtwostarz decay \num{38}
\% of the time~\cite{PDG2022}. The feeddown component \pdf
{\Dm\pip\ellp}{fd} comes from \BuToDstpilnu decays and is taken from signal
MC, in which the \Dstpi final state is produced in \Donez decay \num{45}\% of the time,
in \Dprimeonez decay \num{40}\% of the time, and in \Dtwostarz decay the rest of the time.
The \BToDststlnu background PDF \pdf{\Dm\pip\ellp}{\Dstst, i} is obtained
from 14 different MC samples:
\begin{itemize}
    \item $\decay{\Bu}{\Dbar{}^{**0}\ellp\neul}$ with
$\Dbar{}^{**0} \in(\Dzstarzb, \Donezb, \Dprimeonezb, \Dtwostarzb)$,
    \item $\decay{\Bu}{\Donezb\ellp\neul}$ with $\decay{\Donezb}{\Dzb\pipi}$,
    \item $\BuToDzpipilnu$,
    \item $\BuToDstpipilnu$,
    \item $\decay{\Bd}{\D^{**-}\ellp\neul}$ with $\D^{**-} \in
(\Dzstarm, \Donem, \Dprimeonem, \Dtwostarm)$,
    \item $\decay{\Bd}{\Donem\ellp\neul}$ with $\decay{\Donem}{\Dm\pipi}$,
    \item $\BdToDmpipilnu$,
    \item $\BdToDstpipilnu$.
\end{itemize}
These events constitute background due to
misreconstructed signal candidates, swapping of final state particles between
the \Bsig and \Btag candidates, or events with \decay{\D^{**}}{\D^{
(*)}\piz}. The composition of the different \Dstst states is set to the world
averages of these modes~\cite{PDG2022}. The yields of the \BToDststlnu
background components, $\yield{}{\Dstst, i}$, are calculated as the product of the terms listed
in \cref{tab:fitting:yieldformulas}.
\begin{table}[ht]
\caption{The \BToDststlnu background yields are the product of the scaling factor, the corresponding signal yield, and the efficiency ratio.}
\centering
\resizebox{\textwidth}{!}{%
\begin{tabular}{lSll}
\toprule
\multirow{2}{*}{component}  &   {scaling}            &   {signal}   &   \multirow{2}{*}{efficiency ratio}     \\
                            &   {factor}             &   {yield}    &   \\
\midrule
\BuToDzstarlnu      & 0.62    & \yield{\Dm\pip\ellp}{sig}        & $\frac{N(\text{\BuToDzstarlnu with error in reconstruction as \BuToDmpilnu})}{N(\text{\BuToDzstarlnu correctly reconstructed as \BuToDmpilnu})}$   \\[0.5em]
\BuToDonelnu        & 0.45    & \yield{\Dstarm\pip\ellp}{sig}    & $\frac{N(\text{\BuToDonelnu reconstructed as \BuToDmpilnu})}{N(\text{\BuToDonelnu correctly reconstructed as \BuToDstpilnu})}$   \\[0.5em]
\BuToDoneprimelnu   & 0.4     & \yield{\Dstarm\pip\ellp}{sig}    & $\frac{N(\text{\BuToDoneprimelnu reconstructed as \BuToDmpilnu})}{N(\text{\BuToDoneprimelnu correctly reconstructed as \BuToDstpilnu})}$   \\[0.5em]
\BuToDtwostarlnu    & 0.38    & \yield{\Dm\pip\ellp}{sig}        & $\frac{N(\text{\BuToDtwostarlnu with error in reconstruction as \BuToDmpilnu})}{N(\text{\BuToDtwostarlnu correctly reconstructed as \BuToDmpilnu})}$   \\[0.5em]
\BuToDonelnu (\DonezToDpipi)   & 0.55    & \yield{\Dz\pipi\ellp}{sig}     & $\frac{N(\text{\BuToDonelnu reconstructed as \BuToDmpilnu})}{N(\text{\BuToDonelnu correctly reconstructed as \BuToDzpipilnu})}$   \\[0.5em]
\BuToDzpipilnu      & 0.45    & \yield{\Dz\pipi\ellp}{sig}       & $\frac{N(\text{\BuToDzpipilnu reconstructed as \BuToDmpilnu})}{N(\text{\BuToDzpipilnu correctly reconstructed as \BuToDzpipilnu})}$   \\[0.5em]
\BuToDstpipilnu     & 1       & \yield{\Dstarz\pipi\ellp}{sig}   & $\frac{N(\text{\BuToDstpipilnu reconstructed as \BuToDmpilnu})}{N(\text{\BuToDstpipilnu correctly reconstructed as \BuToDstpipilnu})}$   \\[0.5em]
\BdToDzstarlnu      & 0.71    & \yield{\Dz\pim\ellp}{sig}        & $\frac{N(\text{\BdToDzstarlnu reconstructed as \BuToDmpilnu})}{N(\text{\BdToDzstarlnu correctly reconstructed as \BdToDzpilnu})}$   \\[0.5em]
\BdToDonelnu        & 0.425   & \yield{\Dstarz\pim\ellp}{sig}    & $\frac{N(\text{\BdToDonelnu reconstructed as \BuToDmpilnu})}{N(\text{\BdToDonelnu correctly reconstructed as \BdToDstpilnu})}$   \\[0.5em]
\BdToDoneprimelnu   & 0.47    & \yield{\Dstarz\pim\ellp}{sig}    & $\frac{N(\text{\BdToDoneprimelnu reconstructed as \BuToDmpilnu})}{N(\text{\BdToDoneprimelnu correctly reconstructed as \BdToDstpilnu})}$   \\[0.5em]
\BdToDtwostarlnu    & 0.29    & \yield{\Dz\pim\ellp}{sig}        & $\frac{N(\text{\BdToDtwostarlnu reconstructed as \BuToDmpilnu})}{N(\text{\BdToDtwostarlnu correctly reconstructed as \BdToDzpilnu})}$   \\[0.5em]
\BdToDonelnu (\DonemToDpipi)   & 0.55    & \yield{\Dm\pipi\ellp}{sig}     & $\frac{N(\text{\BdToDonelnu reconstructed as \BuToDmpilnu})}{N(\text{\BdToDonelnu correctly reconstructed as \BdToDmpipilnu})}$   \\[0.5em]
\BdToDmpipilnu      & 0.45    & \yield{\Dm\pipi\ellp}{sig}       & $\frac{N(\text{\BdToDmpipilnu reconstructed as \BuToDmpilnu})}{N(\text{\BdToDmpipilnu correctly reconstructed as \BdToDmpipilnu})}$   \\[0.5em]
\BdToDstpipilnu     & 1       & \yield{\Dstarm\pipi\ellp}{sig}   & $\frac{N(\text{\BdToDstpipilnu reconstructed as \BuToDmpilnu})}{N(\text{\BdToDstpipilnu correctly reconstructed as \BdToDstpipilnu})}$   \\
\bottomrule
\end{tabular}%
}%
\label{tab:fitting:yieldformulas}
\end{table}
The other \BBbar background is taken from a generic \bToc MC sample with six times the
luminosity of data. Off-resonance data is used to model the continuum
PDF $\pdf{\Dm\pip\ellp}{off}$. The yield of the continuum contribution in the
fit is constrained via the ratio of the on- and off-resonance luminosities.
The ratio is allowed to float in the fit within a Gaussian constraint with a
width of \num{1}\%. This accounts for the uncertainty in the
determination of the luminosity ratio.

The fit model for \BdToDzpilnu is constructed similarly:
\begin{equation}
\begin{split}
\pdf{\Dzb\pim\ellp}{} = \yield{\Dzb\pim\ellp}{sig} \pdf{\Dzb\pim\ellp}{sig} + \frac{\effc{\Dzb\pim\ellp}{MC}}{\effc{\Dzb\pim\ellp}{MC}+\effc{\Dstarzb\pim\ellp}{MC}} \yield{\Dstarzb\pim\ellp}{sig} \pdf{\Dzb\pim\ellp}{fd}\\
+ \sum_{\rm i = 1}^{14} \yield{}{\Dstst, i} \pdf{\Dzb\pim\ellp}{\Dstst, i} + \yield{\Dzb\pim\ellp}{\BBbar} \pdf{\Dzb\pim\ellp}{\BBbar} + \frac{L_{\rm on}}{L_{\rm off}} \pdf{\Dzb\pim\ellp}{off}\\
+ \frac{\epsilon(\text{\BdToDstlnu reconstructed as \BdToDzpilnu})}{\epsilon(\text{\BdToDstlnu reconstructed as \BdToDstlnu})} \yield{\Dstarm\ellp}{sig} \pdf{\Dzb\pim\ellp}{\Dstarm\ellp}.
\end{split}
\end{equation}
The signal composition is \num{71}\% \BdToDzstarlnu and \num{29}\%
\BdToDtwostarlnu decays~\cite{PDG2022}. The feeddown is produced in \BdToDonelnu decay \num{42.5}\%
of the time, in \BdToDoneprimelnu decay \num{47}\% of the time, and in \BdToDtwostarlnu decay \num
{10.5}\% of the time. Compared to the \Bu mode an
additional sixth component is added to account for
misreconstructed \BdToDstlnu candidates that survive the \Dstar veto. The
yield of this component is fixed by the product of the $\BdToDstlnu$ signal
yield \yield{\Dstarm\ellp}{sig} from the fit described in \cref
{sec:fitting:btodstlnu} and the ratio of efficiencies of the \BdToDzpilnu
and \BdToDstlnu selections. 

The \BToDstpilnu fit models consist of only four components as there is
no feeddown. The strategy for modelling background from \BToDststlnu is the same as
for \BToDpilnu. The signal PDF template is obtained from signal MC, in which
the \Dstpi final state is produced in \Done decay, \Dprimeone decay, and \Dtwostar
decay at the same proportions as the feeddown components in \BToDpilnu described above.

For \BToDpipilnu the fit model contains four components (signal, feeddown,
other \BBbar, continuum), while for \BToDstpipilnu only three components are
needed as there is no feeddown. Following the findings of the \babar measurement~\cite
{Lees:2015eya} the signal is assumed to proceed via a \Done resonance for
the \Dpipi modes and via a \Dprimeone resonance for the $\Dstar\pi\pi$ modes.
The \BToDstpipilnu templates are constructed with \SI{30}{bins} in the
range \SIrange{-0.5}{1}{\gev}.

\begin{figure}[ht]
\hspace*{\fill}
\begin{overpic}
[width=0.48\textwidth]{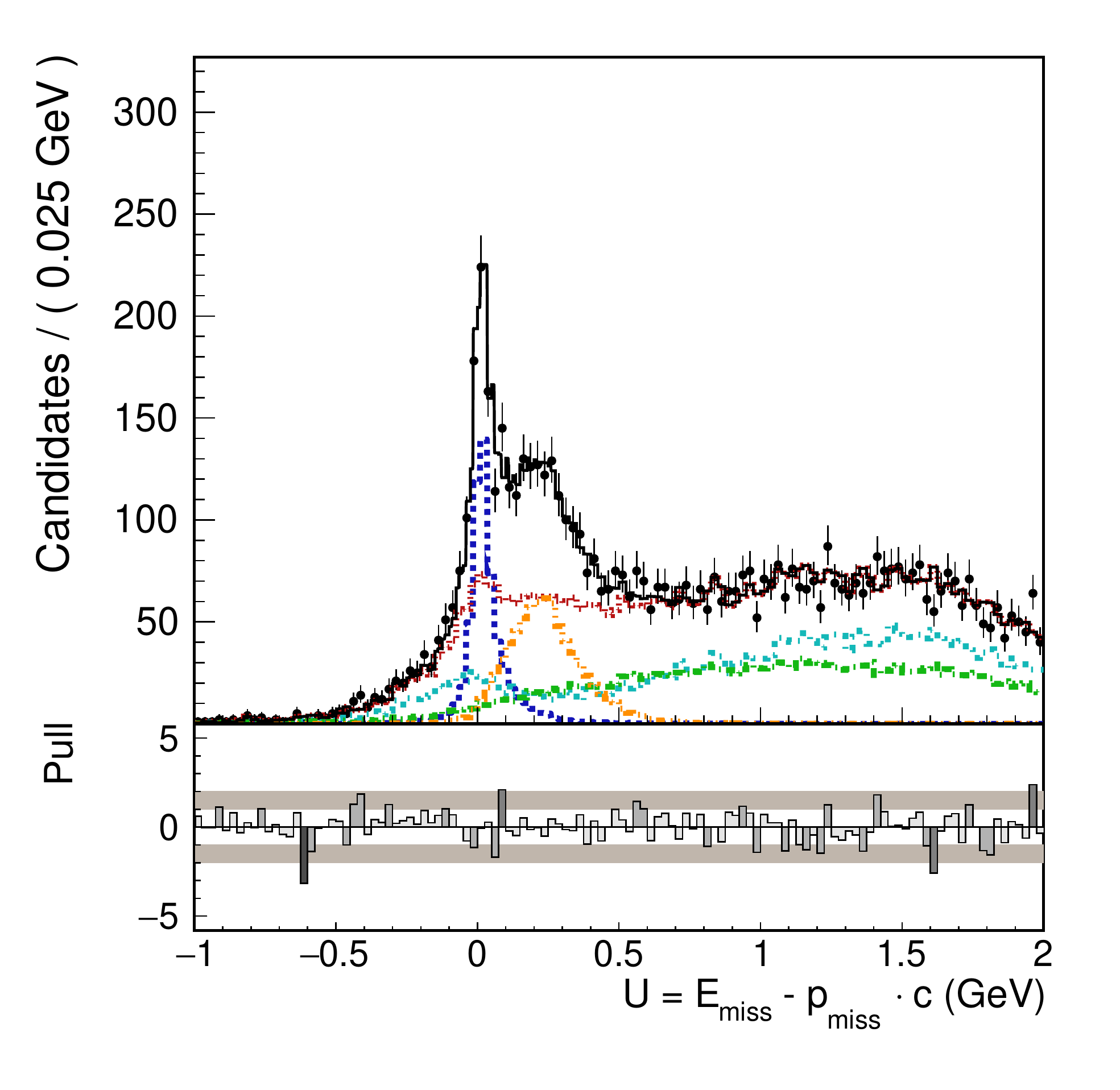}
\put(45,195){Belle}
\put(45,180){\BdToDzpienu}
\put(135,190){\textcolor{black}{\rule{5mm}{2pt}}\,\,\scalefont{0.8}Total}
\put(135,178){\textcolor{nice_blue}{\rule{5mm}{2pt}}\,\,\scalefont{0.8}Signal}
\put(135,166){\textcolor{nice_yellow}{\rule{5mm}{2pt}}\,\,\scalefont{0.8}Feeddown}
\put(135,154){\textcolor{nice_red}{\rule{5mm}{2pt}}\,\,\scalefont{0.8}Background}
\put(145,144){\textcolor{nice_green}{\rule{5mm}{2pt}}\,\,\scalefont{0.8}Continuum}
\put(145,134){\textcolor{nice_turquoise}{\rule{5mm}{2pt}}\,\,\scalefont{0.8}\BBbar}
\end{overpic}
\hfill
\begin{overpic}
[width=0.48\textwidth]{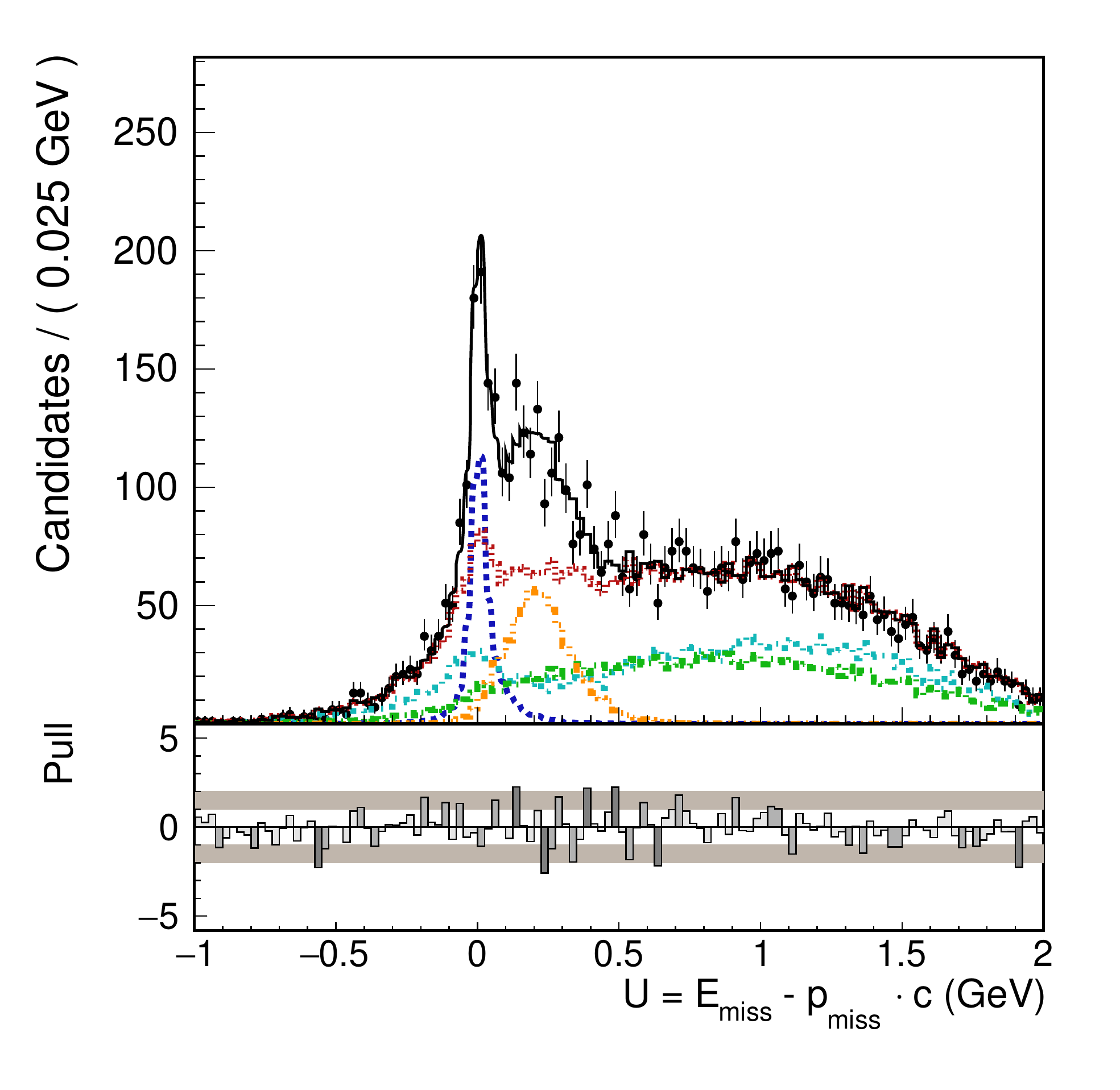}
\put(45,195){Belle}
\put(45,180){\BdToDzpimunu}
\put(135,190){\textcolor{black}{\rule{5mm}{2pt}}\,\,\scalefont{0.8}Total}
\put(135,178){\textcolor{nice_blue}{\rule{5mm}{2pt}}\,\,\scalefont{0.8}Signal}
\put(135,166){\textcolor{nice_yellow}{\rule{5mm}{2pt}}\,\,\scalefont{0.8}Feeddown}
\put(135,154){\textcolor{nice_red}{\rule{5mm}{2pt}}\,\,\scalefont{0.8}Background}
\put(145,144){\textcolor{nice_green}{\rule{5mm}{2pt}}\,\,\scalefont{0.8}Continuum}
\put(145,134){\textcolor{nice_turquoise}{\rule{5mm}{2pt}}\,\,\scalefont{0.8}\BBbar}
\end{overpic}
\hspace*{\fill}
\caption{Distribution of $\Emiss - \pmiss\,c$ of $\BdToDzpienu$ (left) and
 $\BdToDzpimunu$ (right) for the data. The MC shapes, normalized according to
 the result of the fit, are also shown.}
\label{fig:fitting:bdtodpilnu}
\end{figure}

\begin{figure}[ht]
\hspace*{\fill}
\begin{overpic}
[width=0.48\textwidth]{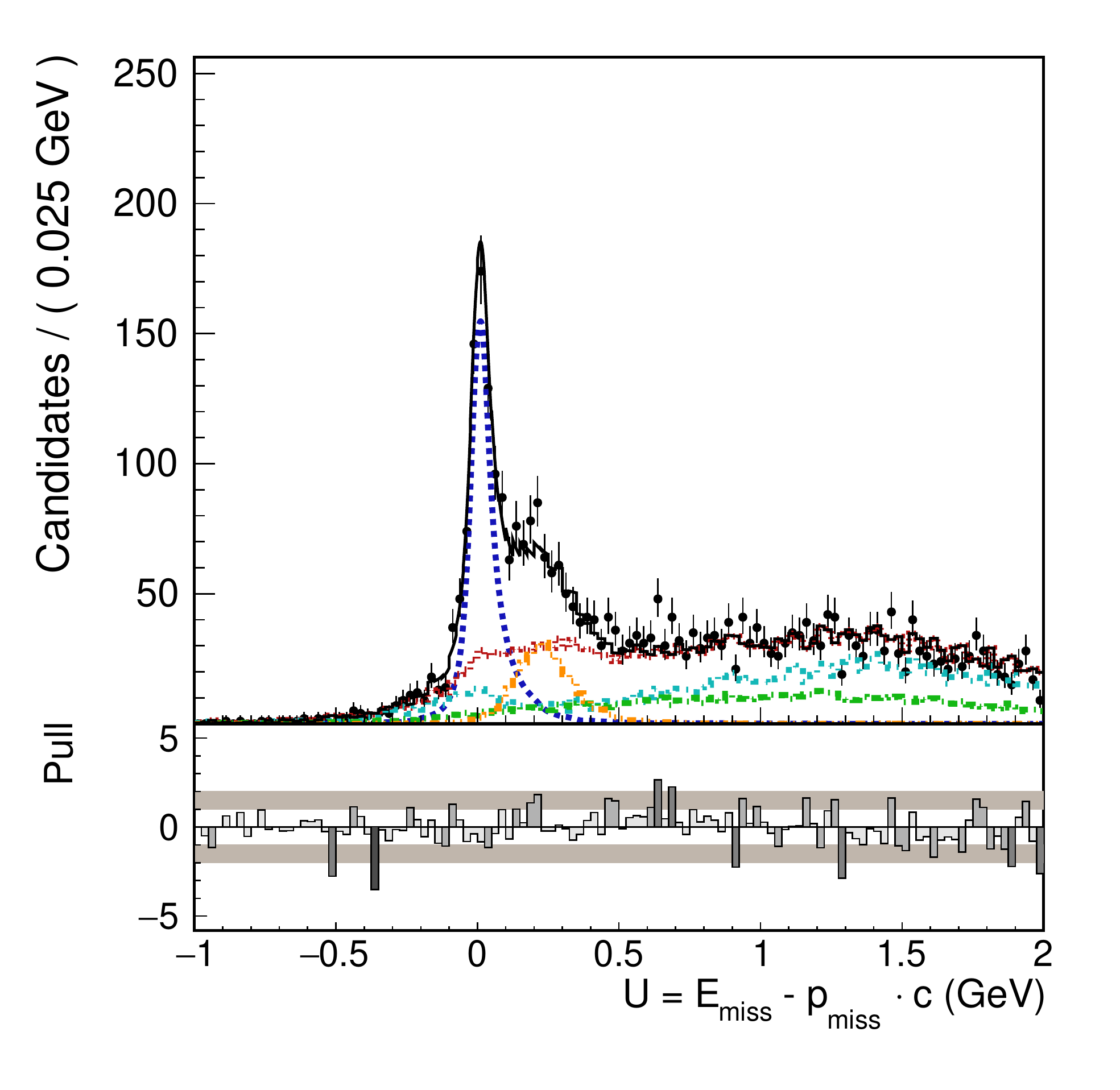}
\put(45,180){Belle}
\put(110,180){\BuToDmpienu}
\put(120,168){\textcolor{black}{\rule{5mm}{2pt}}\,\,\scalefont{0.8}Total}
\put(120,156){\textcolor{nice_blue}{\rule{5mm}{2pt}}\,\,\scalefont{0.8}Signal}
\put(120,144){\textcolor{nice_yellow}{\rule{5mm}{2pt}}\,\,\scalefont{0.8}Feeddown}
\put(120,132){\textcolor{nice_red}{\rule{5mm}{2pt}}\,\,\scalefont{0.8}Background}
\put(130,122){\textcolor{nice_green}{\rule{5mm}{2pt}}\,\,\scalefont{0.8}Continuum}
\put(130,112){\textcolor{nice_turquoise}{\rule{5mm}{2pt}}\,\,\scalefont{0.8}\BBbar}
\end{overpic}
\hfill
\begin{overpic}
[width=0.48\textwidth]{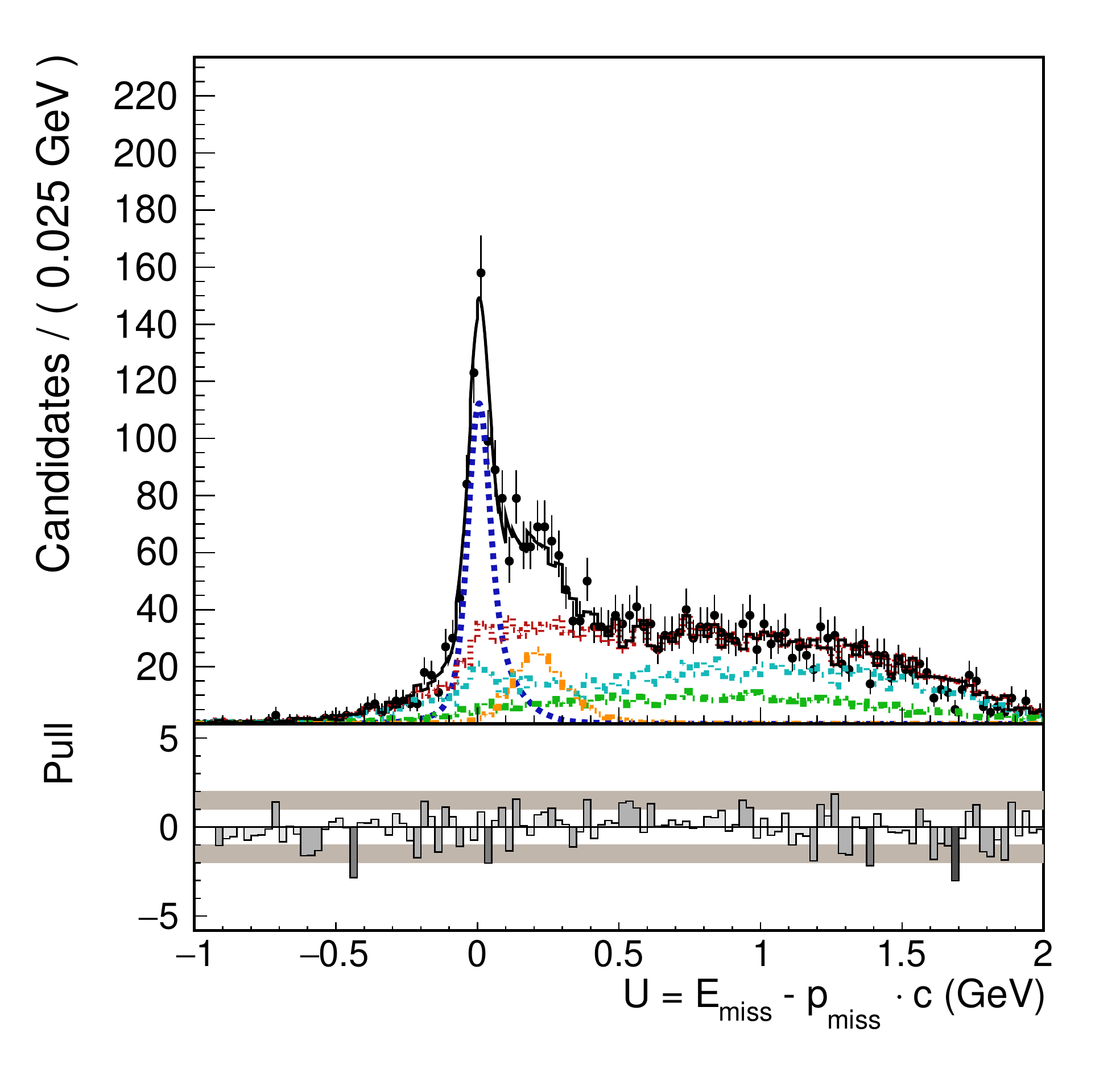}
\put(45,180){Belle}
\put(110,180){\BuToDmpimunu}
\put(120,168){\textcolor{black}{\rule{5mm}{2pt}}\,\,\scalefont{0.8}Total}
\put(120,156){\textcolor{nice_blue}{\rule{5mm}{2pt}}\,\,\scalefont{0.8}Signal}
\put(120,144){\textcolor{nice_yellow}{\rule{5mm}{2pt}}\,\,\scalefont{0.8}Feeddown}
\put(120,132){\textcolor{nice_red}{\rule{5mm}{2pt}}\,\,\scalefont{0.8}Background}
\put(130,122){\textcolor{nice_green}{\rule{5mm}{2pt}}\,\,\scalefont{0.8}Continuum}
\put(130,112){\textcolor{nice_turquoise}{\rule{5mm}{2pt}}\,\,\scalefont{0.8}\BBbar}
\end{overpic}
\hspace*{\fill}
\caption{Distribution of $\Emiss - \pmiss\,c$ of $\BuToDmpienu$ (left) and
 $\BuToDmpimunu$ (right) for the data. The MC shapes, normalized according to
 the result of the fit, are also shown.}
\label{fig:fitting:butodpilnu}
\end{figure}

\begin{figure}[ht]
\hspace*{\fill}
\begin{overpic}
[width=0.48\textwidth]{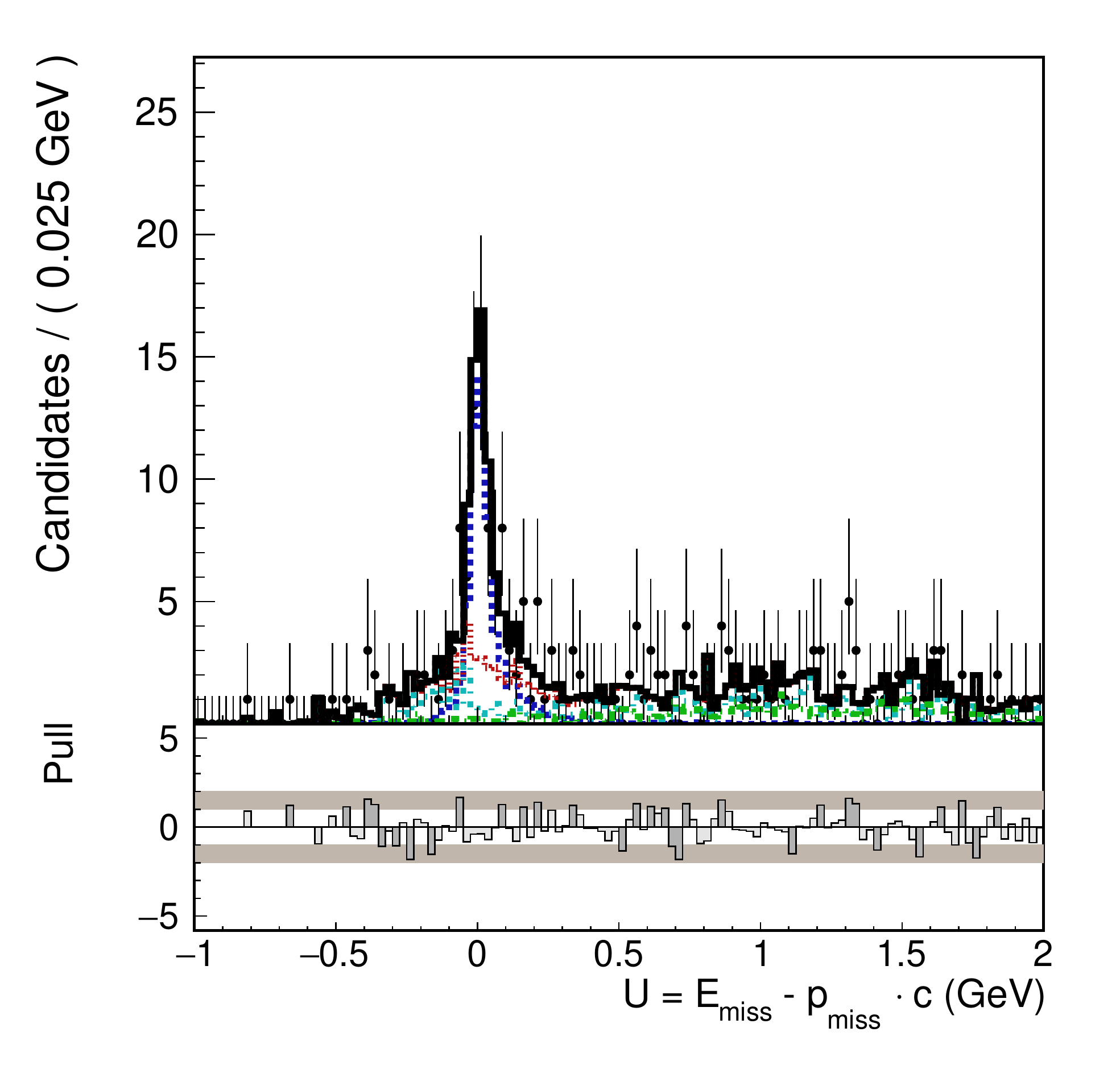}
\put(45,180){Belle}
\put(110,180){\BdToDstpienu}
\put(120,165){\textcolor{black}{\rule{5mm}{2pt}}\,\,Total}
\put(120,150){\textcolor{nice_blue}{\rule{5mm}{2pt}}\,\,Signal}
\put(120,135){\textcolor{nice_red}{\rule{5mm}{2pt}}\,\,Background}
\put(130,123){\textcolor{nice_green}{\rule{5mm}{2pt}}\,\,\scalefont{0.8}Continuum}
\put(130,111){\textcolor{nice_turquoise}{\rule{5mm}{2pt}}\,\,\scalefont{0.8}\BBbar}
\end{overpic}
\hfill
\begin{overpic}
[width=0.48\textwidth]{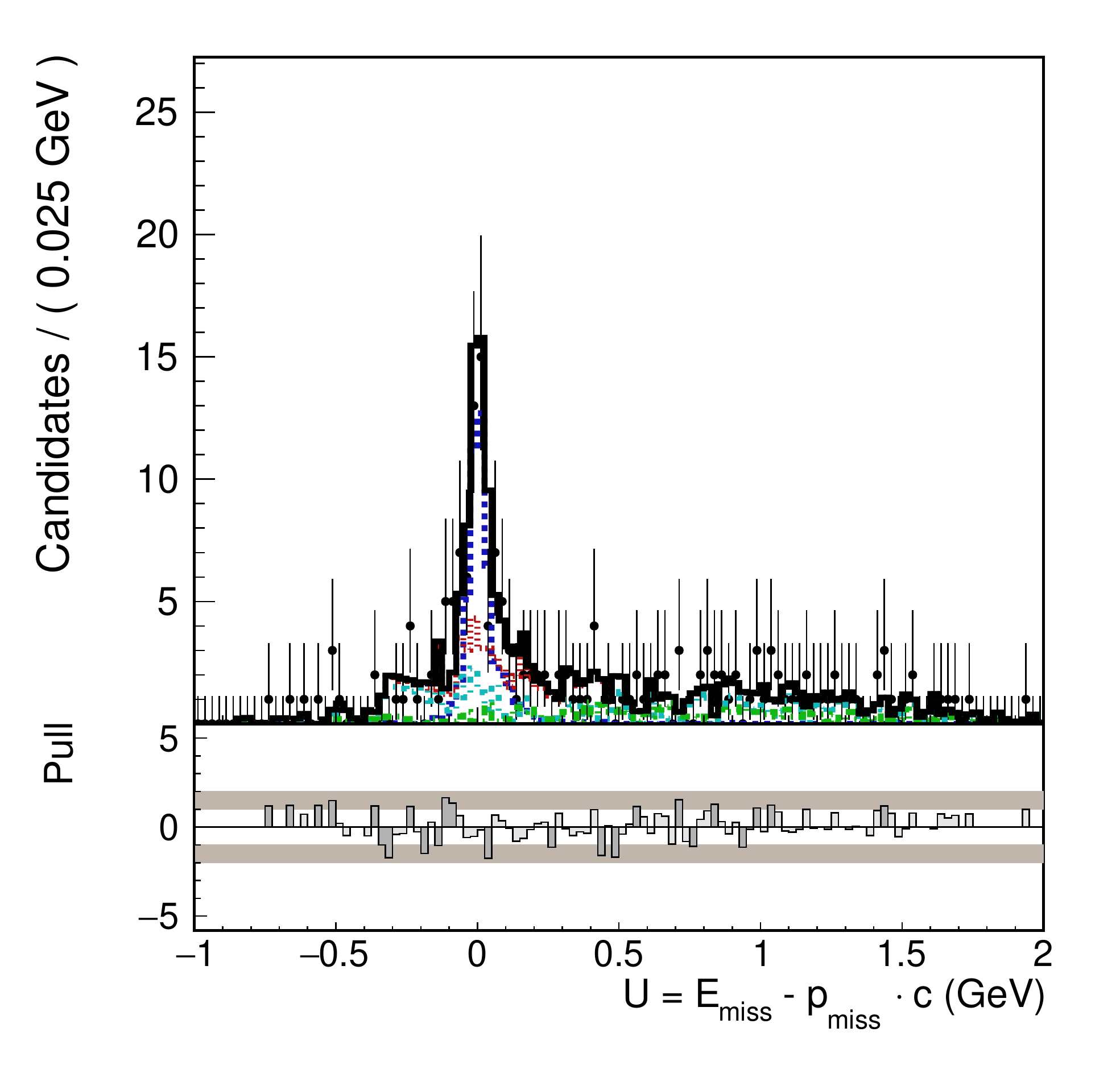}
\put(45,180){Belle}
\put(110,180){\BdToDstpimunu}
\put(120,165){\textcolor{black}{\rule{5mm}{2pt}}\,\,Total}
\put(120,150){\textcolor{nice_blue}{\rule{5mm}{2pt}}\,\,Signal}
\put(120,135){\textcolor{nice_red}{\rule{5mm}{2pt}}\,\,Background}
\put(130,123){\textcolor{nice_green}{\rule{5mm}{2pt}}\,\,\scalefont{0.8}Continuum}
\put(130,111){\textcolor{nice_turquoise}{\rule{5mm}{2pt}}\,\,\scalefont{0.8}\BBbar}
\end{overpic}
\hspace*{\fill}
\caption{Distribution of $\Emiss - \pmiss\,c$ of $\BdToDstpienu$ (left) and
 $\BdToDstpimunu$ (right) for the data. The MC shapes, normalized according
 to the result of the fit, are also shown.}
\label{fig:fitting:bdtodstpilnu}
\end{figure}

\begin{figure}[ht]
\hspace*{\fill}
\begin{overpic}
[width=0.48\textwidth]{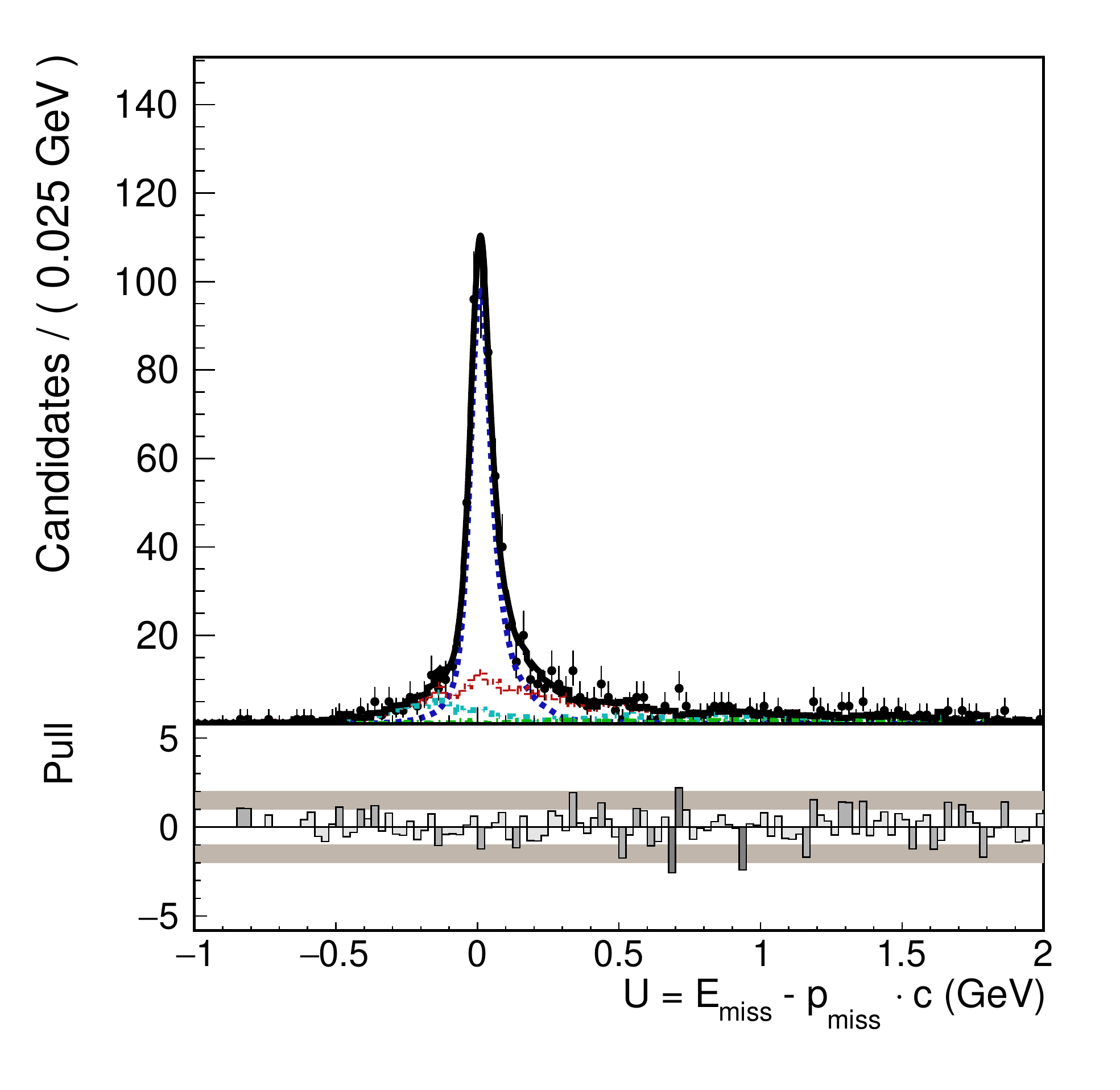}
\put(45,180){Belle}
\put(105,180){\BuToDstpienu}
\put(120,165){\textcolor{black}{\rule{5mm}{2pt}}\,\,Total}
\put(120,150){\textcolor{nice_blue}{\rule{5mm}{2pt}}\,\,Signal}
\put(120,135){\textcolor{nice_red}{\rule{5mm}{2pt}}\,\,Background}
\put(130,123){\textcolor{nice_green}{\rule{5mm}{2pt}}\,\,\scalefont{0.8}Continuum}
\put(130,111){\textcolor{nice_turquoise}{\rule{5mm}{2pt}}\,\,\scalefont{0.8}\BBbar}
\end{overpic}
\hfill
\begin{overpic}
[width=0.48\textwidth]{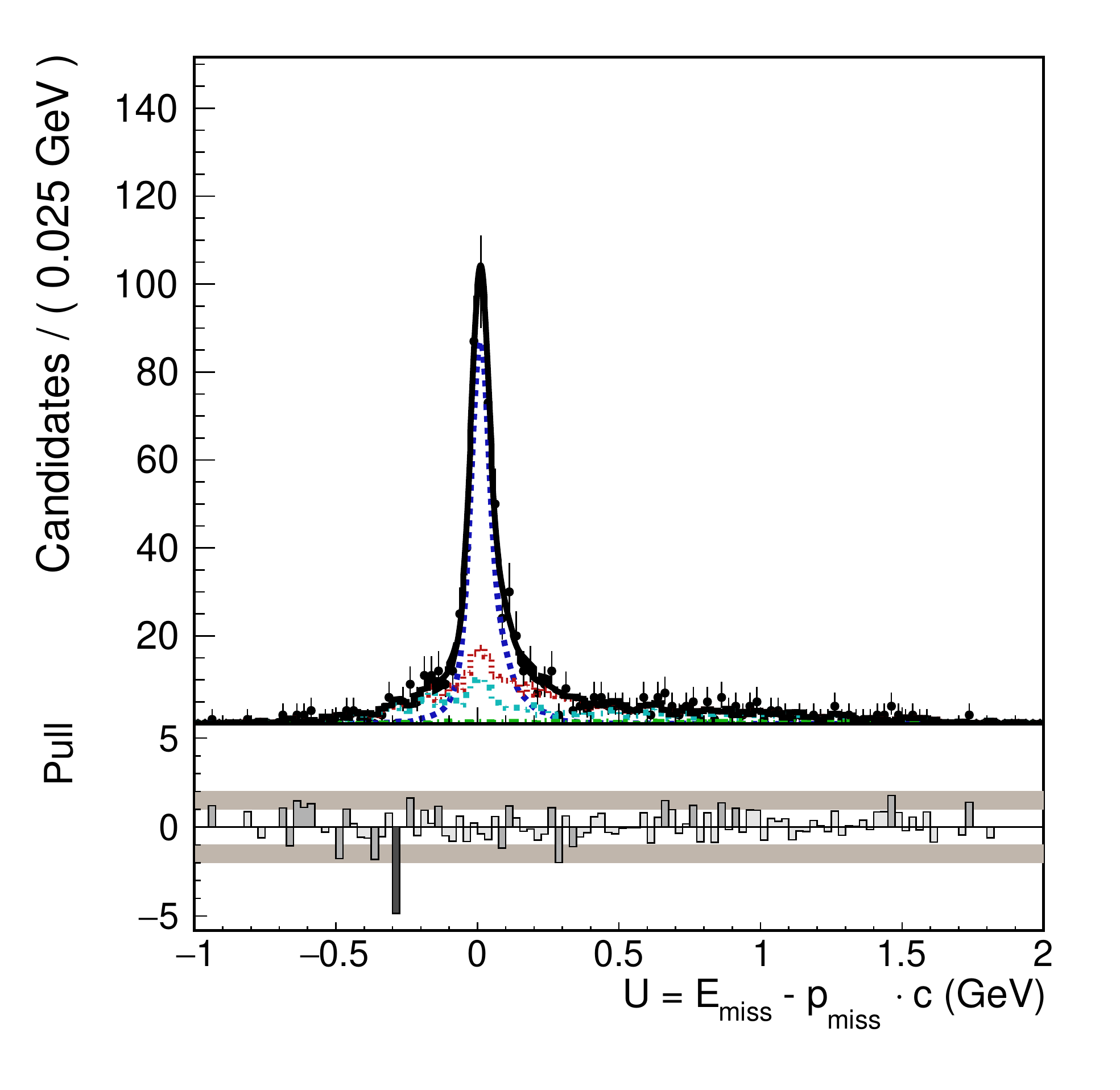}
\put(45,180){Belle}
\put(105,180){\BuToDstpimunu}
\put(120,165){\textcolor{black}{\rule{5mm}{2pt}}\,\,Total}
\put(120,150){\textcolor{nice_blue}{\rule{5mm}{2pt}}\,\,Signal}
\put(120,135){\textcolor{nice_red}{\rule{5mm}{2pt}}\,\,Background}
\put(130,123){\textcolor{nice_green}{\rule{5mm}{2pt}}\,\,\scalefont{0.8}Continuum}
\put(130,111){\textcolor{nice_turquoise}{\rule{5mm}{2pt}}\,\,\scalefont{0.8}\BBbar}
\end{overpic}
\hspace*{\fill}
\caption{Distribution of $\Emiss - \pmiss\,c$ of $\BuToDstpienu$ (left) and
 $\BuToDstpimunu$ (right) for the data. The MC shapes, normalized according
 to the result of the fit, are also shown.}
\label{fig:fitting:butodstpilnu}
\end{figure}

\begin{figure}[ht]
\hspace*{\fill}
\begin{overpic}
[width=0.48\textwidth]{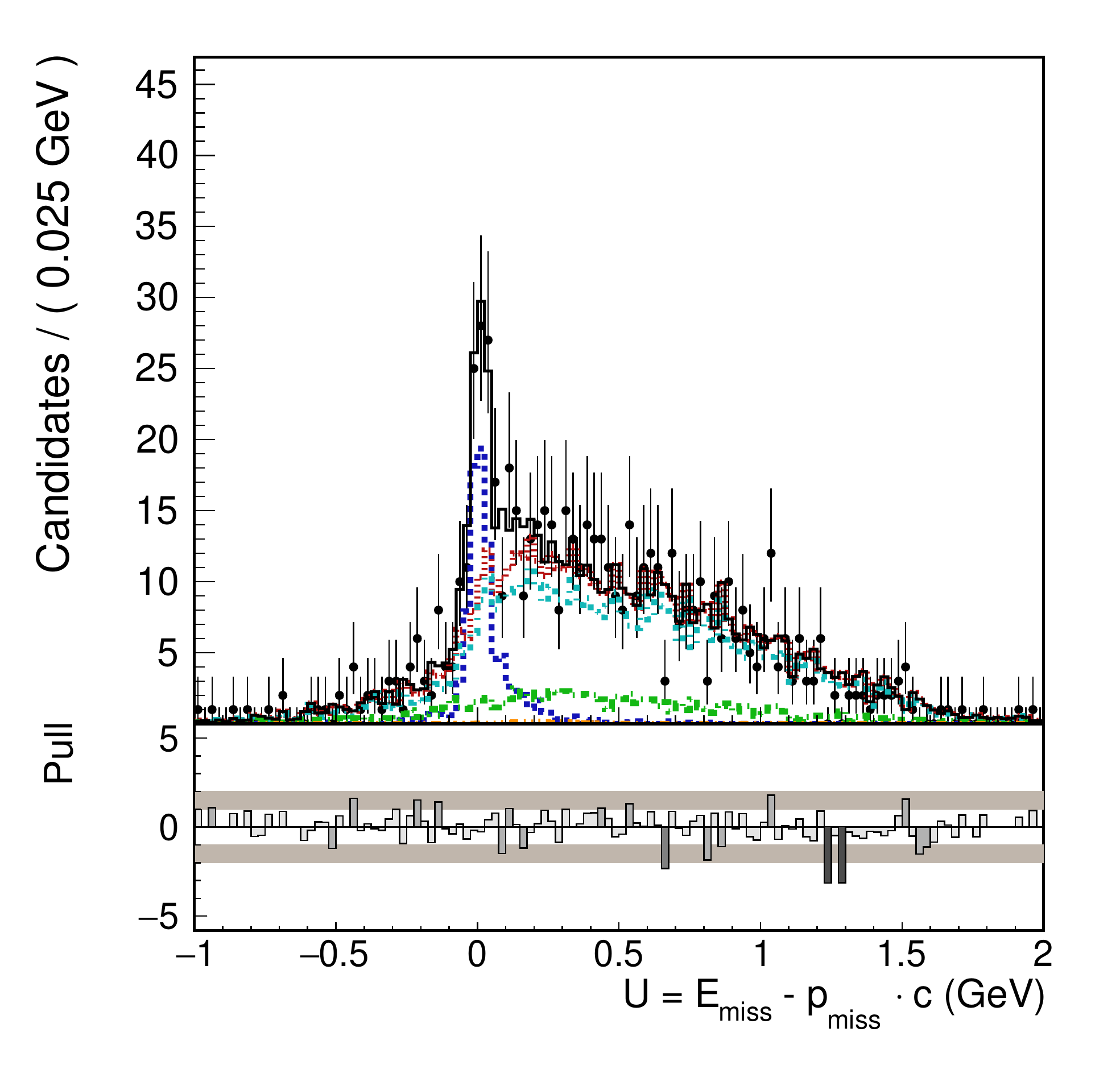}
\put(45,195){Belle}
\put(45,180){\scalefont{0.8}\BdToDmpipienu}
\put(135,190){\textcolor{black}{\rule{5mm}{2pt}}\,\,\scalefont{0.8}Total}
\put(135,178){\textcolor{nice_blue}{\rule{5mm}{2pt}}\,\,\scalefont{0.8}Signal}
\put(135,166){\textcolor{nice_yellow}{\rule{5mm}{2pt}}\,\,\scalefont{0.8}Feeddown}
\put(135,154){\textcolor{nice_red}{\rule{5mm}{2pt}}\,\,\scalefont{0.8}Background}
\put(145,144){\textcolor{nice_green}{\rule{5mm}{2pt}}\,\,\scalefont{0.8}Continuum}
\put(145,134){\textcolor{nice_turquoise}{\rule{5mm}{2pt}}\,\,\scalefont{0.8}\BBbar}
\end{overpic}
\hfill
\begin{overpic}
[width=0.48\textwidth]{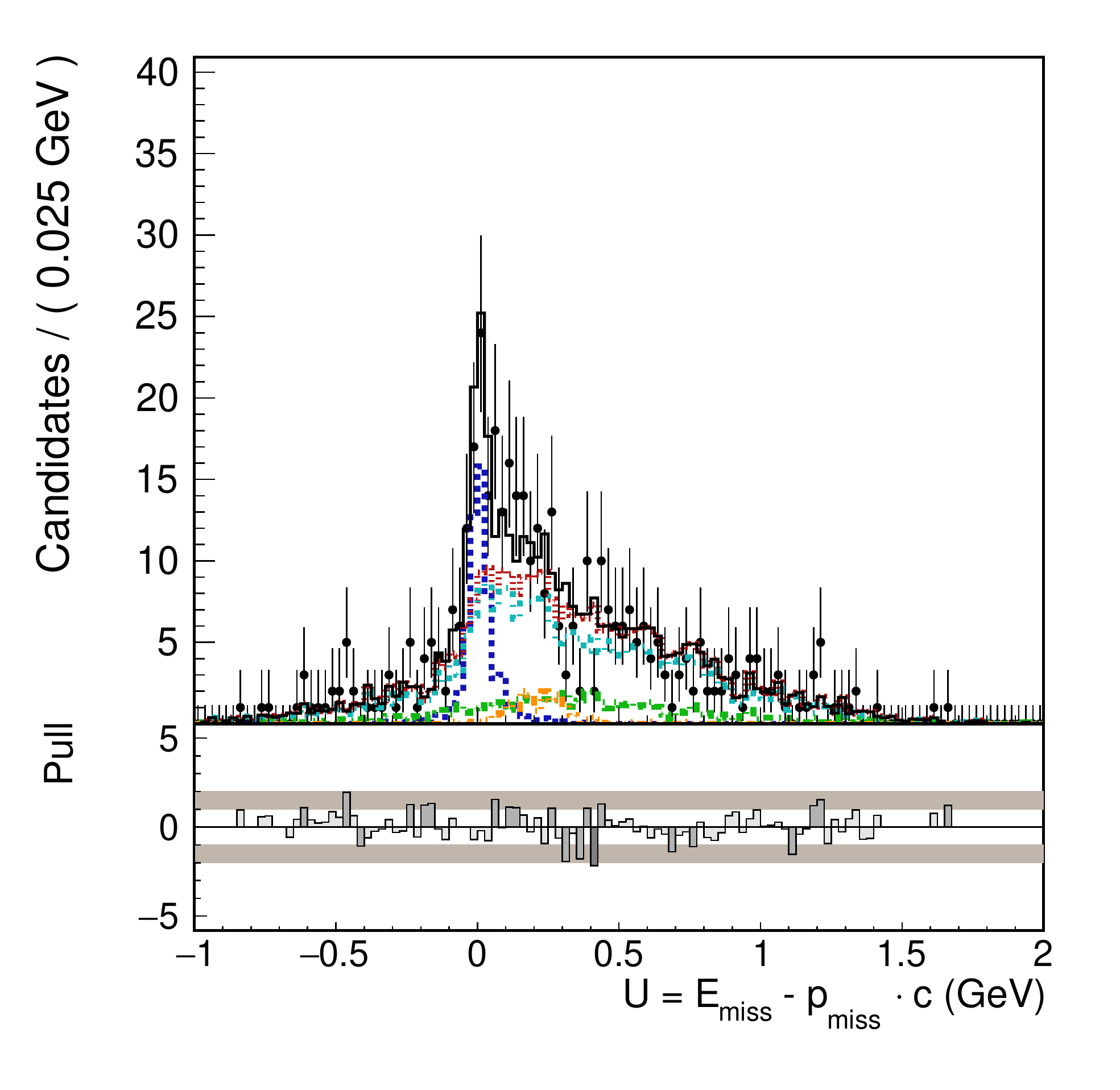}
\put(45,195){Belle}
\put(45,180){\scalefont{0.8}\BdToDmpipimunu}
\put(135,190){\textcolor{black}{\rule{5mm}{2pt}}\,\,\scalefont{0.8}Total}
\put(135,178){\textcolor{nice_blue}{\rule{5mm}{2pt}}\,\,\scalefont{0.8}Signal}
\put(135,166){\textcolor{nice_yellow}{\rule{5mm}{2pt}}\,\,\scalefont{0.8}Feeddown}
\put(135,154){\textcolor{nice_red}{\rule{5mm}{2pt}}\,\,\scalefont{0.8}Background}
\put(145,144){\textcolor{nice_green}{\rule{5mm}{2pt}}\,\,\scalefont{0.8}Continuum}
\put(145,134){\textcolor{nice_turquoise}{\rule{5mm}{2pt}}\,\,\scalefont{0.8}\BBbar}
\end{overpic}
\hspace*{\fill}
\caption{Distribution of $\Emiss - \pmiss\,c$ of $\BdToDmpipienu$ (left) and
 $\BdToDmpipimunu$ (right) for the data. The MC shapes, normalized according
 to the result of the fit, are also shown.}
\label{fig:fitting:bdtodpipilnu}
\end{figure}

\begin{figure}[ht]
\hspace*{\fill}
\begin{overpic}
[width=0.48\textwidth]{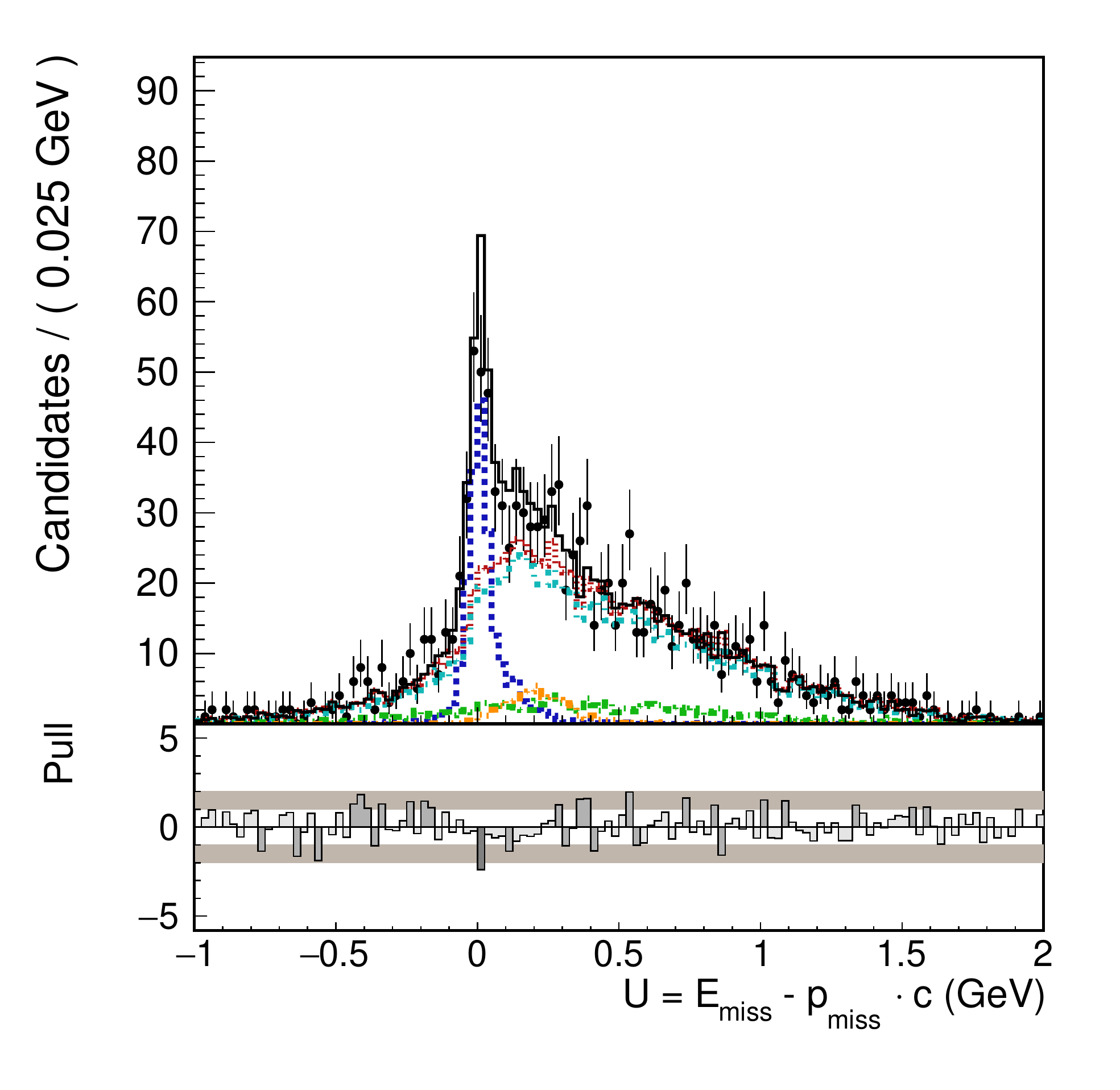}
\put(45,180){Belle}
\put(110,180){\scalefont{0.8}\BuToDzpipienu}
\put(120,168){\textcolor{black}{\rule{5mm}{2pt}}\,\,\scalefont{0.8}Total}
\put(120,156){\textcolor{nice_blue}{\rule{5mm}{2pt}}\,\,\scalefont{0.8}Signal}
\put(120,144){\textcolor{nice_yellow}{\rule{5mm}{2pt}}\,\,\scalefont{0.8}Feeddown}
\put(120,132){\textcolor{nice_red}{\rule{5mm}{2pt}}\,\,\scalefont{0.8}Background}
\put(130,122){\textcolor{nice_green}{\rule{5mm}{2pt}}\,\,\scalefont{0.8}Continuum}
\put(130,112){\textcolor{nice_turquoise}{\rule{5mm}{2pt}}\,\,\scalefont{0.8}\BBbar}
\end{overpic}
\hfill
\begin{overpic}
[width=0.48\textwidth]{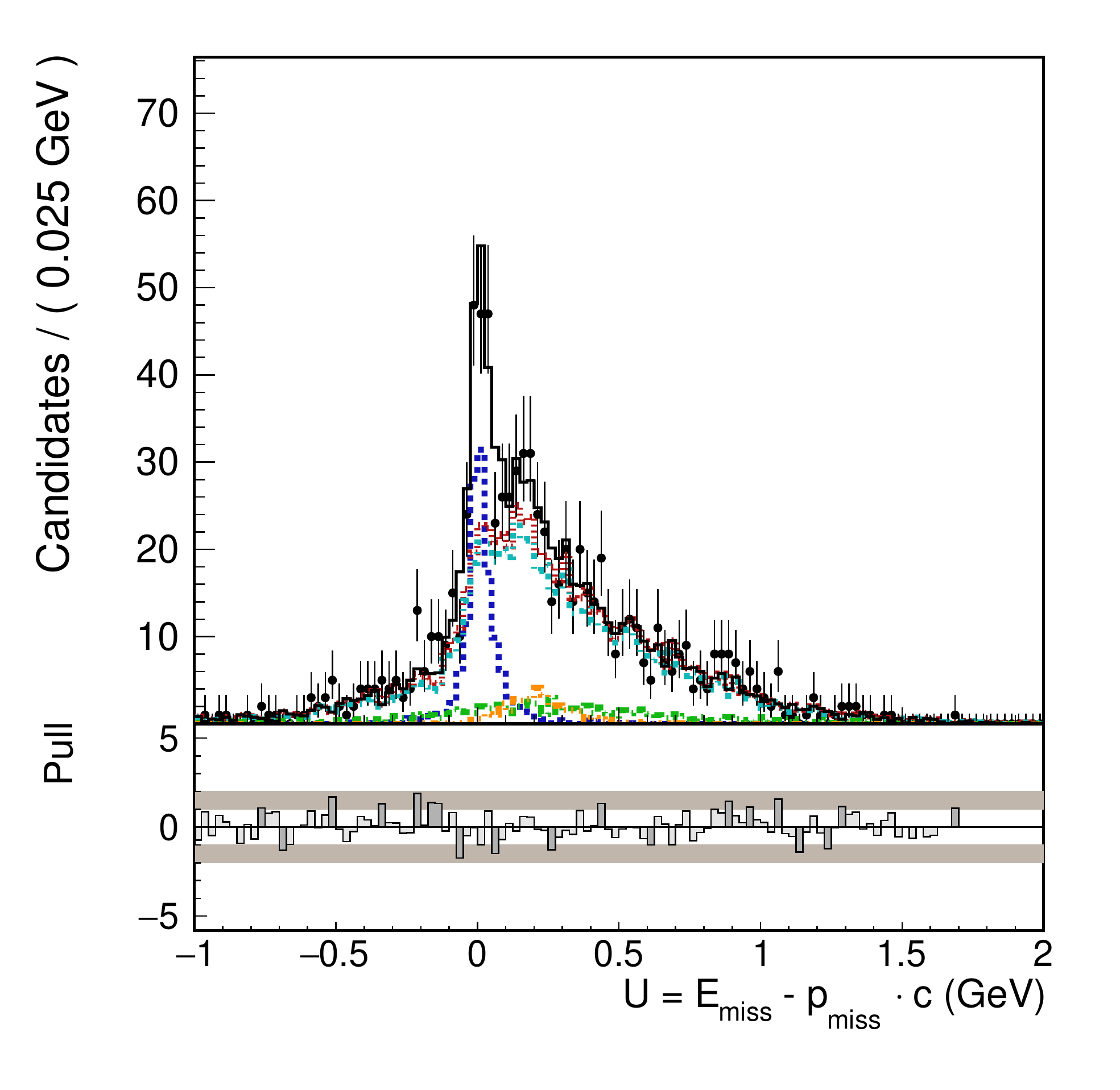}
\put(45,180){Belle}
\put(110,180){\scalefont{0.8}\BuToDzpipimunu}
\put(120,168){\textcolor{black}{\rule{5mm}{2pt}}\,\,\scalefont{0.8}Total}
\put(120,156){\textcolor{nice_blue}{\rule{5mm}{2pt}}\,\,\scalefont{0.8}Signal}
\put(120,144){\textcolor{nice_yellow}{\rule{5mm}{2pt}}\,\,\scalefont{0.8}Feeddown}
\put(120,132){\textcolor{nice_red}{\rule{5mm}{2pt}}\,\,\scalefont{0.8}Background}
\put(130,122){\textcolor{nice_green}{\rule{5mm}{2pt}}\,\,\scalefont{0.8}Continuum}
\put(130,112){\textcolor{nice_turquoise}{\rule{5mm}{2pt}}\,\,\scalefont{0.8}\BBbar}
\end{overpic}
\hspace*{\fill}
\caption{Distribution of $\Emiss - \pmiss\,c$ of $\BuToDzpipienu$ (left) and
 $\BuToDzpipimunu$ (right) for the data. The MC shapes, normalized according
 to the result of the fit, are also shown.}
\label{fig:fitting:butodpipilnu}
\end{figure}

\begin{figure}[ht]
\hspace*{\fill}
\begin{overpic}
[width=0.48\textwidth]{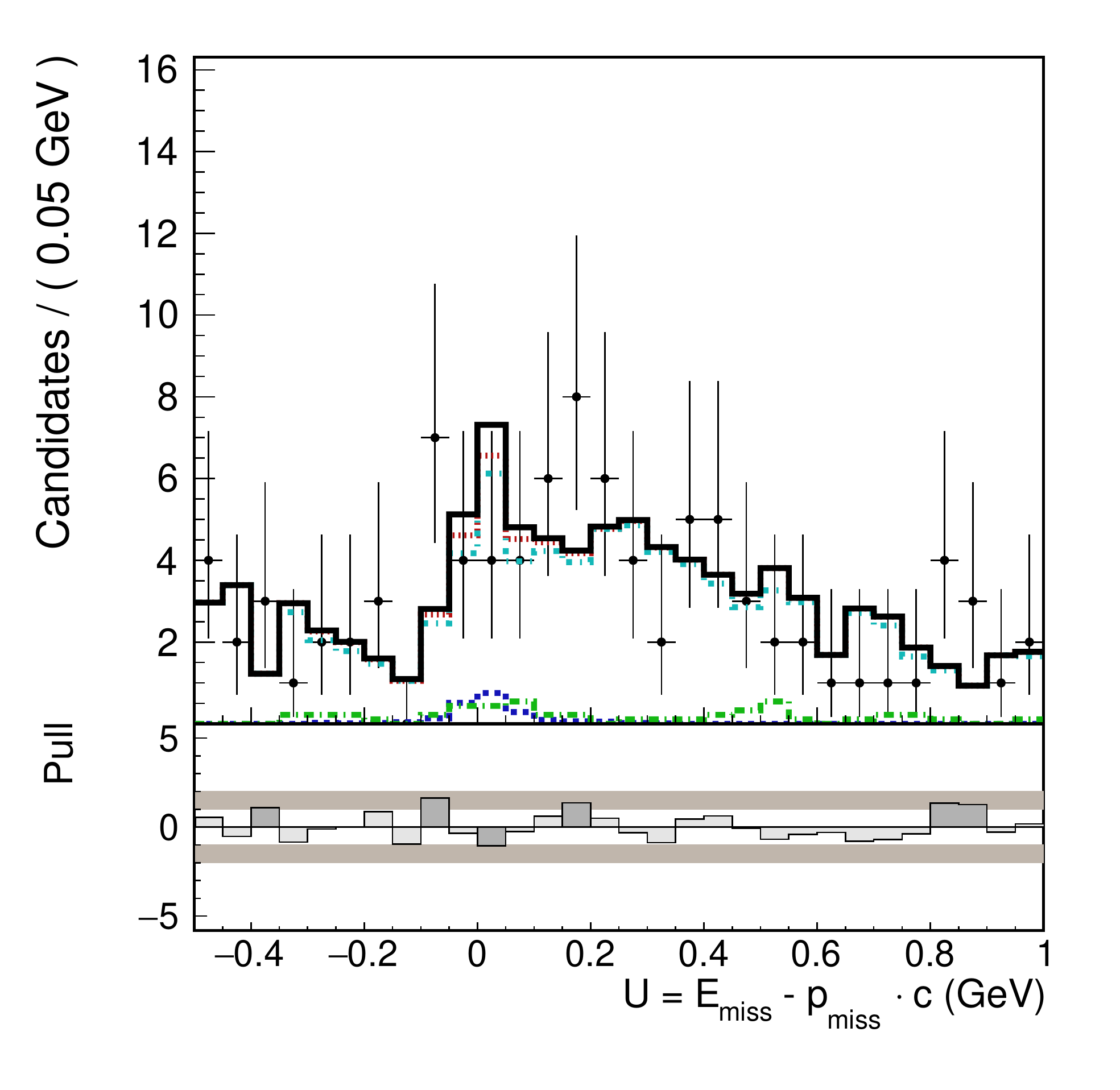}
\put(45,195){Belle}
\put(45,180){\scalefont{0.8}\BdToDstpipienu}
\put(135,180){\textcolor{black}{\rule{5mm}{2pt}}\,\,\scalefont{0.8}Total}
\put(135,165){\textcolor{nice_blue}{\rule{5mm}{2pt}}\,\,\scalefont{0.8}Signal}
\put(135,150){\textcolor{nice_red}{\rule{5mm}{2pt}}\,\,\scalefont{0.8}Background}
\put(145,135){\textcolor{nice_green}{\rule{5mm}{2pt}}\,\,\scalefont{0.8}Continuum}
\put(145,120){\textcolor{nice_turquoise}{\rule{5mm}{2pt}}\,\,\scalefont{0.8}\BBbar}
\end{overpic}
\hfill
\begin{overpic}
[width=0.48\textwidth]{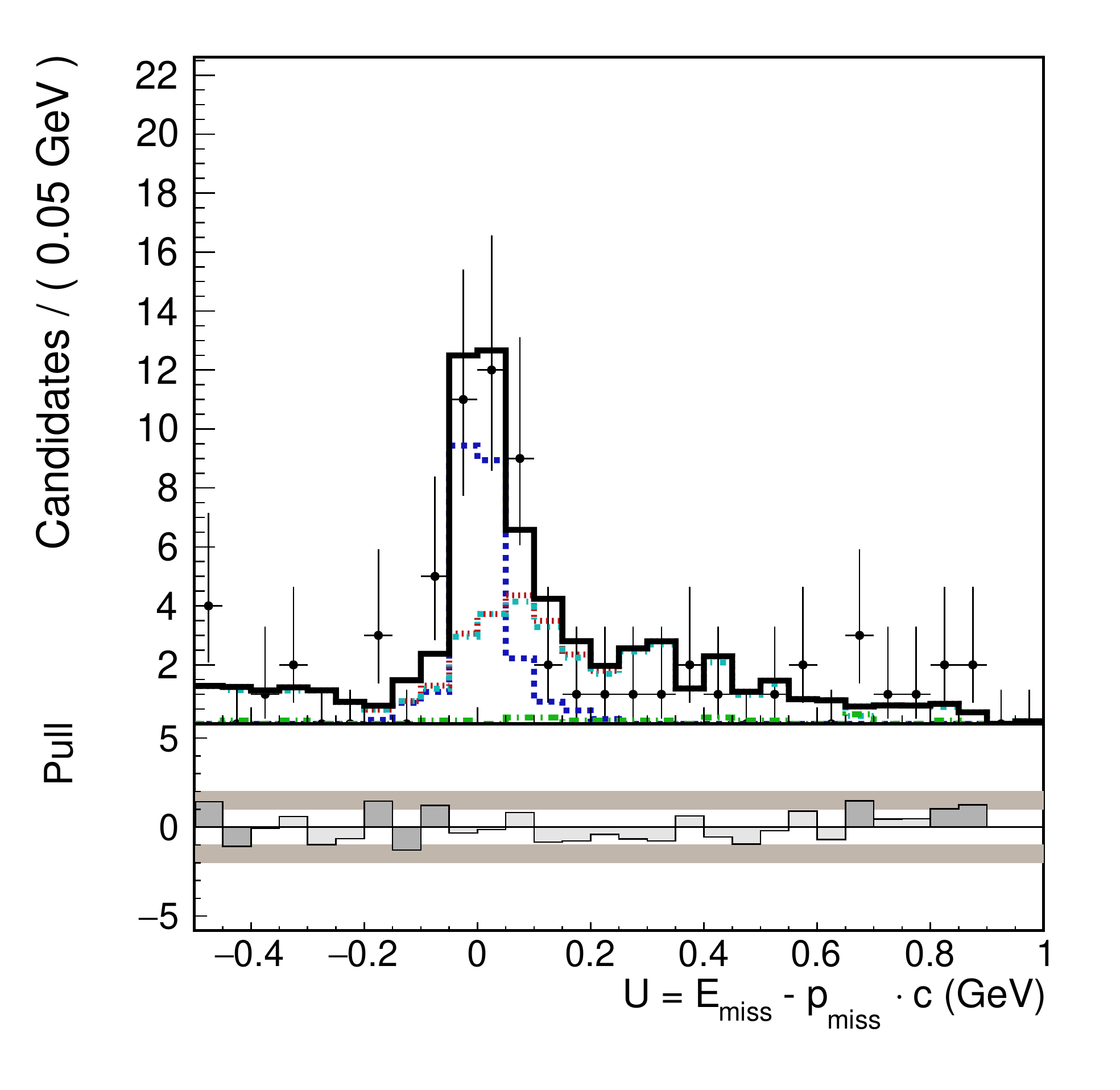}
\put(45,195){Belle}
\put(45,180){\scalefont{0.8}\BdToDstpipimunu}
\put(135,180){\textcolor{black}{\rule{5mm}{2pt}}\,\,\scalefont{0.8}Total}
\put(135,165){\textcolor{nice_blue}{\rule{5mm}{2pt}}\,\,\scalefont{0.8}Signal}
\put(135,150){\textcolor{nice_red}{\rule{5mm}{2pt}}\,\,\scalefont{0.8}Background}
\put(145,135){\textcolor{nice_green}{\rule{5mm}{2pt}}\,\,\scalefont{0.8}Continuum}
\put(145,120){\textcolor{nice_turquoise}{\rule{5mm}{2pt}}\,\,\scalefont{0.8}\BBbar}
\end{overpic}
\hspace*{\fill}
\caption{Distribution of $\Emiss - \pmiss\,c$ of $\BdToDstpipienu$ (left) and
 $\BdToDstpipimunu$ (right) for the data. The MC shapes, normalized according
 to the result of the fit, are also shown.}
\label{fig:fitting:bdtodstpipilnu}
\end{figure}

\begin{figure}[ht]
\hspace*{\fill}
\begin{overpic}
[width=0.48\textwidth]{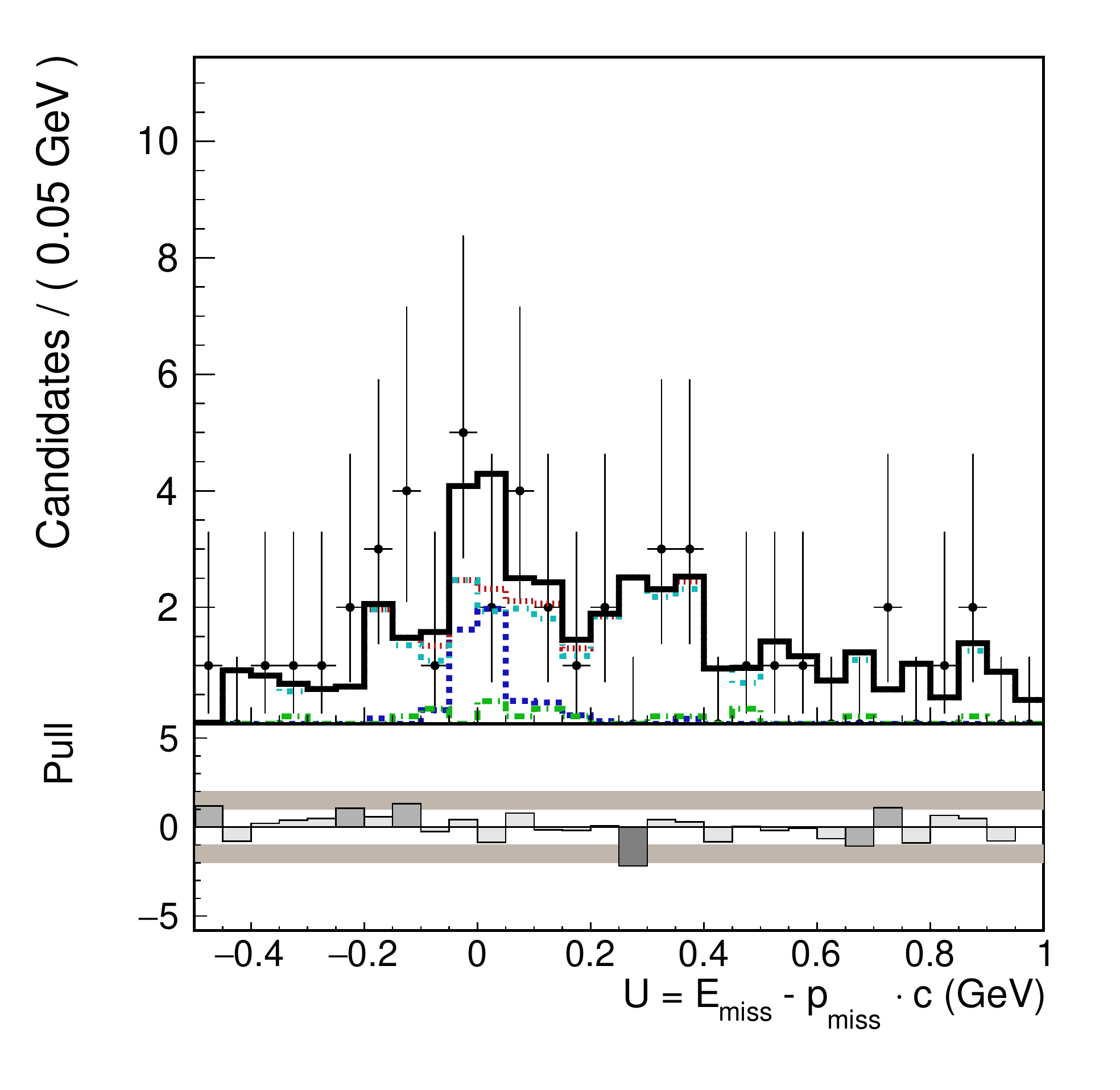}
\put(45,195){Belle}
\put(45,180){\scalefont{0.8}\BuToDstpipienu}
\put(135,180){\textcolor{black}{\rule{5mm}{2pt}}\,\,\scalefont{0.8}Total}
\put(135,165){\textcolor{nice_blue}{\rule{5mm}{2pt}}\,\,\scalefont{0.8}Signal}
\put(135,150){\textcolor{nice_red}{\rule{5mm}{2pt}}\,\,\scalefont{0.8}Background}
\put(145,135){\textcolor{nice_green}{\rule{5mm}{2pt}}\,\,\scalefont{0.8}Continuum}
\put(145,120){\textcolor{nice_turquoise}{\rule{5mm}{2pt}}\,\,\scalefont{0.8}\BBbar}
\end{overpic}
\hfill
\begin{overpic}
[width=0.48\textwidth]{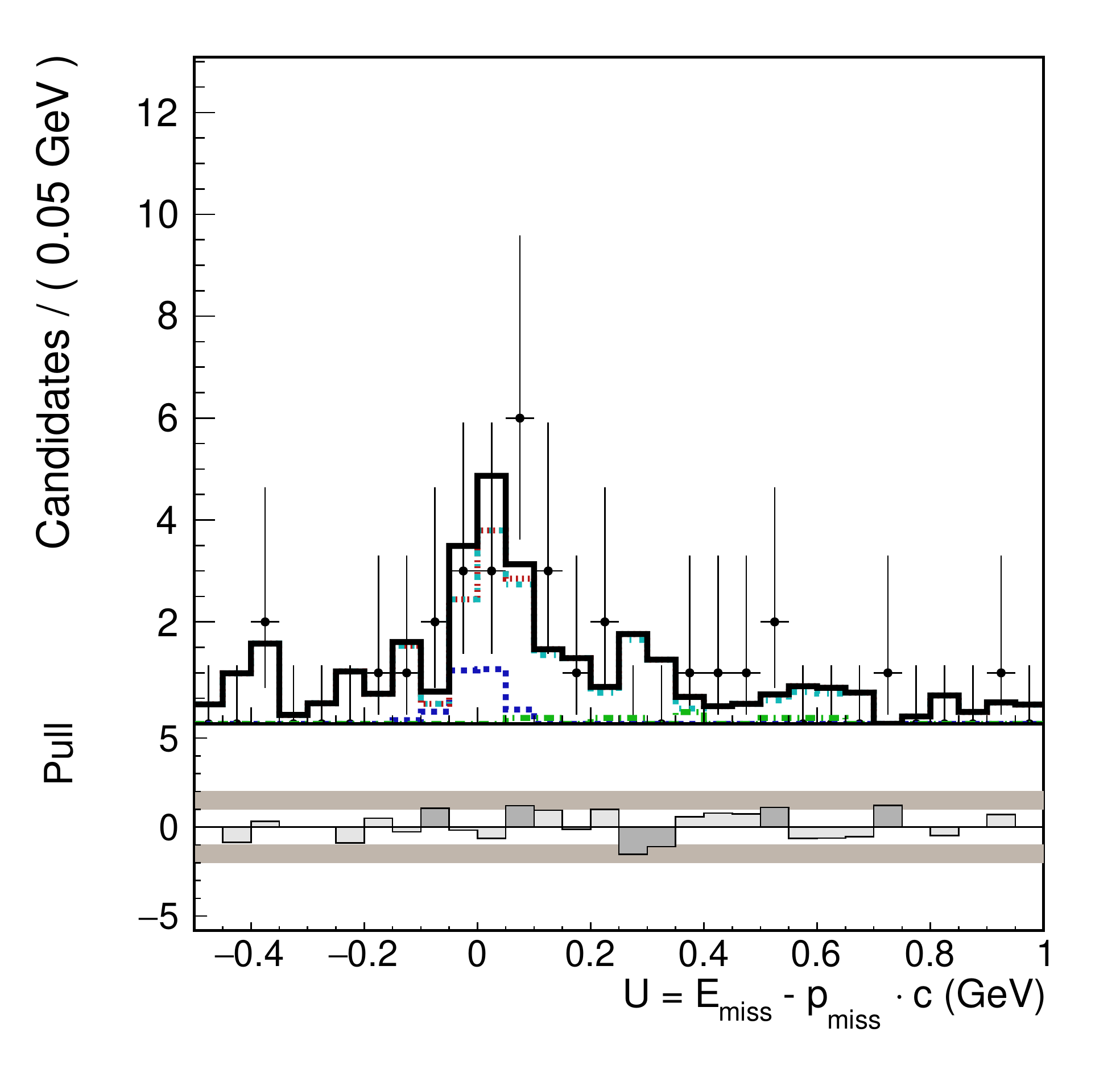}
\put(45,195){Belle}
\put(45,180){\scalefont{0.8}\BuToDstpipimunu}
\put(135,180){\textcolor{black}{\rule{5mm}{2pt}}\,\,\scalefont{0.8}Total}
\put(135,165){\textcolor{nice_blue}{\rule{5mm}{2pt}}\,\,\scalefont{0.8}Signal}
\put(135,150){\textcolor{nice_red}{\rule{5mm}{2pt}}\,\,\scalefont{0.8}Background}
\put(145,135){\textcolor{nice_green}{\rule{5mm}{2pt}}\,\,\scalefont{0.8}Continuum}
\put(145,120){\textcolor{nice_turquoise}{\rule{5mm}{2pt}}\,\,\scalefont{0.8}\BBbar}
\end{overpic}
\hspace*{\fill}
\caption{Distribution of $\Emiss - \pmiss\,c$ of $\BuToDstpipienu$ (left) and
 $\BuToDstpipimunu$ (right) for the data. The MC shapes, normalized according
 to the result of the fit, are also shown.}
\label{fig:fitting:butodstpipilnu}
\end{figure}

The plots of the data and fit results are shown in \crefrange
{fig:fitting:bdtodpilnu}{fig:fitting:butodstpipilnu}. The signal and
background yields are summarized in \cref
{tab:fitting:datayields}.

\clearpage

\begin{table}[ht]
\caption{Fitted \BToDorDstpilnu and \BToDorDstpipilnu signal and background
 yields in the full \belle data sample. The quoted uncertainties are statistical only.}
\centering
\begin{tabular}{lSSS[table-format = 2.0(2)]S[table-format = 2.0(2)]}
\toprule
                        &   \multicolumn{2}{c}{Signal}          &   \multicolumn{2}{c}{Background}      \\
                        &   {electron mode} &   {muon mode}     &   {electron mode} &   {muon mode}     \\
\midrule
\BdToDzpilnu            &   570\pm35        &   433\pm34        &   2641\pm80       &   2190\pm78       \\
\BuToDmpilnu            &   721\pm32        &   569\pm31        &   1329\pm53       &   1302\pm54       \\
\BdToDstpilnu           &   798\pm44        &   690\pm43        &   77\pm12         &   64\pm11         \\
\BuToDstpilnu           &   787\pm35        &   634\pm34        &   172\pm19        &   242\pm23        \\
\BdToDmpipilnu          &   88\pm14         &   58\pm12         &   452\pm26        &   271\pm21        \\
\BuToDzpipilnu          &   196\pm20        &   132\pm18        &   852\pm37        &   603\pm33        \\
\BdToDstpipilnu         &   3\pm10          &   41\pm10         &   86\pm11         &   {$41\pm\phantom{1}8$}          \\
\BuToDstpipilnu         &   57\pm15         &   38\pm14         &  {$37\pm\phantom{1}7$} &   {$26\pm\phantom{1}6$}          \\
\bottomrule
\end{tabular}
\label{tab:fitting:datayields}
\end{table}

\section{Systematic uncertainties}
\label{sec:systematics}

The systematic uncertainties mainly arise from the fit modeling, the
uncertainty on the branching fraction values of the normalization mode
$\BToDstlnu$ and the charm modes, and the hadron PID. For the two-pion modes
there are additional sizable systematic uncertainties from the BDT and from the limited size of the MC sample used to
calculate the signal efficiency of the selection. The various considered
sources of systematic uncertainties are described below. Their numerical
values are summarized in \cref
{tab:systematics:btodpilnu,tab:systematics:btodpipilnu}.

\paragraph{MC statistics fit model:}
To account for the finite size of the MC samples used to produce the PDF
templates, alternative fit PDFs are created by varying the bin contents of
each PDF template according to a Poisson distribution. This is done \num
{1000} times, and after each variation the fit to the collision data is
performed with the new set of templates. It is checked that the pull
distributions are unbiased, where the pull is defined as the difference
between the yields using the varied fit PDF and the nominal yields divided by
the statistical uncertainty of the new yields. The spread of the new signal
yields (about \num{1}\% for the one-pion modes, \numrange{5}{20}\%
for the two-pion modes) is used as an estimate for the systematic
uncertainty.

\paragraph{MC statistics signal efficiency:}
The uncertainty on the calculated signal efficiency ratios in \cref
{tab:selection:efficiencyratio} due to the finite size of the MC samples is
propagated to the branching fractions and ratios, and assigned as systematic
uncertainty.

\paragraph{Charm branching ratios:}
To estimate the uncertainty due to the uncertainties on the branching ratios
of the charm decays, we sample each charm branching ratio \num{10000} times
from a Gaussian distribution with mean and width that equal to the PDG central
value and uncertainty~\cite{PDG2022}. It is assumed that the branching
fractions for different \PD modes are independent. For each sampled set
of \PD branching fractions, the new sum of branching fractions is
calculated for the signal and normalization channels. The reconstruction
efficiency is taken into account via the relative abundance of the modes. The
ratio of the sums is calculated and the spread of the resulting distribution
assigned as systematic uncertainty.

\paragraph{Signal \texorpdfstring{\BToDststlnu}{B -> D** l nu} composition:}
The signal PDF $U$ shapes slightly vary for different intermediate \Dstst
states. Therefore, the overall $U$ shape depends on the \Dstst composition.
To estimate the signal branching-fraction uncertainty due to the
uncertainties in the \Dstst composition, we generate the $U$ distribution
using the template of one \Dstst state and then fit with the nominal signal
template described in \cref{sec:fitting:btodststlnu} whose composition is
taken from Ref.~\cite{PDG2022}. The largest average difference between the
generated and fitted signal yields among the tested \Dstst scenarios, which
varies between \num{0.4}\% and \num{0.8}\%, depending on the
mode, is assigned as a systematic uncertainty.

\paragraph{Lepton PID:}
By using \decay{\gamma\gamma}{\ellell} processes, lepton PID efficiency factors in
kinematic ranges of the momentum and polar angle have been calculated (Chapter~5.4 of Ref.~\cite{BFactories}), which correct for the difference between the selection efficiency
in data and MC. The systematic uncertainties on the PID efficiency factors
account for the method itself and for a possible effect from a hadronic
environment, which is determined using inclusive \decay{B}{\JPsi X} decays.
To propagate the uncertainties to the branching
fractions we sample lepton correction factors for each kinematic bin using a
Gaussian around the nominal value with a width corresponding to the
uncertainty of the correction factor. The average correction factor over all
truth-matched signal events as well as the average correction factor over all
truth-matched candidates of the normalization channels are calculated. The
spread of the distribution of the ratio of the two means is taken as the
systematic uncertainty due to lepton identification. This procedure is
performed separately for each of the \Dstst states, and the largest
uncertainty per \PB and $\D^{(*)}$ mode among all \BToDststlnu modes is assigned as
the systematic uncertainty.

\paragraph{Charged hadron PID:}
Similar to the study for the lepton PID, correction factors for the hadron PID
selection requirements are sampled in bins of the momentum and polar angle to
evaluate the systematic uncertainty on the branching fraction due to the
uncertainties in the determination of the correction factors using
inclusive \Dstar samples (Chapter~5.4 of Ref.~\cite{BFactories}). The average correction factors of
the signal and normalization samples are calculated, then divided by each
other, and the spread of the resulting distribution of ratios is interpreted as the
systematic uncertainty for the hadron PID. Similar to the lepton PID described
above, the largest value over the possible \Dstst states is assigned as the
final systematic uncertainty.

\paragraph{Tracking efficiency:}
For each signal and normalization mode the average track multiplicity over the
various \PD modes is determined in simulation. The difference between the
signal and normalization mode average track multiplicity is multiplied by \num
{0.35}\% (Chapter~15.1.1.2 of Ref.~\cite{BFactories}) and the result is taken as systematic uncertainty
due to tracking efficiency differences between data and MC. For low-momentum
tracks ($\pt < \SI{200}{\mevc}$) an additional tracking-related systematic
uncertainty is calculated. Using a \BdToDstpi sample the slow pion efficiency
is determined in six momentum bins for data and MC (Chapter~15.1.1.2 of Ref.~\cite{BFactories}). The
relative uncertainty of the ratio between the data and MC efficiencies is
taken as systematic uncertainty due to low-momentum tracking. The two
tracking-related systematic uncertainties are added in quadrature.

\paragraph{\texorpdfstring{\piz}{pi0} efficiency:}
The \piz efficiency differs between data and MC. The effect is corrected in
the calculation of the signal efficiency and the uncertainty on the ratio
between the data and MC efficiency of about \num{2.4}\% (Chapter~15.1.4 of Ref.~\cite{BFactories})
is propagated to the systematic uncertainty of the branching
fraction measurement. First, the average \piz multiplicity for each signal
and normalization mode is determined and the difference between the signal
and normalization values is calculated. This difference is multiplied by the
aforementioned uncertainty to obtain the systematic uncertainty due to the \piz
efficiency data-MC ratio.

\paragraph{\texorpdfstring{\BToDstlnu}{B -> D* l nu} and \texorpdfstring{\BToDststlnu}{B -> D** l nu} form factors:}
The $\BToDststlnu$ MC samples are generated with the ISGW2 model~\cite
{Scora:1995ty}. A more accurate description can be achieved with the LLSW
model~\cite{Leibovich:1997em}. To estimate the systematic uncertainty due to
using the ISGW2 model two-dimensional form factor weights in $\omega = \frac
{m_B^2 + m_\Dstst^2 - q^2}{2 m_B m_\Dstst}$, with the masses of the \PB meson
$m_B$ and the \Dstst system $m_\Dstst$, and the four-momentum transfer squared to the
lepton-neutrino system \qsq, and the cosine of the angle between the charged
lepton and the $\PD$ meson $\cos\theta_l$ are determined. These weights are
calculated separately for decays via \Dzstar, \Done, \Dprimeone,
and \Dtwostar mesons. The $U$ distribution is generated using the nominal
ISGW2-based templates and fit with signal and feeddown templates that are
reweighted with the form factor weights described above. The average
difference between the fitted and generated yields over \num{1000} iterations
of generating and fitting is calculated and divided by the generated yield
($f_{\rm sig}$).

Similarly, the simulation of the $\BToDorDstlnu$ modes is based on heavy
quark effective theory (HQET)~\cite{HQET}. A reweighting in the momentum
transfer and the momentum of the charged lepton is applied to account for
outdated values of the CLN~\cite{CLN} form factor parameters $\rho^2$, $R_1$,
and $R_2$. The $U$ distribution is generated with the nominal HQET-based
templates and fit with the reweighted templates. The difference between the
fitted and generated yields divided by the generated yield is calculated ($f_
{\rm norm}$).

The difference of the ratio $f_{\rm sig}$ / $f_{\rm norm}$ from unity is taken
as the systematic uncertainty due to the form factors.

\paragraph{\texorpdfstring{\BR(\BToDstlnu)}{BR(B -> D* l nu)}:}
The PDG average of the branching ratio of the normalization mode \BuToDstlnu
is $\mathcal{B}(\BuToDstlnu) = \num{5.58\pm0.22e-2}$~\cite
{PDG2022}, introducing a systematic uncertainty of \num{3.9}\%. The corresponding PDG average
for the \Bd mode is $\mathcal{B}(\BdToDstlnu) = \num{4.97\pm0.12e-2}$~\cite
{PDG2022}, which introduces a systematic uncertainty of \num{2.4}\%.

\paragraph{BDT:}
The BDT to suppress continuum background in $\BToDorDstpipilnu$ is trained
with signal MC and off-resonance data. Differences in the input variable
distributions between the signal simulation and signal events in real data
might introduce a bias in the calculation of the signal efficiency. To
estimate the associated uncertainty, the BDT output is calculated for the
cross-check and normalization modes $\BToDlnu$ and $\BToDstlnu$. The same
requirement on the BDT output as for the \Dpipi signal-candidate selection is
applied for these \BToDorDstlnu modes and the fit to the \BToDorDstlnu sample
described in \cref{sec:fitting:btodstlnu} is performed. The ratio between the
\BToDorDstlnu yield of this fit and the yield obtained without the BDT
requirement is considered a data-based efficiency of the BDT requirement.
This efficiency is compared with the signal MC efficiency of
the \BToDorDstlnu samples. The largest relative difference between the data- and
MC-based efficiencies among the \BToDlnu
and \BToDstlnu values is taken as the BDT-related systematic uncertainty.
This procedure assumes that the BDT, which uses variables of the \Btag meson
reconstruction and event-shape variables, is mostly independent of the \Bsig
meson reconstruction.

\begin{table}[ht]
\caption{Relative systematic uncertainties (in \si{\percent}) in the determination of the $\BToDorDstpilnu$ branching fractions.}
\centering
\resizebox{\textwidth}{!}{%
\begin{tabular}{lSSSS}
\toprule
                            &  {\BdToDzpilnu}  &  {\BuToDmpilnu}  &  {\BdToDstpilnu}  &  {\BuToDstpilnu}   \\
\midrule
MC statistics: fit model    &       1.0        &       0.7        &        1.1        &      0.7           \\
MC statistics: efficiency   &       0.6        &       0.5        &        0.7        &      0.7           \\
Charm branching ratios      &       1.0        &       1.4        &        1.1        &      1.2           \\
Signal \Dstst composition   &       0.5        &       0.5        &        0.4        &      0.8           \\
Lepton PID                  &       0.1        &       0.1        &        0.1        &      0.1           \\
Charged hadron PID          &       0.2        &       1.1        &        0.2        &      2.0           \\
Tracking efficiency         &       0.2        &       0.4        &        0.4        &      0.6           \\
\piz efficiency             &       0.1        &       0.3        &        0.2        &      0.2           \\
\BToDstlnu / \BToDststlnu form factors     &       0.3        &       0.1        &        0.2        &      0.1           \\
\midrule
sum                         &       1.7        &       2.2        &        1.8        &      2.8           \\
\bottomrule
\BR(\BToDstlnu)             &       2.4        &       3.9        &        2.4        &      3.9           \\
sum incl. \BR(\BToDstlnu)   &       2.9        &       4.5        &        3.0        &      4.8           \\
\bottomrule
\end{tabular}%
}%
\label{tab:systematics:btodpilnu}
\end{table}

\begin{table}[ht]
\caption{Relative systematic uncertainties (in \si{\percent}) in the determination of the $\BToDorDstpipilnu$ branching fractions.}
\centering
\resizebox{\textwidth}{!}{%
\begin{tabular}{lSSSS}
\toprule
                            &   {\BdToDmpipilnu}  &   {\BuToDzpipilnu} &   {\BdToDstpipilnu}  &   {\BuToDstpipilnu}  \\
\midrule
MC statistics: fit model    &       7.0           &       4.3          &      19.8            &        10.5          \\
MC statistics: efficiency   &       2.4           &       1.8          &       3.4            &         3.6          \\
Charm branching ratios      &       1.4           &       0.9          &       1.0            &         0.9          \\
BDT                         &       3.9           &       2.4          &       2.2            &         2.5          \\
Lepton PID                  &       0.3           &       0.3          &       0.3            &         0.4          \\
Charged hadron PID          &       1.7           &       2.5          &       0.5            &         1.1          \\
Tracking efficiency         &       0.3           &       0.5          &       0.3            &         0.5          \\
\piz efficiency             &       0.3           &       0.1          &       0.2            &         0.2          \\
\BToDstlnu form factors     &       0.1           &       0.1          &       0.1            &         0.1          \\
\midrule
sum                         &       8.7           &       6.8          &      22.3            &        11.6          \\
\midrule
\BR(\BToDstlnu)             &       2.4           &       3.9          &       2.4            &         3.9          \\
sum incl. \BR(\BToDstlnu)   &       9.0           &       7.8          &      22.4            &        12.0          \\
\bottomrule
\end{tabular}%
}%
\label{tab:systematics:btodpipilnu}
\end{table}

\section{Branching fraction results}

The weighted average branching fraction ratios are calculated based on the
total uncertainties. The calculation takes into account that some component uncertainties are
correlated between the electron and muon mode. The results and
the ratios between the electron and muon mode branching fractions are listed
in \cref{tab:inclusive:resultsRatio}.
\begin{table}[ht]
\caption{Branching fraction ratio results and ratios between electron and muon decay modes with statistical and systematic uncertainties. The denominator for the branching fraction ratios is \BdToDstlnu for the \Bd modes and \BuToDstlnu for the \Bu modes.}
\centering
\setlength{\tabcolsep}{20pt}
\begin{tabular}{lcc}
\toprule
Decay mode    & Branching fraction ratio [\si{\percent}]  & $\electron/\muon$ ratio                       \\
\midrule
\BdToDzpilnu &  \num[parse-numbers=false]{\phantom{1}\BRBdToDzpilRatio\pm\BRBdToDzpilstatRatio\stat\pm\BRBdToDzpilsystRatio\syst} & \num[parse-numbers=false]{1.13\pm0.11\stat}       \\
\BuToDmpilnu &  \num[parse-numbers=false]{\phantom{1}\BRBuToDmpilRatio\pm\BRBuToDmpilstatRatio\stat\pm\BRBuToDmpilsystRatio\syst} & \num[parse-numbers=false]{1.07\pm0.08\stat}       \\
\BdToDstpilnu & \num[parse-numbers=false]{\BRBdToDstpilRatio\pm\BRBdToDstpilstatRatio\stat\pm\BRBdToDstpilsystRatio\syst} & \num[parse-numbers=false]{0.98\pm0.08\stat}     \\
\BuToDstpilnu & \num[parse-numbers=false]{\phantom{1}\BRBuToDstpilRatio\pm\BRBuToDstpilstatRatio\stat\pm\BRBuToDstpilsystRatio\syst} & \num[parse-numbers=false]{1.06\pm0.08\stat}      \\
\BdToDmpipilnu &  \num[parse-numbers=false]{\phantom{1}\BRBdToDmpipilRatio\pm\BRBdToDmpipilstatRatio\stat\pm\BRBdToDmpipilsystRatio\syst} & \num[parse-numbers=false]{1.18\pm0.31\stat}   \\
\BuToDzpipilnu &  \num[parse-numbers=false]{\phantom{1}\BRBuToDzpipilRatio\pm\BRBuToDzpipilstatRatio\stat\pm\BRBuToDzpipilsystRatio\syst} & \num[parse-numbers=false]{1.23\pm0.21\stat}   \\
\BdToDstpipilnu & \num[parse-numbers=false]{\phantom{1}\BRBdToDstpipilRatio\pm\BRBdToDstpipilstatRatio\stat\pm\BRBdToDstpipilsystRatio\syst} & \num[parse-numbers=false]{0.06\pm0.21\stat}  \\
\BuToDstpipilnu & \num[parse-numbers=false]{\phantom{1}\BRBuToDstpipilRatio\pm\BRBuToDstpipilstatRatio\stat\pm\BRBuToDstpipilsystRatio\syst} & \num[parse-numbers=false]{1.1\phantom{1}\pm0.5\phantom{1}\stat}  \\
\bottomrule
\end{tabular}
\label{tab:inclusive:resultsRatio}
\end{table}

The results are the most precise determinations of these branching fraction
ratios to date (except for \BdToDstpipilnu). All values are compatible with the previous world averages.
The electron and muon values are compatible with each other within one
standard deviation apart from those for \BdToDstpipilnu. The p-value
of the hypothesis that the latter are compatible is \num{0.5}\%~\footnote{The deviation of the ratios from unity
cannot naively be interpreted in terms of standard deviations.}.

The branching fraction ratios are converted into absolute branching fractions
by multiplying them with the branching fraction of \BToDstlnu. The results
are listed in \cref{tab:inclusive:results}.

\begin{table}[ht]
\caption{Branching fraction results with statistical and systematic uncertainties.}
\centering
\setlength{\tabcolsep}{20pt}
\begin{tabular}{lc}
\toprule
Decay mode      &   Branching fraction [\si{\percent}]  \\
\midrule
\BdToDzpilnu    &   \num[parse-numbers=false]{\BRBdToDzpil\pm\BRBdToDzpilstat\stat\pm\BRBdToDzpilsyst\syst} \\
\BuToDmpilnu    &   \num[parse-numbers=false]{\BRBuToDmpil\pm\BRBuToDmpilstat\stat\pm\BRBuToDmpilsyst\syst} \\
\BdToDstpilnu   &   \num[parse-numbers=false]{\BRBdToDstpil\pm\BRBdToDstpilstat\stat\pm\BRBdToDstpilsyst\syst} \\
\BuToDstpilnu   &   \num[parse-numbers=false]{\BRBuToDstpil\pm\BRBuToDstpilstat\stat\pm\BRBuToDstpilsyst\syst} \\
\BdToDmpipilnu  &   \num[parse-numbers=false]{\BRBdToDmpipil\pm\BRBdToDmpipilstat\stat\pm\BRBdToDmpipilsyst\syst} \\
\BuToDzpipilnu  &   \num[parse-numbers=false]{\BRBuToDzpipil\pm\BRBuToDzpipilstat\stat\pm\BRBuToDzpipilsyst\syst} \\
\BdToDstpipilnu &   \num[parse-numbers=false]{\BRBdToDstpipil\pm\BRBdToDstpipilstat\stat\pm\BRBdToDstpipilsyst\syst} \\
\BuToDstpipilnu &   \num[parse-numbers=false]{\BRBuToDstpipil\pm\BRBuToDstpipilstat\stat\pm\BRBuToDstpipilsyst\syst} \\
\bottomrule
\end{tabular}
\label{tab:inclusive:results}
\end{table}

\section{Exclusive \texorpdfstring{\BToDststlnu}{B -> D** l nu} branching fractions}

Using the sPlot technique~\cite{Pivk:2004ty} with the implementation of Ref.~\cite{Cows}, signal weights are assigned to
each event based on the fit to the $U$ distribution. This allows the
background contribution to the $m(\Dpi)$, $m(\Dstpi)$, and $m(\Dpipi)$ distributions to be statistically
subtracted, and the signal-only distribution to be studied.
We perform weighted unbinned maximum likelihood fits to the invariant mass
distributions. The uncertainty calculation is based on Ref.~\cite{Langenbruch}.

\begin{figure}[ht]
\hspace*{\fill}
\begin{overpic}
[width=0.48\textwidth]{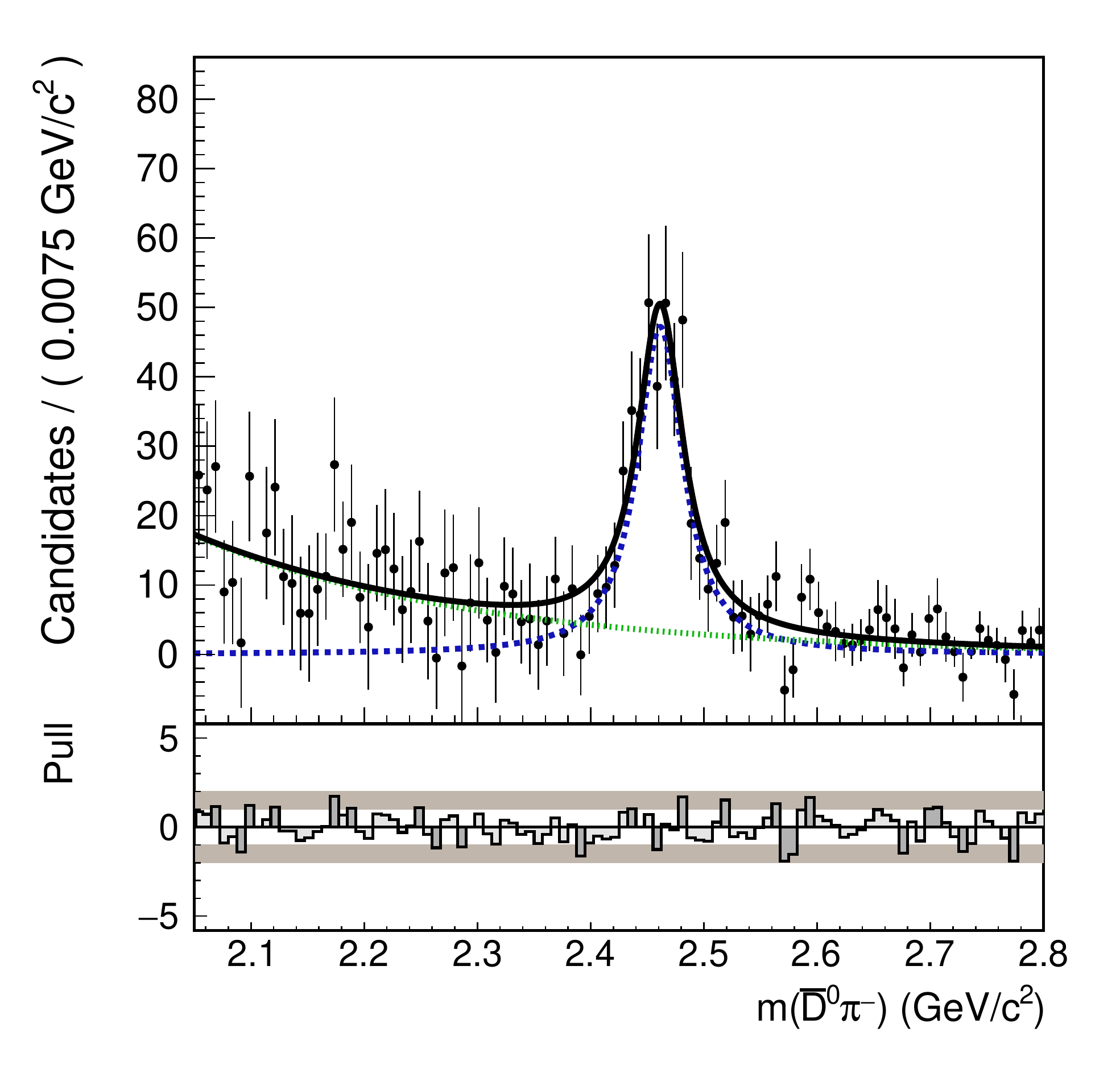}
\put(45,190){Belle}
\put(105,190){\BdToDzpilnu}
\put(50,175){\textcolor{black}{\rule{5mm}{2pt}}\,\,\scalefont{0.8}Total}
\put(50,160){\textcolor{nice_blue}{\rule{5mm}{2pt}}\,\,\scalefont{0.8}\Dtwostarm}
\put(50,145){\textcolor{nice_green}{\rule{5mm}{2pt}}\,\,\scalefont{0.8}other \Dzb\pim}
\end{overpic}
\hfill
\begin{overpic}
[width=0.48\textwidth]{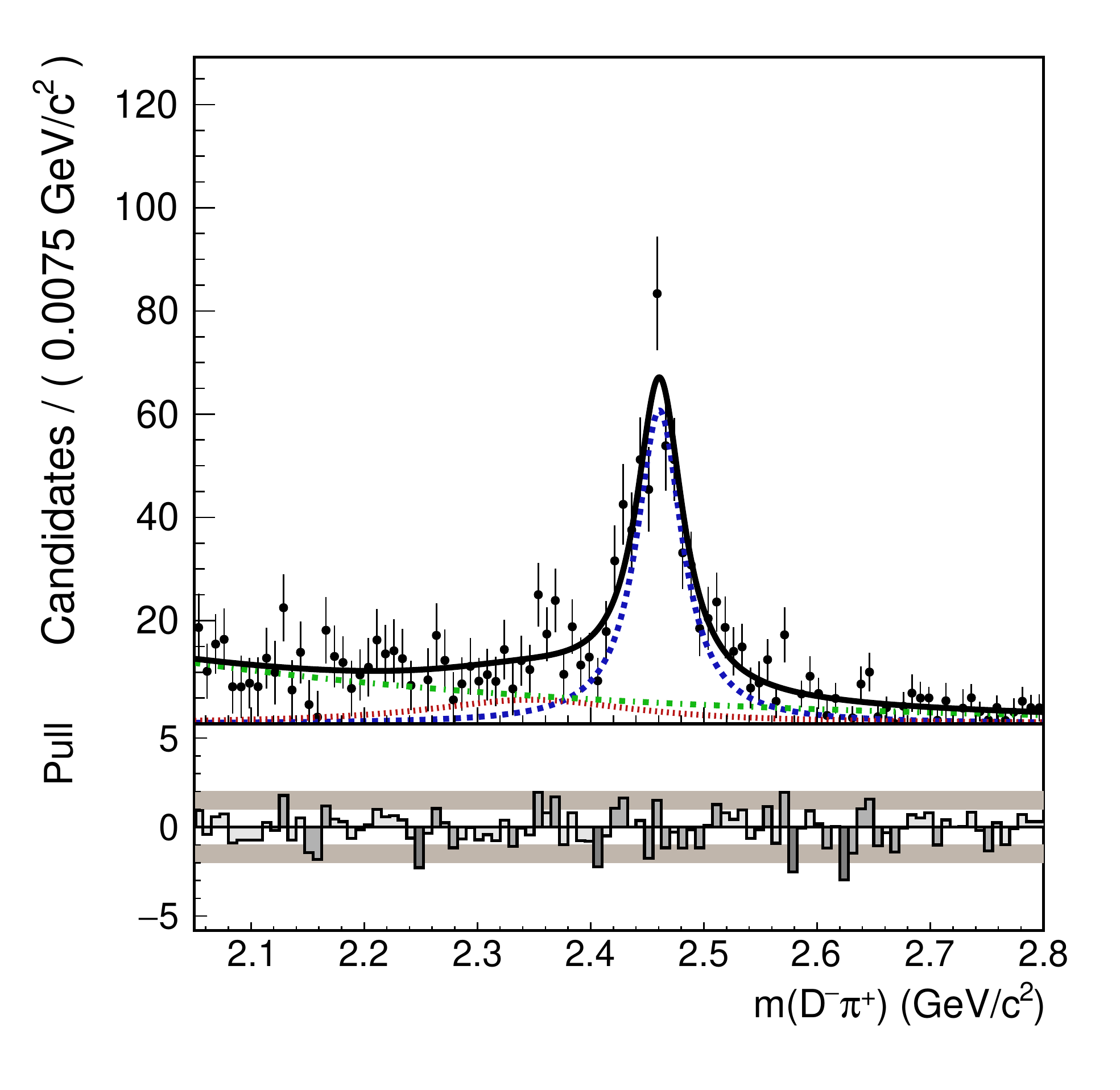}
\put(45,190){Belle}
\put(105,190){\BuToDmpilnu}
\put(50,175){\textcolor{black}{\rule{5mm}{2pt}}\,\,\scalefont{0.8}Total}
\put(50,160){\textcolor{nice_blue}{\rule{5mm}{2pt}}\,\,\scalefont{0.8}\Dtwostarzb}
\put(50,145){\textcolor{nice_red}{\rule{5mm}{2pt}}\,\,\scalefont{0.8}\Dzstarzb}
\put(50,130){\textcolor{nice_green}{\rule{5mm}{2pt}}\,\,\scalefont{0.8}other \Dm\pip}
\end{overpic}
\hspace*{\fill}
\caption{Invariant $m(\Dpi)$ mass distribution of \BdToDzpilnu
 (left) and \BuToDmpilnu (right) reconstruction after applying signal
 weights determined from a fit of the $U$ distribution using the sPlot
 technique.}
\label{fig:exclusive:mdpi}
\end{figure}

For the \BToDpilnu modes the PDG reports decays via
the \Dzstar and \Dtwostar resonances. These two contributions are
parametrized with Breit-Wigner functions that are convolved with a Gaussian
distribution. The width of the Gaussian is fixed from simulations to \SI{3.4}
{\MeVcc}. The peak position and width of the \Dzstar and \Dtwostar resonances
are allowed to float in the fit. However, they are constrained within
Gaussian distributions using their world averages and corresponding
uncertainties~\cite{PDG2022}. In a second fit the peak positions and widths
are fixed to the results from the first fit. The difference in the
statistical uncertainties between the two fits is used to single out the
uncertainty introduced by the Gaussian constraint. It is interpreted as a
systematic uncertainty. The weighted $m(\Dpi)$ distribution (see \cref
{fig:exclusive:mdpi}) shows that a third component must be added to the fit
model. Here, we choose an exponential distribution.
The yields, which are listed in \cref{tab:exclusive:dpi},
are converted into branching fractions using \cref{eq:overview:BRratio}. The
statistical uncertainty is extracted directly from the fit, while the
systematic uncertainty is the sum in quadrature of the relative uncertainties
of the inclusive branching fractions reported in \cref
{tab:inclusive:results} and the uncertainties introduced by the limited
knowledge of the \Dstst peak positions and width described above. In the fit to the $m(\Dzb\pim)$ distribution the yield of the \Dzstarm
component is compatible with zero. Therefore, instead of calculating a branching fraction, an upper limit at \num{90}\% confidence level (CL) is set. We create \num{2000} new data samples by bootstrapping~\cite{Bootstrap} the original data (randomly selecting events, each with its corresponding weight, while allowing repetition of the events). The $\Dzb\pim$ mass fit is performed for each sample. The \num{90}\% CL upper limit on the yield is the value that is higher than that found in \num{90}\% of the samples in which a positive \Dzstarm yield is obtained. This yield is then converted into the upper limit.
\begin{table}[ht]
\caption{Fitted \Dstst yields, statistical significances, and branching fractions for the \Dpi final state. The statistical significance is calculated as $\mathcal{S} = \sqrt{2\Delta \mathcal{L}}$, where $\Delta \mathcal{L}$ is the difference between the log-likelihood value of the nominal fit and of a fit with the signal yield fixed to zero.}
\centering
\setlength{\tabcolsep}{10pt}
\begin{tabular}{lSSc}
\toprule
                                                    &   {yield}   & $\mathcal{S}$   &   branching fraction [\si{\%}]     \\
\midrule
\BdToDzstarlnu with \decay{\Dzstarm}{\Dzb\pim}      &   {-}       & {-}   &   \num{<0.044} at \num{90}\% CL \\
\BdToDtwostarlnu with \decay{\Dtwostarm}{\Dzb\pim}  &   457\pm45  & 25.2  &   \num[parse-numbers=false]{0.157\pm0.015\stat\pm0.005\syst} \\
other \BdToDzpilnu                                  &   547\pm45  & {-}   &   -   \\
\BuToDzstarlnu with \decay{\Dzstarzb}{\Dm\pip}      &   180\pm72  & 3.9   &   \num[parse-numbers=false]{0.054\pm0.022\stat\pm0.005\syst} \\
\BuToDtwostarlnu with \decay{\Dtwostarzb}{\Dm\pip}  &   590\pm39  & 24.9  &   \num[parse-numbers=false]{0.163\pm0.011\stat\pm0.007\syst} \\
other \BuToDmpilnu                                  &   520\pm70  & {-}   &   -   \\
\bottomrule
\end{tabular}
\label{tab:exclusive:dpi}
\end{table}
The results for the decays via the \Dtwostar resonance are compatible with the
world averages. They constitute the most precise measurements of these
branching fractions to date. On the other hand, the value for \BR
(\BuToDzstarlnu) $\times$ \BR(\decay{\Dzstarzb}{\Dm\pip}) is significantly
smaller than previous measurements. This applies even more so to the \Bd
mode, where no contribution could be found in this analysis.

\begin{figure}[ht]
\hspace*{\fill}
\begin{overpic}
[width=0.48\textwidth]{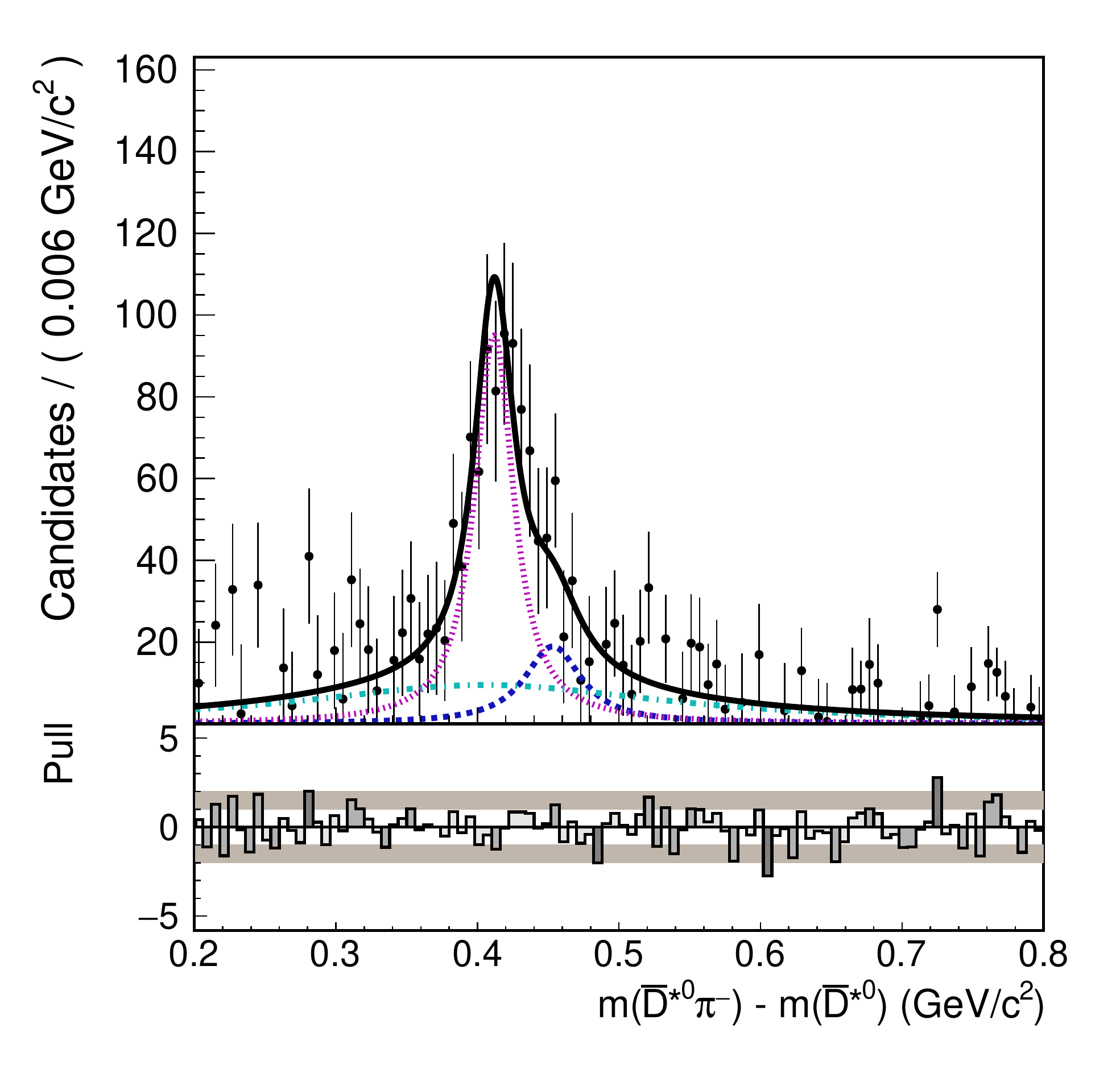}
\put(45,190){Belle}
\put(105,190){\BdToDstpilnu}
\put(115,175){\textcolor{black}{\rule{5mm}{2pt}}\,\,\scalefont{0.8}Total}
\put(115,160){\textcolor{nice_purple}{\rule{5mm}{2pt}}\,\,\scalefont{0.8}\Donem}
\put(115,145){\textcolor{nice_turquoise}{\rule{5mm}{2pt}}\,\,\scalefont{0.8}\Dprimeonem}
\put(115,130){\textcolor{nice_blue}{\rule{5mm}{2pt}}\,\,\scalefont{0.8}\Dtwostarm}
\end{overpic}
\hfill
\begin{overpic}
[width=0.48\textwidth]{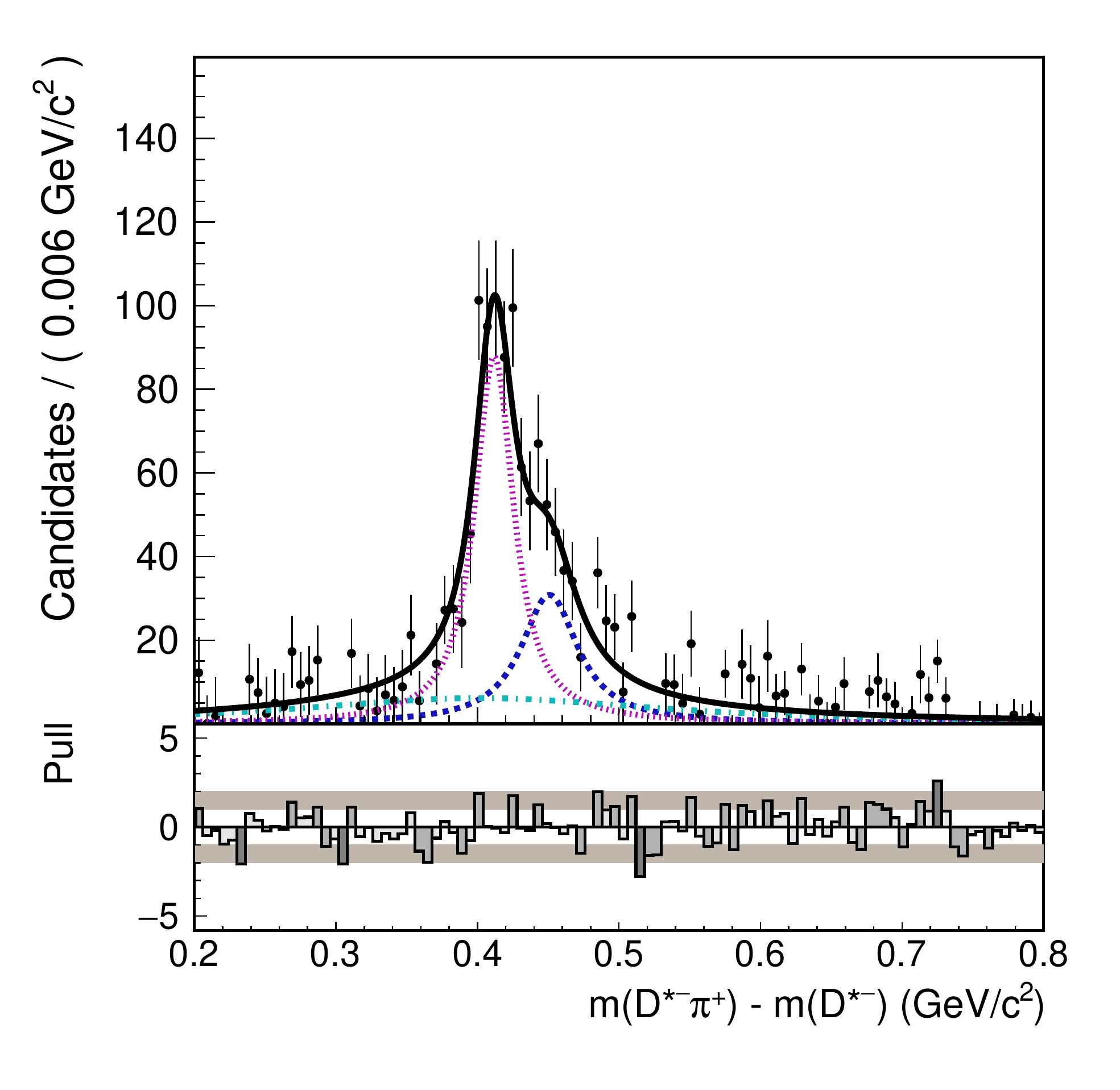}
\put(45,190){Belle}
\put(105,190){\BuToDstpilnu}
\put(115,175){\textcolor{black}{\rule{5mm}{2pt}}\,\,\scalefont{0.8}Total}
\put(115,160){\textcolor{nice_purple}{\rule{5mm}{2pt}}\,\,\scalefont{0.8}\Donezb}
\put(115,145){\textcolor{nice_turquoise}{\rule{5mm}{2pt}}\,\,\scalefont{0.8}\Dprimeonezb}
\put(115,130){\textcolor{nice_blue}{\rule{5mm}{2pt}}\,\,\scalefont{0.8}\Dtwostarzb}
\end{overpic}
\hspace*{\fill}
\caption{Distribution of the mass difference $m(\Dstpi) - m(\Dstar)$ of \BdToDstpilnu
 (left) and \BuToDstpilnu (right) reconstruction after applying signal
 weights determined from a fit of the $U$ distribution using the sPlot
 technique.}
\label{fig:exclusive:mdstpi}
\end{figure}

Three \Dstst resonances are known for the \Dstpi final
state, \Done, \Dprimeone, and \Dtwostar. The three components are
parametrized with Breit-Wigner functions convolved with a Gaussian. The shape
parameters of the two narrow resonances \Done and \Dtwostar are constrained within Gaussian distributions to their world averages~\cite{PDG2022}, while the peak position and width of the broad \Dprimeone resonance is fixed to its world average. Instead of fitting
the $m(\Dstpi)$ mass directly the invariant mass of the \Dstar is subtracted. This allows to conveniently incorporate the feeddown component as well. By subtracting the invariant mass of the \D meson from $m(\Dpi)$ the peaks align.
We perform the fit in the range \SIrange{0.2}{0.8}{\GeVcc}. The data and
the overlaid fit projections are shown in \cref{fig:exclusive:mdstpi}.
The yields of the three components and the resulting branching fractions are
listed in \cref{tab:exclusive:dstpi}. The systematic
uncertainty is dominated by the shape uncertainties. It is determined by fitting twice, once with the shape parameters floating and once fixed.
\begin{table}[ht]
\caption{Fitted \Dstst yields, statistical significances, and branching fractions for the \Dstpi final state.}
\centering
\setlength{\tabcolsep}{10pt}
\begin{tabular}{lS[table-format = 3.0(3)]Sc}
\toprule
                                                          &   {yield}   & $\mathcal{S}$    &   branching fraction [\si{\%}]     \\
\midrule
\BdToDonelnu with \decay{\Donem}{\Dstarzb\pim}            &   866\pm142 & 25.3 &   \num[parse-numbers=false]{0.306\pm0.050\stat\pm0.028\syst} \\
\BdToDoneprimelnu with \decay{\Dprimeonem}{\Dstarzb\pim}  &   523\pm173 & 17.3 &   \num[parse-numbers=false]{0.206\pm0.068\stat\pm0.025\syst} \\
\BdToDtwostarlnu with \decay{\Dtwostarm}{\Dstarzb\pim}    &   145\pm114 & 4.4  &   \num[parse-numbers=false]{0.051\pm0.040\stat\pm0.010\syst} \\
\BuToDonelnu with \decay{\Donezb}{\Dstarm\pip}            &   {$698\pm\phantom{1}65$}  & 24.2 &   \num[parse-numbers=false]{0.249\pm0.023\stat\pm0.014\syst} \\
\BuToDoneprimelnu with \decay{\Dprimeonezb}{\Dstarm\pip}  &   {$353\pm\phantom{1}93$}  & 13.3 &   \num[parse-numbers=false]{0.138\pm0.036\stat\pm0.008\syst} \\
\BuToDtwostarlnu with \decay{\Dtwostarzb}{\Dstarm\pip}    &   {$382\pm\phantom{1}74$}  & 11.8  &   \num[parse-numbers=false]{0.137\pm0.026\stat\pm0.009\syst} \\
\bottomrule
\end{tabular}
\label{tab:exclusive:dstpi}
\end{table}
The results for the decays via the narrower \Done and \Dtwostar resonances are
compatible with previous measurements and the world averages. For the decay
via the wider \Dprimeone resonance the branching fractions are measured \num
{35}\% (\num{50}\%) lower than the world average in the \Bd (\Bu) mode.

\begin{figure}[ht]
\hspace*{\fill}
\begin{overpic}
[width=0.48\textwidth]{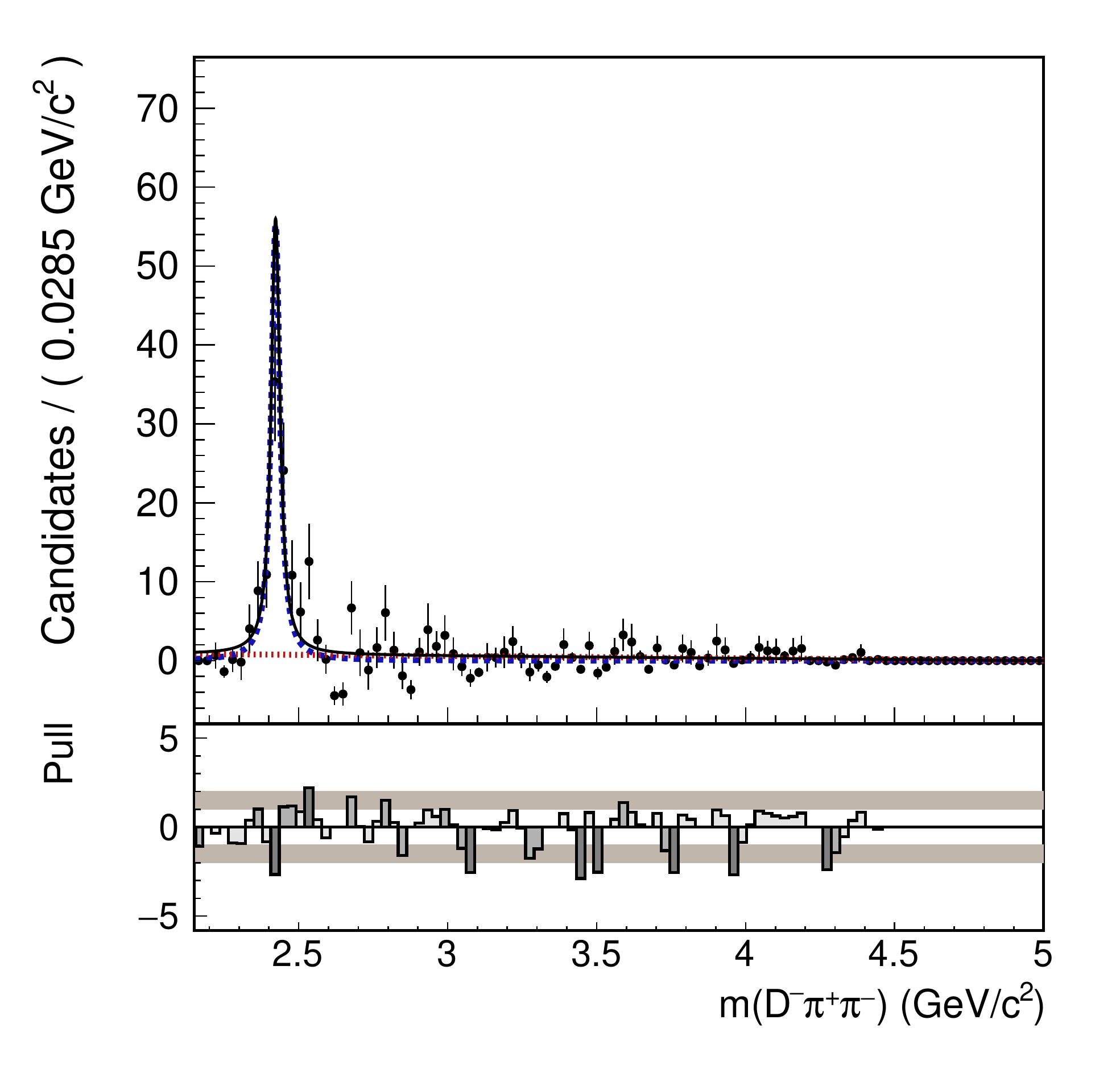}
\put(45,190){Belle}
\put(105,190){\BdToDmpipilnu}
\put(115,175){\textcolor{black}{\rule{5mm}{2pt}}\,\,\scalefont{0.8}Total}
\put(115,160){\textcolor{nice_blue}{\rule{5mm}{2pt}}\,\,\scalefont{0.8}\Donem}
\put(115,145){\textcolor{nice_red}{\rule{5mm}{2pt}}\,\,\scalefont{0.8}other \Dm\pipi}
\end{overpic}
\hfill
\begin{overpic}
[width=0.48\textwidth]{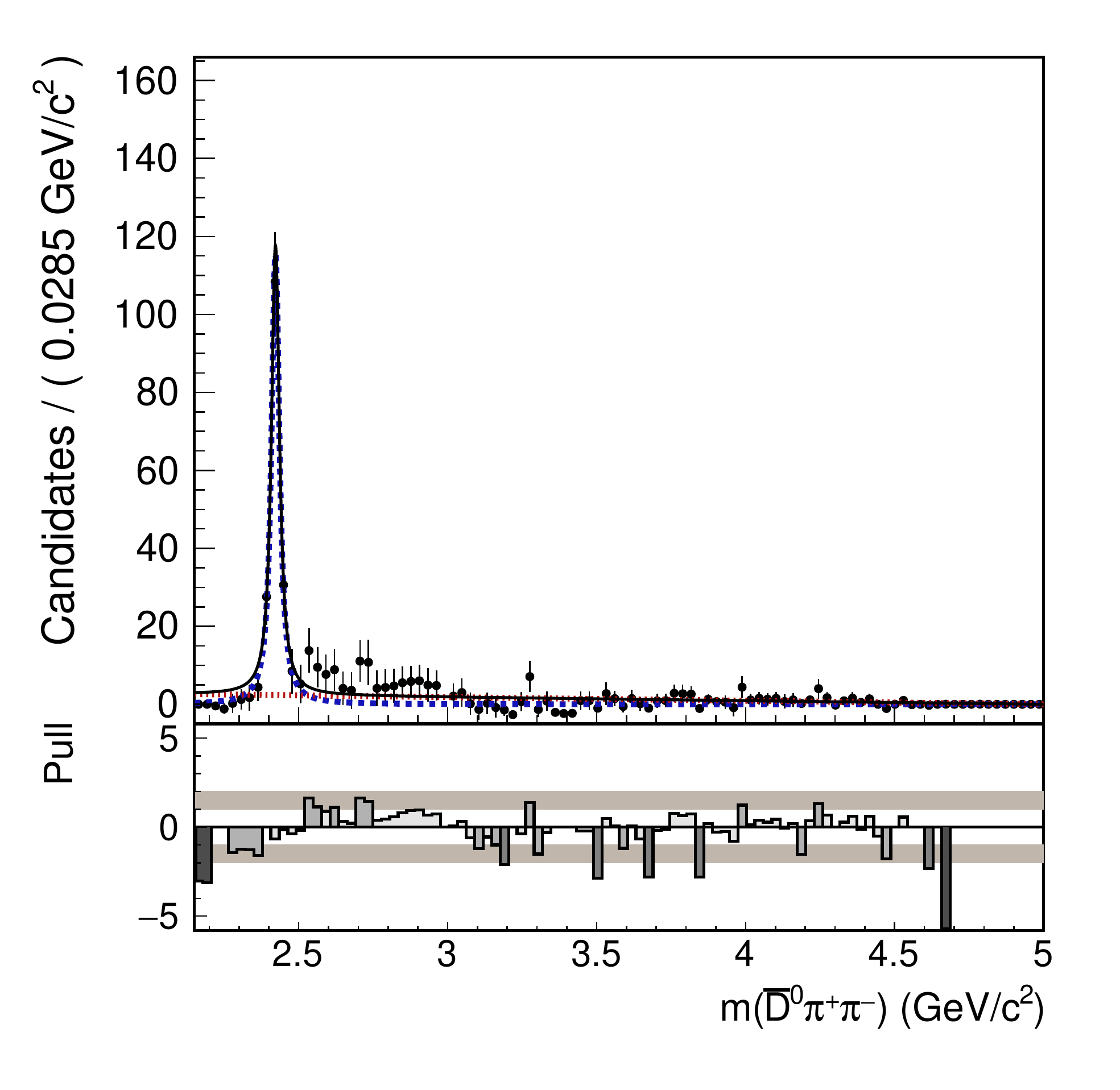}
\put(45,190){Belle}
\put(105,190){\BuToDzpipilnu}
\put(115,175){\textcolor{black}{\rule{5mm}{2pt}}\,\,\scalefont{0.8}Total}
\put(115,160){\textcolor{nice_blue}{\rule{5mm}{2pt}}\,\,\scalefont{0.8}\Donezb}
\put(115,145){\textcolor{nice_red}{\rule{5mm}{2pt}}\,\,\scalefont{0.8}other \Dzb\pipi}
\end{overpic}
\hspace*{\fill}
\caption{Invariant $m(\Dpipi)$ mass distribution of \BdToDmpipilnu
 (left) and \BuToDzpipilnu (right) reconstruction after applying signal
 weights determined from a fit of the $U$ distribution using the sPlot
 technique. A fit function consisting of a Gaussian and a first-order polynomial is overlaid.}
\label{fig:exclusive:mdpipi_fit}
\end{figure}

The weighted unbinned maximum likelihood fit to the $m(\Dpipi)$ distribution
is performed in the range \SIrange{2.15}{5}{\GeVcc} (see \cref
{fig:exclusive:mdpipi_fit}). Initially, the fit model consists of a single Gaussian and
a first-order polynomial.
The fitted peak position and width are compatible with the \Done resonance for the \Bd
and \Bu modes. Therefore, the Gaussian component is interpreted as \decay
{\Bd}{\Donem\ellp\neul} with \DonemToDpipi and \decay{\Bu}
{\Donezb\ellp\neul} with \DonezbToDpipi, respectively. The peaking component
is replaced with a Breit-Wigner function convolved with a Gaussian. The
shape parameters of the Breit-Wigner are set to the PDG values, but allowed
to float within a Gaussian constraint. We find \num{103
(13)} events for the \Bd mode and \num{197(20)} events for the \Bu mode. By
comparing the log-likelihood with a fit, in which the \Done yield is fixed to
zero, the statistical significance is determined to be \num{17.3} for the \Bd
mode and \num{25.1} for the \Bu mode. The
remaining signal events (\num{42(13)} events in the \Bd mode and \num{131(20)} events in the \Bu mode), which are parametrized with the polynomial, can
either be a non-resonant decay process or a decay via a very broad resonance,
such as the \Dzstar or \Dprimeone. However, with our statistical power we can
only state that there must be at least one additional process besides the
decay via the \Done resonance, but cannot characterize it further. The \Done
yields are converted into the following branching fractions:
\begin{align}
\BR(\BdToDonelnu) \times \BR(\DonemToDpipi) &= \num[parse-numbers=false]{(0.102\pm0.013\stat\pm0.009\syst)}\%\\
\BR(\BuToDonelnu) \times \BR(\DonezbToDpipi) &= \num[parse-numbers=false]{(0.105\pm0.011\stat\pm0.008\syst)}\%
\end{align}
This is the first observation of these decay modes.

\section{Conclusion}

In conclusion, using hadronic tagging, we have measured the \BToDorDstpilnu
and \mbox{\BToDorDstpipilnu} branching fractions, achieving the highest precision
to date (except for \BdToDstpipilnu). These results were obtained from a data
sample that contains $772 \times 10^6 B\bar{B}$ pairs collected near the
$\FourS$ resonance with the Belle detector at the KEKB asymmetric energy
$\epem$ collider. All values are compatible with the previous world averages.
Furthermore, the mass spectra of the hadronic final state particles were
studied after statistically subtracting the background contributions. We have
extracted several exclusive \BToDststlnu branching fractions including the
first observations of \BToDonelnu with \DoneToDpipi.

\newpage


\section{Acknowledgments}

This work, based on data collected using the Belle detector, which was
operated until June 2010, was supported by
the Ministry of Education, Culture, Sports, Science, and
Technology (MEXT) of Japan, the Japan Society for the
Promotion of Science (JSPS), and the Tau-Lepton Physics
Research Center of Nagoya University;
the Australian Research Council including grants
DP180102629, 
DP170102389, 
DP170102204, 
DE220100462, 
DP150103061, 
FT130100303; 
Austrian Federal Ministry of Education, Science and Research (FWF) and
FWF Austrian Science Fund No.~P~31361-N36;
the National Natural Science Foundation of China under Contracts
No.~11675166,  
No.~11705209;  
No.~11975076;  
No.~12135005;  
No.~12175041;  
No.~12161141008; 
Key Research Program of Frontier Sciences, Chinese Academy of Sciences (CAS), Grant No.~QYZDJ-SSW-SLH011; 
the Ministry of Education, Youth and Sports of the Czech
Republic under Contract No.~LTT17020;
the Czech Science Foundation Grant No. 22-18469S;
Horizon 2020 ERC Advanced Grant No.~884719 and ERC Starting Grant No.~947006 ``InterLeptons'' (European Union);
the Carl Zeiss Foundation, the Deutsche Forschungsgemeinschaft, the
Excellence Cluster Universe, and the VolkswagenStiftung;
the Department of Atomic Energy (Project Identification No. RTI 4002) and the Department of Science and Technology of India;
BSF and ISF (Israel);
the Istituto Nazionale di Fisica Nucleare of Italy;
National Research Foundation (NRF) of Korea Grant
Nos.~2016R1\-D1A1B\-02012900, 2018R1\-A2B\-3003643,
2018R1\-A6A1A\-06024970, RS\-2022\-00197659,
2019R1\-I1A3A\-01058933, 2021R1\-A6A1A\-03043957,
2021R1\-F1A\-1060423, 2021R1\-F1A\-1064008, 2022R1\-A2C\-1003993;
Radiation Science Research Institute, Foreign Large-size Research Facility Application Supporting project, the Global Science Experimental Data Hub Center of the Korea Institute of Science and Technology Information and KREONET/GLORIAD;
the Polish Ministry of Science and Higher Education and
the National Science Center;
the Ministry of Science and Higher Education of the Russian Federation, Agreement 14.W03.31.0026, 
and the HSE University Basic Research Program, Moscow; 
University of Tabuk research grants
S-1440-0321, S-0256-1438, and S-0280-1439 (Saudi Arabia);
the Slovenian Research Agency Grant Nos. J1-9124 and P1-0135;
Ikerbasque, Basque Foundation for Science, Spain;
the Swiss National Science Foundation;
the Ministry of Education and the Ministry of Science and Technology of Taiwan;
and the United States Department of Energy and the National Science Foundation.
These acknowledgements are not to be interpreted as an endorsement of any
statement made by any of our institutes, funding agencies, governments, or
their representatives.
We thank the KEKB group for the excellent operation of the
accelerator; the KEK cryogenics group for the efficient
operation of the solenoid; and the KEK computer group and the Pacific Northwest National
Laboratory (PNNL) Environmental Molecular Sciences Laboratory (EMSL)
computing group for strong computing support; and the National
Institute of Informatics, and Science Information NETwork 6 (SINET6) for
valuable network support.

\clearpage

\appendix
\section{Additional fit results}
\label{appendix}

\begin{table}[ht]
\caption{Fitted mean and width of Gaussian used to smear signal $U$ templates.}
\centering
\begin{tabular}{lSSS[table-format = 2.1(2)]S[table-format = 2.1(2)]}
\toprule
            & \multicolumn{2}{c}{Mean [\si{\mev}]}  & \multicolumn{2}{c}{$\sigma$ [\si{\mev}]}  \\
                  & {Electron mode} & {Muon mode}   & {Electron mode} & {Muon mode} \\
\midrule
\BdToDmlnu        & 2.7\pm1.0       & 1.4\pm0.9     & 7.7\pm2.1       & 9.7\pm2.1   \\
\BuToDzlnu        & 0.7\pm0.7       & -0.3\pm0.7    & 10.7\pm1.8      & 9.7\pm2.5   \\
\BdToDstlnu       & 1.3\pm0.8       & 1.4\pm0.8     & 9.3\pm3.3       & 10.5\pm1.7  \\
\BuToDstlnu       & -0.4\pm1.8      & -0.1\pm1.5    & 22.5\pm3.2      & 16.8\pm2.8  \\
\BdToDzpilnu      & 9.9\pm0.8       & 2.4\pm2.9     & 0.5\pm0.5       & 4\pm6       \\
\BuToDmpilnu      & 5.4\pm2.4       & 3.4\pm3.3     & 13\pm5          & 23\pm6      \\
\BuToDstpilnu     & 5.1\pm2.8       & 7.1\pm3.0     & 15\pm5          & 15\pm5      \\
\bottomrule
\end{tabular}
\label{tab:smearing}
\end{table}


\begin{thebibliography}{99}
\bibitem{CC}
Throughout this paper, 
the inclusion of the charge conjugate mode decay is implied
unless otherwise stated.
\end{thebibliography}
\end{document}